\documentclass[lettersize,journal]{IEEEtran}
\usepackage{amsmath,amsfonts}
\usepackage{algorithmic}
\usepackage{algorithm}
\usepackage{array}
\usepackage[caption=false,font=normalsize,labelfont=sf,textfont=sf]{subfig}
\usepackage{textcomp}
\usepackage{stfloats}
\usepackage{url}
\usepackage{verbatim}
\usepackage{graphicx}
\usepackage{cite}
\usepackage{multirow} 
\usepackage{makecell}
\usepackage{booktabs}
\usepackage{colortbl}
\usepackage[table ]{ xcolor}
\usepackage{graphicx}
\usepackage{epstopdf}
\usepackage{orcidlink}
\usepackage{amssymb}
\usepackage{stfloats}
\usepackage{bbding}
\usepackage{color}
\hypersetup{
	colorlinks=true,
	linkcolor=black,
	citecolor=black
}
\usepackage{stfloats}
\usepackage{float}
\usepackage{xcolor}
\usepackage{footnote}
\usepackage{CJKutf8}
\usepackage{boldline}
\usepackage{tikz} 
\usepackage{wasysym}
\usepackage{supertabular}
\usepackage{longtable}
\usepackage{xcolor}
\usepackage{tikz}
\newcommand*\emptycirc[1][1ex]{\tikz\draw (0,0) circle (#1);} 
\newcommand*\halfcirc[1][1ex]{%
	\begin{tikzpicture}
		\draw[fill] (0,0)-- (90:#1) arc (90:270:#1) -- cycle ;
		\draw (0,0) circle (#1);
\end{tikzpicture}}
\newcommand*\fullcirc[1][1ex]{\tikz\fill (0,0) circle (#1);}

\begin{document}

\title{Generative AI-Empowered Secure Communications\\ in Space-Air-Ground Integrated Networks:\\ A Survey and Tutorial}

\author{Chenbo Hu, Ruichen Zhang,~\IEEEmembership{Member,~IEEE},  Bo Li,~\IEEEmembership{Member,~IEEE}, Xu Jiang,~\IEEEmembership{Member,~IEEE},\\ 
	Nan Zhao,~\IEEEmembership{Senior Member,~IEEE}, Marco Di Renzo,~\IEEEmembership{Fellow,~IEEE}, Dusit Niyato,~\IEEEmembership{Fellow,~IEEE}, \\Arumugam Nallanathan,~\IEEEmembership{Fellow,~IEEE}, and George K. Karagiannidis,~\IEEEmembership{Fellow,~IEEE}
	
	\vspace*{-2em}
	%
	\thanks{Chenbo Hu, Bo Li, and Xu Jiang are with the School of Information Science and Engineering, Harbin Institute of Technology (Weihai), Weihai 264209, China (e-mails: huchenbo@stu.hit.edu.cn; libo1983@hit.edu.cn; xjiang@hit.edu.cn).
		
	Ruichen Zhang and Dusit Niyato are with the College of Computing and Data Science, Nanyang Technological University, Singapore (e-mails: ruichen.zhang@ntu.edu.sg; dniyato@ntu.edu.sg).	
			
	Nan Zhao is with the School of Information and Communication Engineering, Dalian University of Technology, Dalian 116024, China (e-mail: zhaonan@dlut.edu.cn).
	
	Marco Di Renzo is with Université Paris-Saclay, CNRS, Centrale
	Supélec, Laboratoire des Signaux et Systèmes, 91192 Gif-sur-Yvette, France,
	and also with King’s College London, Centre for Telecommunications
	Research—Department of Engineering, WC2R 2LS London, U.K. (e-mail:
	marco.di-renzo@universite-paris-saclay.fr; marco.di\_renzo@kcl.ac.uk).
	
	Arumugam Nallanathan is with the School of Electronic Engineering and Computer Science, Queen Mary University of London, London and also with the Department of Electronic Engineering, Kyung Hee University, Yongin-si, Gyeonggi-do 17104, Korea (e-mail: a.nallanathan@qmul.ac.uk).
	
	George K. Karagiannidis is with  Department of Electrical and Computer Engineering, Aristotle University of Thessaloniki, 54124 Thessaloniki, Greece (e-mail: geokarag@auth.gr).
	}
}

%

\maketitle

\begin{abstract}
\textbf{
Space-air-ground integrated networks (SAGINs) face unprecedented security challenges due to their inherent characteristics, such as multidimensional heterogeneity and dynamic topologies. These characteristics fundamentally undermine conventional security methods and traditional artificial intelligence (AI)-driven solutions.
Generative AI (GAI) is a transformative approach that can safeguard SAGIN security by synthesizing data, understanding semantics, and making autonomous decisions.
This survey fills existing review gaps by examining GAI-empowered secure communications across SAGINs.
First, we introduce secured SAGINs and highlight GAI’s advantages over traditional AI for security defenses.
Then, we explain how GAI mitigates failures of authenticity, breaches of confidentiality, tampering of integrity, and disruptions of availability across the physical, data link, and network layers of SAGINs.
Three step-by-step tutorials discuss how to apply GAI to solve specific problems using concrete methods, emphasizing its generative paradigm beyond traditional AI.
Finally, we outline open issues and future research directions, including lightweight deployment, adversarial robustness, and cross-domain governance, to provide major insights into GAI’s role in shaping next-generation SAGIN security.
}
\end{abstract}


\vspace{-1.5em}
\section{Introduction}
\subsection{Toward GAI-Enabled SAGIN Security}
Over the past decade, the advancement of 5G and the prospects for 6G have led to significant progress in space-air-ground integrated networks (SAGINs), including the extensive deployment of low earth orbit (LEO) satellite constellations (e.g., Starlink and OneWeb), the development of high-altitude platform station technology, and the enhancement through software-defined networking (SDN) and network function virtualisation (NFV) technologies \cite{refIntro_A_2}. 
This multi-layer heterogeneous architecture aims to provide global three-dimensional seamless coverae by 2030, fulfilling the rigorous demands of future applications such as intelligent transportation, telemedicine, and smart cities, characterised by high reliability, ultra-low latency, and extensive connectivity. \cite{refIntro_SAGIN}.

SAGIN communication faces numerous intricate security threats stemming from distinct problems, including multi-dimensional heterogeneity, dynamic topology, and resource limitations, in contrast to conventional single-domain networks (e.g., terrestrial and aerial networks) \cite{refI_B_1}.
This significantly enhances the complexity of security measures implemented for such cross-domain systems. 
The ephemeral connectivity of cross-domain nodes makes static authentication susceptible to spoofing attacks. 
Furthermore, coordinated assaults across the physical, data link, and network layers are being escalated. 
Resource-limited nodes find it challenging to facilitate high-intensity real-time integrity verification. 
Consequently, the fundamental security needs of SAGIN communication—authenticity, confidentiality, integrity, and availability—are susceptible to risks of imbalance. 
Thus, guaranteeing secure communications in SAGINs is of paramount importance \cite{ref_security_sagin1} \cite{ref_security_sagin2}\cite{refII_A_1}\cite{refII_A_8}.

However, security methods, as encryption and intrusion detection, frequently prove inadequate in the dynamic and heterogeneous SAGINs. 
Conventional artificial intelligence (AI) methodologies, such as deep learning (DL) and deep reinforcement learning (DRL), offer improved functionalities for anomaly detection, threat identification, and adaptive security strategies; however, they demonstrate considerable shortcomings in the context of SAGIN security. 
It necessitates substantial quantities of labelled samples for training; however, SAGIN attack samples are limited and demonstrate non-independent and identically distributed (non-IID) traits, resulting in a significant decline in detection accuracy. 
Architecturally, unimodal models like convolutional neural networks (CNNs) and recurrent neural networks (RNNs) are inadequate in effectively capturing cross-domain attack features.
Moreover, inadequate dynamic flexibility required regular retraining to address emerging threats, hence failing to satisfy real-time anti-jamming demands. 
These constraints hinder the efficacy of conventional AI in safeguarding SAGINs \cite{refI_B_2}.

Generative AI (GAI), as a prominent subset of AI, presents an innovative framework for addressing security limitations in SAGINs, using its robust skills in data production, semantic comprehension, and autonomous decision-making \cite{refIntro_1}.
In contrast to conventional AI, which primarily analyses, interprets, and classifies data to address specific issues, GAI has superior capability in analysing and recording intricate multidimensional data distributions, facilitating the creation of highly correlated new data (e.g., photos, text, video) \cite{refIntro_4}.
Specifically, although traditional AI depends on past data to identify known risks, GAI may generate variations of unknown attacks by analysing data distributions, thereby mitigating the lack of labelled data.
Moreover, transformer models of GAI utilise multi-head attention techniques to correlate cross-domain traffic, hence identifying multi-domain attack patterns overlooked by unimodal CNNs.
It utilises real-time dynamic defence via prompt engineering (PE) and in-context learning (ICL) to enable millisecond-level anti-jamming responses. 
In summary, GAI not only mitigates the intrinsic deficiencies of static defences and conventional AI but also establishes a proactive-adaptive defence system specifically designed for heterogeneous SAGINs through privacy-preserving synthesis, cross-domain attack simulation, high-fidelity reconstruction, and semantic strategy generation.

This paper offers thorough insights into how GAI can efficiently tackle the substantial security difficulties posed by SAGINs.

\subsection{Related Surveys and Contributions}

\begin{table*}[t]
	\vspace*{-1em}
	\renewcommand{\arraystretch}{2}
	\centering
	\caption{Summary of Related Surveys, Where \  \fullcirc,\ \halfcirc,\ and \emptycirc \ Represent Comprehensive Review, Partial Review, and Not Review, Respectively}
	\setlength{\tabcolsep}{0.8pt}
	\renewcommand{\arraystretch}{1}
		\begin{tabular}{|p{0.8cm} |p{12cm}|p{1.2cm}|p{1.2cm}|p{1.2cm}|}
			\rowcolor{gray!20}
			\Xhline{0.7pt}
			\makecell[c]{\textbf{{Ref.}}}
			&\makecell[c]{\textbf{{Contribution}}} 
			&\makecell[c]{\textbf{{SAGIN}}} 
			&\makecell[c]{\textbf{{Security}}} 
			&\makecell[c]{\textbf{{GAI}}} \\	
			\Xhline{0.7pt}
			\makecell[c]{\cite{refI_B_1}}&\makecell[l]{A survey of security threats, attack methods, and defense schemes in SAGINs, addressing the unique \\vulnerabilities of heterogeneous SAGINs and explicitly covering traditional AI-enabled defenses.}  &\makecell[c]{  \fullcirc}
			& \makecell[c]{\fullcirc}& \makecell[c]{\emptycirc}
			\\
			\Xhline{0.7pt}	
					
			\makecell[c]{\cite{refI_B_2}}&\makecell[l]{A survey of AI applications across satellite systems, covering use cases, hardware implementation \\challenges, and future directions, with explicit focus on traditional AI with little coverage of GAI.}   &\makecell[c]{  \halfcirc}
			& \makecell[c]{\halfcirc}& \makecell[c]{\halfcirc}
			\\	
			\Xhline{0.7pt}	
			
			\makecell[c]{\cite{refI_B_3}}&\makecell[l]{A review of AI applications, including generative models for security enhancement, across diverse\\ satellite communication challenges such as anti-jamming, interference management.}   &\makecell[c]{  \halfcirc}
			& \makecell[c]{\halfcirc}& \makecell[c]{\halfcirc}
			\\
			\Xhline{0.7pt}
						
			\makecell[c]{\cite{refI_B_4}}&\makecell[l]{A review of AI applications and challenges in 6G UAV-satellite  networks, systematically classifying \\and comparing solutions for security risks including traditional AI and GAI.}   &\makecell[c]{  \halfcirc}
			& \makecell[c]{\halfcirc}& \makecell[c]{\halfcirc}
			\\	
			\Xhline{0.7pt}
			
			\makecell[c]{\cite{refI_B_5}}&\makecell[l]{
			A survey of GAI applications for physical layer communication security, covering defense mechanisms\\ like RF authentication, anomaly detection, and anti-jamming.}   &\makecell[c]{  \emptycirc}
		& \makecell[c]{\halfcirc}& \makecell[c]{\fullcirc}
			\\
			\Xhline{0.7pt}	
			
			\makecell[c]{\cite{refI_B_6}}&\makecell[l]{
			A survey of physical layer security for low-altitude economy networks, highlighting the application \\of GAI for enhancing anti-eavesdropping strategies, anomaly detection, and optimizing defenses.}    &\makecell[c]{  \halfcirc}
		& \makecell[c]{\fullcirc}& \makecell[c]{\halfcirc}
			\\	
			\Xhline{0.7pt}	
			
			\makecell[c]{\cite{refI_B_7}}&\makecell[l]{A survey of GAI models addressing security issues in physical layer communications, covering \\confidentiality, authentication, availability, resilience, and integrity.}  &\makecell[c]{  \emptycirc}
			& \makecell[c]{\fullcirc}& \makecell[c]{\fullcirc}
			\\
			\Xhline{0.7pt}
			
			\makecell[c]{\cite{refI_B_8}}&\makecell[l]{A review of the fundamentals, applications, and challenges of GAI in mobile networking, with a\\ specific focus on  advancing security solutions (e.g., intrusion detection and jamming mitigation).}   &\makecell[c]{  \halfcirc}
			& \makecell[c]{\halfcirc}& \makecell[c]{\fullcirc}
			\\	
			\Xhline{0.7pt}
			
			\makecell[c]{\cite{refI_B_9}}&\makecell[l]{A survey on deploying AIGC services in mobile edge-cloud networks, covering architectures,\\ generative models, and  use cases, with a focus on low-latency, privacy, and resource efficiency.} &\makecell[c]{  \halfcirc}
			& \makecell[c]{\halfcirc}& \makecell[c]{\fullcirc}
			\\
			\Xhline{0.7pt}	
			
			\makecell[c]{\cite{refI_B_10}}&\makecell[l]{A review of GAI applications within the IoEV, highlighting its critical role in addressing security \\threats like adversarial attacks, and cyber-physical anomaly detection.}  &\makecell[c]{  \halfcirc}
			& \makecell[c]{\halfcirc}& \makecell[c]{\fullcirc}
			\\	
			\Xhline{0.7pt}	
			
			\makecell[c]{\cite{refI_B_11}}&\makecell[l]{A survey on deep learning in anomaly detection, highlighting the critical role of generative models for \\both data augmentation to and enhanced anomaly identification in security-critical domains like IoT.} &\makecell[c]{  \emptycirc}
			& \makecell[c]{\fullcirc}& \makecell[c]{\fullcirc}
			\\
			\Xhline{0.7pt}
			
			\makecell[c]{\cite{refI_B_12}}&\makecell[l]{A review of privacy and security concerns in GAI from, highlighting GAI's potential applications in\\ solving security problems, such as anomaly detection and cybersecurity threat identification.}  &\makecell[c]{  \emptycirc}
			& \makecell[c]{\fullcirc}& \makecell[c]{\fullcirc}
			\\	
			\Xhline{0.7pt}
			
			\makecell[c]{\cite{refI_B_13}}&\makecell[l]{A survey on security and privacy challenges in generative data from AIGC, highlighting the\\ applications of GAI in developing methods such as data synthesis and multimodal defense frameworks.}  &\makecell[c]{  \halfcirc}
			& \makecell[c]{\fullcirc}& \makecell[c]{\fullcirc}
			\\
			\Xhline{0.7pt}
			
			\makecell[c]{\cite{refI_B_14}}&\makecell[l]{A survey of architectures, applications, and security challenges of LLM-based edge intelligence,\\ highlighting GAI's role in optimizing security solutions e.g., threat detection.}   &\makecell[c]{  \emptycirc}
			& \makecell[c]{\halfcirc}& \makecell[c]{\fullcirc}
			\\	
			\Xhline{0.7pt}

			\makecell[c]{\textbf{Our}\\ \textbf{paper}}&\makecell[l]{Focus on how GAI can address various security issues in SAGINs and discuss the development\\ potential of GAI in SAGIN secure communications compared to traditional AI.}   &\makecell[c]{ \fullcirc}
			& \makecell[c]{\fullcirc}& \makecell[c]{\fullcirc}
			\\	
			\Xhline{0.7pt}
			
		\end{tabular}
	\label{table_related_surveys}
\end{table*}

The security challenges and AI-driven methodologies for SAGINs and its sub-networks have garnered significant interest, as demonstrated in Table \ref{table_related_surveys}.
The research in \cite{refI_B_1} offered an extensive examination of security threats, attack methodologies, and defensive strategies, focussing on the distinct vulnerabilities in heterogeneous SAGINs and partially addressing conventional AI-based defence mechanisms.
Since that time, specialised AI review functions for SAIGNs or their sub-networks have been developing.
For instance, in \cite{refI_B_2}, the utilisation of AI in satellite communications was thoroughly examined, concentrating mostly on conventional AI techniques for various security issues while providing minimal insight into GAI methodologies. 
The research in \cite{refI_B_3} specifically investigated AI applications, encompassing generative models for security improvement, in various satellite communication issues like beam hopping, anti-jamming, channel modelling, and space-air-ground integration.
The authors of \cite{refI_B_4} thoroughly delineated the uses, problems, and future possibilities of AI in 6G UAV-satellite communication networks, carefully categorising and contrasting several AI solutions for security threats, encompassing classical AI and GAI.

Furthermore, certain surveys highlight the GAI for secure physical layer communication.  The research in \cite{refI_B_5} examined GAI applications for the security of physical layer communication, encompassing defence strategies such as authentication, anomaly detection, and anti-jamming, while also tackling new dangers posed by GAI-driven adversarial attacks.
\cite{refI_B_6} examined secure physical layer communication methods for low-altitude economic networking, emphasising the novel application of GAI to improve anti-eavesdropping tactics, anomaly detection, and the optimisation of security measures.
Moreover, the authors in \cite{refI_B_7} presented an exhaustive survey of GAI models that tackle security concerns in physical layer communications, encompassing confidentiality, authentication, availability, resilience, and integrity.

Turning to mobile networks, \cite{refI_B_8} comprehensively reviewed the fundamentals, applications, and challenges of GAI in mobile and wireless networking, partially advancing wireless security solutions (e.g., intrusion detection, jamming mitigation, and generative steganography).
\cite{refI_B_9} investigated the AI-generated content (AIGC) services in mobile edge-cloud networks, including architectures, generative models, implementation challenges, and real-world use cases, with a focus on low-latency, privacy, and resource efficiency. 
Targeting internet of electric vehicle networks (IoEVs), \cite{refI_B_10} categorized GAI applications across four layers (battery, electric vehicle, grid, and security), highlighting its critical role in addressing security challenges like adversarial attacks and cyber-physical anomaly detection.
\cite{refI_B_11} reviewed deep learning advancements in anomaly detection, highlighting the critical role of generative models for both data augmentation to address class imbalance and enhanced anomaly identification in security-critical domains like IoT and cybersecurity.

Furthermore, \cite{refI_B_12} analyzed privacy and security concerns in GAI from five key perspectives (user, ethical, regulatory, technological, and institutional), while also highlighted GAI's potential applications in solving security problems, such as anomaly detection, cybersecurity threat identification, and enhancing autonomous system safety.
\cite{refI_B_13} systematically reviewed security and privacy challenges in generative data from AIGC, analyzed through the lens of core information security properties (i.e., privacy, controllability, authenticity, and compliance), while emphasizing the transformative applications of GAI in developing innovative countermeasures such as synthetic data synthesis, multimodal defense frameworks, and adaptive security mechanisms.
The study in \cite{refI_B_14} explored architectures, applications, and security challenges of large language model (LLM)-based edge intelligence, highlighting GAI's role in optimizing threat detection, vulnerability management, and automated security solutions for resource-constrained environments.

Unlike existing surveys and tutorials, which focus primarily on GAI applications in sub-networks of SAGINs or on a specific type of GAI-enabled security defense technique, our survey
provides a detailed classification of attacks and GAI defenses across the entire SAGIN, considering security requirements and network architectures.
Our work bridges an essential gap in current research by providing a comprehensive analysis of the status and potential of GAI applications in SAGINs, as well as a full comparison with traditional AI.
Our work explains how GAI enhances SAGIN security compared to traditional AI, a topic that has been previously under-explored.
The contributions of our survey are as follows:
\begin{itemize}
	\item We present the architectures of SAGINs and the security challenges from a security perspective. Meanwhile, we outline the basic GAI models and summarize their suitability for solving security issues and advantages over traditional AI.
	
	\item We investigate four categories of security problems that may be encountered in SAGINs, namely authenticity failures, confidentiality breaches, integrity tampering, and availability disruptions, and outline the corresponding GAI-based approaches, categorized by physical layer, data link layer, and network layer.
	
	\item We implement three tutorials that are employed to detail how GAI addresses the multi-class security issues in SAGINs, highlighting the advantages of GAI compared to traditional AI.
	
	\item We discuss the open issues and future research directions from the perspectives of resource constraints and lightweight deployment, adversarial robustness and trustworthy mechanisms, cross-domain coordination and governance compliance, respectively.
	
\end{itemize}

\begin{figure}[!t]
	\centering
	\includegraphics[width=3.5in]{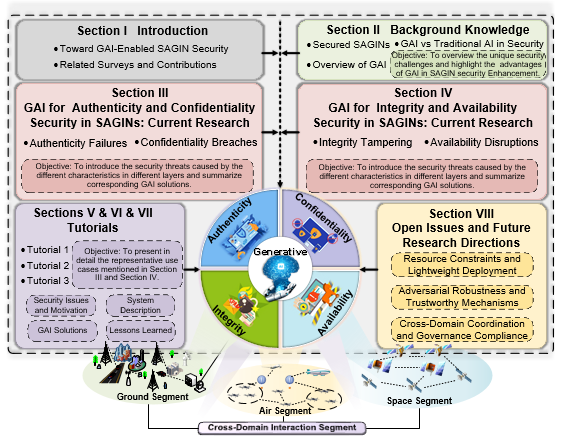}
	\caption{Organization structure of this paper.}
	\label{fig_outline}
\end{figure}

The remainder of this work is outlined in Fig. \ref{fig_outline}.
From a perspective of security, we introdcue the architectures of SAGINs, overview the fundamental concepts of GAI, and compare the traditional AI with GAI in Section II. 
In Section III, a comprehensive exploration of GAI for authenticity failures and confidentiality breaches is presented. In addition, Section IV reviews the GAI solutions for integrity tampering and availability disruptions.
In Sections V, VI, and VII, we conduct and analyze potential tutorials.
Furthermore, we discuss open issues and discuss future research directions in Section VIII followed by Section IX as conclusion.
For visibility, Table \ref{table_acronyms} lists the main acronyms quoted in this survey.

\begin{table*}[t]
	\vspace*{-2em}
	\renewcommand{\arraystretch}{2}
	\centering
	\caption{List of Main Abbreviations.}
	\label{table0}
		\setlength{\tabcolsep}{1pt}
	\renewcommand{\arraystretch}{0.8}
	\begin{tabular}{c|p{5.3cm}||c|p{5cm}}
		\Xhline{1pt}
		\rowcolor{gray!20}
		\multirow{1}{*}{\textbf{{Abbreviation}}} &\makecell[c]{\textbf{Description}}&\multirow{1}{*}{\textbf{Abbreviation}} & \makecell[c]{\textbf{Description}}\\
		
		\Xhline{1pt}
		\makecell[c]{ AAV} & \makecell[l]{Autonomous Aerial Vehicle}&
		\makecell[c]{IDS} & \makecell[l]{Intrusion Detection System} \\
		
		\Xhline{0.5pt}
		\makecell[c]{ADS-B} & \makecell[l]{Automatic Dependent Surveillance-Broadcast}&
		\makecell[c]{IoT} & \makecell[l]{Internet of Thing} \\
		
		\Xhline{0.5pt}
		\makecell[c]{AI} & \makecell[l]{Artificial Intelligence} &
		\makecell[c]{ISL} &\makecell[l]{Inter-Satellite Link} \\
		
		\Xhline{0.5pt}
		\makecell[c]{AIGC} & \makecell[l]{AI-Generated Content} &
		\makecell[c]{LLM} & \makecell[l]{Large Language Model}\\
		
		\Xhline{0.5pt}
		\makecell[c]{ARP} & \makecell[l]{Address Resolution Protocol} &
		\makecell[c]{LSTM} &\makecell[l]{Long Short-Term Memory}\\
		
		\Xhline{0.5pt}
		\makecell[c]{BDS} & \makecell[l]{Beidou Navigation Satellite}&
		\makecell[c]{MAC} & \makecell[l]{Media Access Control} \\
		
		\Xhline{0.5pt}
		\makecell[c]{CIR} & \makecell[l]{Channel Impulse Response}&
		\makecell[c]{MAD} & \makecell[l]{Malicious Attack Detector} \\
		
		\Xhline{0.5pt}
		\makecell[c]{CNN } & \makecell[l]{Convolutional Neural Network} &
		\makecell[c]{MAVLink} &\makecell[l]{Micro Air Vehicle Link} \\
		
		\Xhline{0.5pt}
		\makecell[c]{CSI} & \makecell[l]{Channel State Information} &
		\makecell[c]{MDP} & \makecell[l]{Markov Decision Process}\\
		
		\Xhline{0.5pt}
		\makecell[c]{ DDoS} & \makecell[l]{Distributed Denial of Service} &
		\makecell[c]{MEC} &\makecell[l]{Multi-Access Edge Computing}\\
		\Xhline{0.5pt}
		
		\makecell[c]{DL} & \makecell[l]{Deep Learning}&
		\makecell[c]{MITM} & \makecell[l]{Man-in-the-Middle} \\
		\Xhline{0.5pt}
		
		\makecell[c]{DNN} & \makecell[l]{Deep Neural Network}&
		\makecell[c]{ML} & \makecell[l]{Machine Learning} \\
		
		\Xhline{0.5pt}
		\makecell[c]{DoS } & \makecell[l]{Denial-of-Service} &
		\makecell[c]{MTL} &\makecell[l]{Multi-Task Learning} \\
		
		\Xhline{0.5pt}
		\makecell[c]{DP} & \makecell[l]{Differential Privacy} &
		\makecell[c]{RNN} & \makecell[l]{Recurrent Neural Network}\\
		
		\Xhline{0.5pt}
		\makecell[c]{DRL } & \makecell[l]{Deep Reinforcement Learning} &
		\makecell[c]{RF} &\makecell[l]{Radio Frequency}\\
		
		\Xhline{0.5pt}
		\makecell[c]{GAI} & \makecell[l]{Generative AI}&
		\makecell[c]{SAGIN} & \makecell[l]{Space-Air-Ground Integrated Network} \\
		\Xhline{0.5pt}
		
		\makecell[c]{GAN} & \makecell[l]{Generative Adversarial Network}&
		\makecell[c]{SDR} & \makecell[l]{Software-Defined Radio} \\
		
		\Xhline{0.5pt}
		\makecell[c]{GDM } & \makecell[l]{Generative Diffusion Model} &
		\makecell[c]{SVM} &\makecell[l]{Support Vector Machine} \\
		
		\Xhline{0.5pt}
		\makecell[c]{GNSS} & \makecell[l]{Global Navigation Satellite System} &
		\makecell[c]{TBM} & \makecell[l]{Transformer-Based Model}\\
		
		\Xhline{0.5pt}
		\makecell[c]{GPS } & \makecell[l]{Global Positioning system} &
		\makecell[c]{VAE} &\makecell[l]{Variational Autoencoder}\\

		\Xhline{1pt}	
	\end{tabular}
	\label{table_acronyms}
\end{table*}

\section{Background Knowledge}
In this section, we first introduce the SAGINs from a security perspective, including components and security challenges. Subsequently, we provide an overview of GAI and compare its advantages over traditional AI in addressing SAGIN security issues.

\subsection{Secured SAGINs}
The distinctive characteristics of SAGINs, featured by self-organization, heterogeneity, and time-variability, introduce unprecedented challenges\cite{refII_A_1}. Among them, security vulnerabilities caused by architectural weaknesses, exposed nodes, and unsecured communication links pose critical threats to four fundamental communication security requirements: authenticity, confidentiality, integrity, and availability\cite{refI_B_1}, \cite{refII_A_4}.
\begin{itemize}
	\item \textit{Authenticity:} Authenticity ensures the legitimacy of network entities and the trustworthiness of data sources in SAGINs, and prevents unauthorized access through multi-layered authentication mechanisms\cite{refII_A_Authentication}. 
	Current implementations employ hardware-based identification (e.g., media access control (MAC) address), protocol-layer authentication (e.g., digital certificates), and biometrically-enhanced dynamic credentials\cite{refII_A_5}.
	
	\item \textit{Confidentiality:} Confidentiality ensures that data transmission is accessible exclusively to authorized users. Since the transmitted data contains information about the user's behavior, malicious attackers can indirectly infer sensitive information using unintentionally leaked available messages \cite{refII_A_8}. 
	
	\item \textit{Integrity:} Integrity ensures data accuracy and reliability during transmission, storage, and processing. The attack against integrity is less intense but more sophisticated.
	Attack surfaces extend beyond simple original data modification to include protocol-level interference that disrupts information exchange processes from within network layers \cite{refII_A_9}. 
	
	\item \textit{Availability:} Availability ensures authorized users to access the wireless network whenever and wherever they request it, even under adversarial conditions or system failures \cite{refII_A_10}. Attackers attempt to delay, block or even interrupt transmissions, thus rendering network resources unavailable.
	
\end{itemize}

The following discusses the security challenges and their impact on security requirements across four segments.

\subsubsection{Secured Cross-Domain Interaction Segment}
The cross-domain interaction mechanism consists of dynamic protocol stacks, multi-domain gateways, and intelligent control planes that coordinate the three heterogeneous segments: space, air, and ground \cite{refII_A_11}, \cite{refII_A_12}. 
This framework serves as a vital facilitator for multi-domain coordination in SAGINs \cite{refIntro_A_2}, \cite{refII_A_13}, encountering two main security challenges: protocol heterogeneity \cite{refI_B_1} and vulnerabilities associated with delay-sensitive handovers \cite{refII_A_14}. 
The transient connectivity of dynamic nodes, such as satellites and autonomous aerial vehicles, makes static authentication mechanisms vulnerable to spoofing attacks. In these scenarios, adversaries can impersonate legitimate gateways and introduce false routing information, thereby compromising authenticity \cite{refII_A_15},\cite{refII_A_16}. 
Furthermore, confidentiality risks emerge from vulnerabilities in inter-satellite link (ISL) protocols, allowing eavesdroppers to capture cross-domain session keys through compromised key mapping tables at protocol gateways\cite{refII_A_17}. 
Divergent protocol implementations, such as the Consultative Committee for Space Data Systems (CCSDS), Micro Air Vehicle Link (MAVLink) for aerial platforms, and the 3rd Generation Partnership Project (3GPP) for ground networks, increase the vulnerability of data conversion processes to manipulation.  For example, man-in-the-middle (MITM) attacks as described in \cite{refII_A_19} effectively modified AAV control commands during AAV-to-ground switching operations, thereby undermining data integrity.
Additionally, threats to availability arise from the targeted jamming of cross-domain synchronisation signals, potentially disabling positioning systems in AAV-ground synchronisation \cite{refII_A_20}.

\subsubsection{Secured Space Segment}
The space segment includes satellites, ground stations, and space-based Internet of Things (IoT) devices, utilising technologies such as LEO constellations. 
The broadened network exposure presents security risks, including susceptibility to assaults and data manipulation. 
Attackers can spoof global positioning system (GPS) signals by employing publicly accessible ephemeris data to distort satellite navigation. 
Although satellite laser communications provide narrow beams, high-altitude platforms can capture signals with sophisticated optical tracking, hence jeopardising confidentiality \cite{refII_A_23}. 
Furthermore, the constrained computational capabilities of on-board devices impede real-time high-strength integrity checks \cite{refII_A_24}. 
Moreover, adversaries can anticipate satellite trajectories and inundate satellite uplinks with high-power transmitters, significantly diminishing single-beam throughput\cite{refII_A_25}, \cite{refII_A_26}.  Sleep deprivation can be exploited by attackers to intentionally and persistently transmit erroneous control commands\cite{refII_A_26}, \cite{refII_A_27} to deplete network resources.

\subsubsection{Secured Air Segment}
The air segment includes AAVs, airships, aerial base stations, and airborne sensors, distinguished by dynamic mobility and line-of-sight dependencies\cite{refII_A_28}, \cite{refII_A_29}. 
Key security issues include AAV identity spoofing and the interception of wireless signals\cite{refII_A_30}. 
Malicious nodes can replicate AAV MAC addresses to infiltrate networks and seize control of swarms through deceptive link state announcements, compromising authenticity \cite{refII_A_31}. 
In \cite{refII_A_32}, the radio frequency (RF) fingerprint authentication technique was introduced to extract hardware features of node communication modules to mitigate attacks. 
Millimetre-wave links in legitimate AAVs exhibit vulnerabilities regarding the leakage of unencrypted messages \cite{refII_A_33}, \cite{refII_A_34}. The quantum key distribution technique \cite{refII_A_8} shows potential for mitigating decryption risks. 
Furthermore, node mobility enables MITM attacks that alter AAV commands, thereby undermining integrity \cite{refII_A_19}. 
The restricted computational capacity of on-board devices, akin to space nodes, hinders effective real-time integrity assessments \cite{refII_A_28}.
Distributed denial of service (DDoS) attacks, particularly those involving numerous distributed bot-type attacks, inundate the air network, thereby limiting access for legitimate users and impairing availability\cite{refII_A_36}, \cite{refII_A_37}.

\begin{table*}[t]
	\renewcommand{\arraystretch}{2}
	\centering
	\caption{Summary of Security Challenges in SAGINs.}
	\label{table1}
	\setlength{\tabcolsep}{1pt}
	\renewcommand{\arraystretch}{0.7}
	\begin{tabular}{c||c||c||c||c}
		\toprule[1pt]
		\rowcolor{gray!20}
		\multirow{1}{*}{\textbf{Segment}} &	\multirow{1}{*}{\makecell[c]{\textbf{Authenticity}}}&	\multirow{1}{*}{\textbf{Confidentiality}}&	\multirow{1}{*}{\textbf{Integrity}} & 	\multirow{1}{*}{\textbf{Availability}}\\
		
		\midrule
		
		\makecell[c]{\textbf{Cross-Domain} \\ \textbf{Interaction}}  &\makecell[l]{\textbullet{\textbf{ Spoofed Gateways:}}\\ Forge authentication to inject\\ false routing \cite{refII_A_15},\cite{refII_A_16}.}	& \makecell[l]{\textbullet{\textbf{ ISL Eavesdropping:}} \\Intercept  data in protocol\\ conversion \cite{refII_A_17}.}&\makecell[l]{\textbullet{\textbf{ MITM in Handovers:}}\\MITM tampers with control\\ commands\cite{refII_A_19}.}	&\makecell[l]{\textbullet{\textbf{ Cross-Layer Jamming:}}\\Jamming attacks threaten\\ cross-domain signals\cite{refII_A_20}.} \\
		
		\midrule
		
		\makecell[c]{\textbf{Space}}  & \makecell[l]{\textbullet{\textbf{ GPS Spoofing:}}\\Forge commands to \\manipulate signals\cite{refII_A_22}.}&\makecell[l]{\textbullet{\textbf{ Laser Interception:}}\\High-altitude eavesdropping \\for narrow beams\cite{refII_A_23}.}	&\makecell[l]{\textbullet{\textbf{ Real-Time Check Limits:}}\\Limited resources restrict\\ real-time  validation\cite{refII_A_24}.}	&\makecell[l]{\textbullet{\textbf{ High-Power Jamming:}}\\Overwhelm satellite uplinks\\ with high-power signals\cite{refII_A_25}, \cite{refII_A_26}.}	\\	
		\midrule
		
		\makecell[c]{\textbf{Air}}  &\makecell[l]{\textbullet{\textbf{ MAC Cloning:}}\\Hijack networks via spoofed\\ addresses\cite{refII_A_31}.}	&\makecell[l]{\textbullet{\textbf{ mmWave Leakage:}}\\Millimetre-wave links suffer\\ from  leakage\cite{refII_A_33}, \cite{refII_A_34}.}		&\makecell[l]{\textbullet{\textbf{ Path-Switching Injection:}}\\Injecting data when switching\\ paths in AAV networks\cite{refII_A_19}.}	&\makecell[l]{\textbullet{\textbf{ DDoS Flooding:}}\\
			Botnets overload aerial command\\ channels\cite{refII_A_36}, \cite{refII_A_37}.}	\\	
		\midrule
		
		\makecell[c]{\textbf{Ground}}  &\makecell[l]{\textbullet{\textbf{ Rogue BS:}}\\Fake BS harvests user \\data\cite{refII_A_43}.} 	&\makecell[l]{\textbullet{\textbf{ Edge Server Breaches:}}\\Attacked MEC nodes\\ leak sensitive data\cite{refII_A_44}.}	&\makecell[l]{\textbullet{\textbf{ ML Traffic Inference:}}\\ML deduces secrets from\\ encrypted patterns\cite{refII_A_45}.}	&\makecell[l]{\textbullet{\textbf{ Firmware Tampering:}}\\Malware alters IoT device\\ updates\cite{refII_A_46}.}	\\	
		
		\bottomrule[1pt]	
	\end{tabular}
	
	\label{table_challenges_security}
\end{table*}

\subsubsection{Secured Ground Segment}
The ground segment is composed of multiple sub-networks, such as 5G/6G base stations \cite{refII_A_38}, narrowband IoT networks \cite{refII_A_39}, and worldwide interoperability for microwave access (WiMAX) \cite{refII_A_40}. 
Its distributed architectures and legacy protocols (e.g., message queuing telemetry transport (MQTT) \cite{refII_A_41} and constrained application protocol \cite{refII_A_42}) expose critical secure vulnerabilities. 
In terms of undermining authenticity, rogue base stations, masquerade as legitimate base stations to spoof the users and obtain their data \cite{refII_A_43}.  
Confidentiality leaks arise from compromised edge servers.
Adversarial machine learning (ML) models can deduce sensitive industrial IoT data from encrypted multi-access edge computing (MEC) traffic patterns \cite{refII_A_44}. 
Integrity violations occur when malware modifies firmware updates in industrial IoT devices. 
For example, in \cite{refII_A_45}, malware added malicious content to escalate an application's privileges, allowing hackers to exfiltrate sensitive data. 
Furthermore, availability disasters can stem from physical infrastructure damage, such as severed backbone fibre-optic cables \cite{refII_A_46}.

In summary, SAGINs exhibit many new types of security challenges, and each segment confronts its own unique security threats, as summarized in Table \ref{table_challenges_security}.

\vspace{-1em}
\subsection{Overview of GAI}
GAI  is increasingly adopted across various fields of communication networks \cite{refII_B_buchong2}, and its distinctive features make it very suitable for SAGIN security enhancement. Secured GAI models be categorized as follows:

\begin{itemize}
	\item \textit{Variational Autoencoders (VAEs):} VAEs are generative frameworks that integrate DL with probabilistic graphical models, consisting of an encoder, a decoder, and a probabilistic latent space. 
	The encoder compresses sensitive data into a probabilistic latent distribution, thereby enhancing confidentiality through non-linear dimensionality reduction. A joint source channel coding scheme for point-to-point wireless communications was proposed by \cite{refII_B_1}, utilising vector quantised VAE (VQ-VAE) and achieving nearly 90\% accuracy. This scheme applies to anti-eavesdropping within the space segment. Simultaneously, the decoder reconstructs data from latent samples, facilitating integrity verification via anomaly detection.  The AC-VAE framework presented in \cite{refII_B_2} employed active learning alongside contrastive VAE-based models, resulting in F1 scores ranging from 0.68 to 0.96 using merely 3\% of labelled data. This approach prevents dynamic routing table tampering in LEO satellite networks, thereby enhancing the efficiency of data integrity verification. VAEs, while adaptable, demonstrate blurred reconstructions and instability during training in high-dimensional contexts.
	
	\item \textit{Generative Adversarial Networks (GANs):} 
	GANs are unsupervised learning frameworks that model accurate data distributions through the adversarial training between a generator and a discriminator \cite{refII_B_GAN}.  The discriminator detects anomalies by differentiating between legitimate data and adversarial samples. In \cite{refII_B_3}, an FS-GAN algorithm was introduced, leveraging the principles of federated learning (FL), self-supervised learning, and GANs, resulting in an improvement of over 20\% in average classification performance. The enhancement of cross-domain protocol vulnerability detection is anticipated, along with a reduction in the risk of MITM hijacking during the conversion between MAVLink and hypertext transfer protocol (HTTP). The generator creates realistic attack scenarios to strengthen defences proactively.  A mask guided adversarial training (MAGAT) structure was proposed by \cite{refII_B_4}, integrating adversarial training with mask-guided operations to protect against malicious face editing. This architecture can be utilised to pre-train AAV biometric authentication systems, thereby mitigating identity authenticity failures resulting from MAC address forgery. Despite their advantages, GANs experience mode collapse and exhibit limited control over local feature generation, which constrains their ability to synthesise a diverse range of threats.
	
	\item \textit{Generative Diffusion Models (GDMs):} 
	GDMs are score-based generative models that learn data distributions through a bi-directional process of forward diffusion and reverse denoising\cite{refII_B_5}, \cite{refII_B_6}, \cite{refII_B_GDM}. 
	The forward diffusion process progressively injects controlled Gaussian noise into real data until it becomes pure noise\cite{refII_B_7}, whereas the reverse denoising network effectively recover corrupted raw data from adversarial environments iteratively by training neural networks\cite{refII_B_AIGC}. 
	As demonstrated in \cite{refII_B_8}, a reverse denoising inspired DRL algorithm was proposed to balance content reconstruction fidelity and transmission efficiency in mobile AIGC networks. Complementary work in \cite{refII_B_9} presented an approximate message passing algorithm based on the reverse denoising process, which could achieve an improvement in signal reconstruction quality.
	These methods facilitate content reconstruction after data tampering in the space segment.
	While GDMs avoid mode collapse in GANs and exceed VAEs in generation fidelity, their reliance on 1000+ denoising steps limits real-time applications (e.g., real-time video encryption) \cite{refII_B_buchong1}.
	
	\begin{table*}[t]
		\renewcommand{\arraystretch}{2}
		\centering
		\caption{Summary of basic principles, security applications, and features of typical GAI models, Where Green Tick Represents strengths and Red Cross Represents Weaknesses.}
		\label{table2}
		\setlength{\tabcolsep}{2pt}
		\renewcommand{\arraystretch}{0.6}
		\begin{tabular}{c||c||c||c}
			\toprule[1pt]
			\rowcolor{gray!20}
			
			\multirow{1}{*}{\textbf{{Model}}} &\multirow{1}{*}{\textbf{ Principle}}&\multirow{1}{*}{\textbf{Security Applications}} & \multirow{1}{*}{\textbf{Pros \& Cons}}\\
			
			\midrule
			\multirow{1}{*}{ \textbf{VAEs}} & \makecell[l]{Through an encoder and \\ a decoder, the latent space \\ can be learned, and new \\content is generated}&
			\makecell[l]{		
				\textbullet{\textbf{ VQ-VAE for Secure Coding:}}\\A  source channel coding scheme with above 90\% accuracy\cite{refII_B_1}.\\\textbullet{\textbf{ AC-VAE for Anomaly Detection:}}\\Utilize the active learning to improve VAE with less labels\cite{refII_B_2}.
			} & \makecell[l]{\raisebox{-0.4ex}{\textcolor{green}{\CheckmarkBold}}  \; Ideal latent-space anomaly detection\\ \raisebox{-0.4ex}{\textcolor{red}{\XSolidBrush}} \; Blurred reconstructions with weak \\ \quad  \quad  protocol-level tamper detection} \\
			
			\midrule
			\multirow{1}{*}{ \textbf{GANs}} & \makecell[l]{Model accurate data \\distributions through the\\ adversarial training} &
			\makecell[l]{		
				\textbullet{\textbf{ FS-GAN for Anomaly Detection:}}\\Distinguish legitimate data from adversarial samples \cite{refII_B_3}.\\\textbullet{\textbf{ MAGAT Against Malicious Face Editing:}}\\Utilize MaGAT  to defend against malicious face editing\cite{refII_B_4}.
			} &\makecell[l]{\raisebox{-0.4ex}{\textcolor{green}{\CheckmarkBold}}  \; Powerful attack traffic generation \\ \raisebox{-0.4ex}{\textcolor{red}{\XSolidBrush}} \;  Easy mode collapse and  dependence \\ \quad \quad  on empirical parameterization\\ \quad \quad  with poor performance} \\
			
			\midrule
			\multirow{1}{*}{ \textbf{GDMs} } & \makecell[l]{Learn data distributions\\ through forward diffusion\\ and reverse denoising} &
			\makecell[l]{		
				\textbullet{\textbf{ DRL-GDM for Content Recovery:}}\\Balance content recovery and transmission  \cite{refII_B_8}.\\\textbullet{\textbf{ GDMs for Approximate Message Passing:}}\\Utilize the reverse process to acheive the message passing \cite{refII_B_9}.
			} & \makecell[l]{\raisebox{-0.4ex}{\textcolor{green}{\CheckmarkBold}}  \; Robust model and flexible for real-\\ \quad\quad  time spoofing defense\\ \raisebox{-0.4ex}{\textcolor{red}{\XSolidBrush}} \; High energy consumption}\\
			
			\midrule
			\multirow{1}{*}{ \textbf{TBMs}} & \makecell[l]{Self-attention-based \\ process for sequential data} &
			\makecell[l]{		
				\textbullet{\textbf{ TAEF for Anomaly Detection:}}\\Achieve detection on multiple real hyperspectral datasets \cite{refII_B_11}.\\\textbullet{\textbf{ TBMs for Intrusion Detection:}}\\Detect
				sophisticated and dynamic threats\cite{refII_B_13}.
			} &\makecell[l]{\raisebox{-0.4ex}{\textcolor{green}{\CheckmarkBold}}  \; Well cross-modal protocol semantic\\  \quad  \quad  analysis for cross-domain defense \\ \raisebox{-0.4ex}{\textcolor{red}{\XSolidBrush}} \; High computational load with limited\\  \quad  \quad   real-time security in AAVs detection}\\
			
			\midrule
			\multirow{1}{*}{\textbf{ LLMs }} & \makecell[l]{ Generate human-like\\ content through\\ contextual reasoning} &
			\makecell[l]{		
				\textbullet{\textbf{ ChatNet for Eavesdropping Prevention: }}\\Transform natural language into encrypted configurations\cite{refII_B_14}.\\\textbullet{\textbf{ LLMs for Privacy Preserving:}}\\Present a split learning system with the privacy-preserving \cite{refII_B_16}.
			} & \makecell[l]{\raisebox{-0.4ex}{\textcolor{green}{\CheckmarkBold}}  \; Dynamic policy generation and \\ \quad \quad  natural-language threat parsing \\ \raisebox{-0.4ex}{\textcolor{red}{\XSolidBrush}} \; Context window limitations with\\ \quad  \quad  cross-domain attack chain failure} \\
			
			\bottomrule[1pt]	
		\end{tabular}
		\label{table_GAI_models}
	\end{table*}

	\item \textit{Transformer-Based Models (TBMs):} TBMs are DL architectures based on a self-attentive mechanism that process sequential data through an encoder-decoder framework and perform well in natural language processing tasks\cite{refII_B_10}. 
	The multi-head self-attention module captures global dependencies, overcoming the local constraints of CNNs and RNNs. 
	The encoder captures the complex dependencies and analyzes input sequences for security detection, while the decoder generates target sequence using cross-attention. \cite{refII_B_11} developed TAEF, a transformer-autoencoder hybrid for hyperspectral anomaly detection,	which could be adopted in identifying payload tampering that occurs in satellite-ground links and AAV-ground links.
	TBMs also advance physical layer security\cite{refII_B_12} and intrusion detection\cite{refII_B_13}. Despite these strengths, challenges such as high computational demands, strong data dependency, and integration complexity hinder widespread adoption.
	
	\item \textit{LLMs:} 
	LLMs are transformer-based neural networks trained on massive text corpora to understand and generate human-like text through contextual reasoning\cite{refII_B_14}, \cite{refII_B_15}. 
	While LLMs and TBMs are similar, LLMs uniquely rely on massive pre-trained language backbones and contextual reasoning to generate security policies (e.g., cross-domain rules) or parse natural-language threat descriptions. These capabilities are critical for addressing SAGIN’s dynamic multi-domain collaboration and human-in-the-loop adaptation, which generic TBMs lack owing to task-specific rigidity.
	Specifically, LLMs consist of three core components: a pre-trained backbone (e.g., GPT-4 and LLaMA) for general language patterns, domain adaptation techniques (e.g., retrieval-augmented generation) to specialize in vertical fields, and tool interfaces enabling interaction with external systems (e.g., network analyzers). 
	Security applications include natural-language-to-configuration translation, as implemented in ChatNet's encrypted SAGIN policy generation \cite{refII_B_14}, and privacy-preserving sensor data transmission via collaborative mobile agent systems \cite{refII_B_16}.
	Despite these strengths, LLMs remain vulnerable to adversarial prompts and face challenges such as high computational costs and context window limitations\cite{refII_B_17}, \cite{refII_B_18}.	
	
\end{itemize}

In summary, GAI models can be targeted to address authenticity failures, confidentiality breaches, integrity tampering, and availability disruptions in SAGINs through data generation, feature reconstruction or policy optimisation capabilities.
The basic principles, applications in security, and features of these GAI models are concluded in Table \ref{table_GAI_models}.

\subsection{GAI vs Traditional AI in Security}While traditional AI (e.g., discriminative AI) has demonstrated efficacy in addressing conventional security challenges. GAI, a transformative paradigm in AI's evolution, is revolutionizing the field through its unique features, such as data augmentation, scenario simulation, and cross-modal analysis\cite{refII_B_1_1}, \cite{refII_B_1_2}, \cite{refII_B_1_3}. The fundamental distinctions between these paradigms include the core paradigm, architecture, data dependency, and dynamic adaptation, which enable GAI as a superior solution for modern security enhancement.

\begin{itemize}
	\item \textit{Core Paradigm:} Traditional AI relies on discriminative models (classification/regression) to differentiate normal from abnormal patterns. However, its over-reliance on historical data restricts it to known threats, rendering it ineffective against dynamic unknown threats in complex SAGINs such as zero-day attack \cite{refII_B_1_4}.
	For instance, while support vector machines (SVMs) are effective for known threats \cite{refII_B_1_5}, their accuracy drops to only 50\% against unknown threats \cite{refII_B_1_6}. 
	In contrast, GAI leverages generative models to synthesize diverse attack variants by learning underlying data distributions. FS-GAN in \cite{refII_B_3} could generate hybrid attack traffic without labeled data and improving detection accuracy more than 20\%. 
	This capability enables the generation of variant MITM attacks for cross-domain interactions, facilitating pre-training of SAGIN intrusion detection systems.

	\item \textit{Architecture:} Conventional AI depends on established feature engineering and static architectures, such as CNNs and RNNs. This unimodal approach increases information leakage and diminishes cross-domain attack detection accuracy in SAGINs. A single CNN is insufficient to defend against diverse MITM attacks across multiple domains \cite{refII_B_1_7}. In contrast, GAI facilitates multimodal inputs and end-to-end generation through modular designs, such as progressive denoising in GDM and self-attention in TBM, enhancing the accuracy of threat detection via cross-modal correlation. The advantage was demonstrated through a cross-modal transformer (FmFormer) \cite{refII_B_1_8}.  This framework enables the correlation of satellite telemetry, AAV sensor logs, and ground network traffic to detect multi-domain routing hijacking attacks in SAGINs, which single-modality CNNs are unable to identify.
	
	\item \textit{Data Dependency:} Conventional AI depends on substantial labelled datasets, resulting in diminished performance in scenarios with limited data\cite{refII_B_1_10}. Centralised training raises privacy concerns, particularly in widely exposed SAGINs\cite{refII_B_1_11}. GAI addresses these limitations by employing unsupervised and few-shot learning, thereby mitigating label scarcity through the generation of synthetic data. As a result, it exhibits reduced response time in identifying unknown attacks and can effectively balance privacy with model performance through prospective simulation. A dilated convolutional transformer-based GAN (DCT-GAN) for time series anomaly detection was proposed in \cite{refII_B_1_12} to enhance model accuracy and generalisation, facilitating the precise detection of stealthy satellite-side attacks in data-scarce SAGINs.
	
	\item \textit{Dynamic Adaptation:} Traditional AI relies on static models that require periodic retraining to adapt to evolving threats in SAGINs, resulting in delayed responses to emerging risks. 
	Specifically, the underlying data distribution changes over time, which can render ML models trained on historical data obsolete \cite{refII_B_1_13} \cite{refII_B_1_14}.
	GAI, however, achieves real-time adaptability through PE and ICL, which enables milliseconds of real-time anti-jamming and on-demand dynamic optimization. 
	\cite{refII_B_1_15} combined DL with generative model to utilize ICL for dynamically optimized defense against zero-day attacks, gaining a 15\% performance improvement. 
	Besides, ChatGPT was adopted for automated anomaly detection script generation via PE \cite{refII_B_1_16}. 
	Hence, these methodologies enable real-time defense against evolving SAGIN threats without costly model retraining.
\end{itemize}
In summary, GAI has more powerful analytical capabilities and novel generative features than traditional AI, which makes it uniquely suited for tackling complex security challenges in SAGINs.
The differences between them and examples of security enhancement via GAI are summarized in Table \ref{table_GAI_AI}.
\begin{table*}[t]
	\renewcommand{\arraystretch}{2}
	\centering
	\caption{Summary of Differences Between GAI and Traditional AI for Security Enhancement, Where Red Dot Represents the Drawbacks of Traditional AI, Green Dot Represents the Advantages of GAI, and Black Dot Represents the GAI Enhanced Security Examples.}
	\label{table3}
	\setlength{\tabcolsep}{4pt}
	\renewcommand{\arraystretch}{0.1}
	\begin{tabular}{p{1.6cm}||p{4.5cm}||p{4.5cm}||p{5.3cm}}
		\toprule[1pt]
		\rowcolor{gray!20}
		
		\makecell[c]{\textbf{{Aspect}}} &\makecell[c]{\textbf{ Traditional AI}}&\makecell[c]{\textbf{Generative  AI}} & \makecell[c]{\textbf{SAGIN Security Enhancement via GAI}}\\
		
		\midrule
		\makecell[c]{ \textbf{Core} \\ \textbf{Paradigm}} & \makecell[l]{
			\raisebox{-0.7ex}{\scalebox{2}{\textcolor{red}{\textbullet}}} \textbf{Poor Dynamic Threat Detection: \quad \quad}\\
			Discriminative models are limited to\\ classifying known threats.
		}&
		\makecell[l]{
			\raisebox{-0.7ex}{\scalebox{2}{\textcolor{green}{\textbullet}}} \textbf{Superior Various Attack Defense:}\\
			Generative models  create diverse \\samples to enhance robustness.
		} & \makecell[l]{
			\raisebox{-0.7ex}{\scalebox{2}{\textbullet}}{\textbf{ FS-GAN for Various Attack Generation:}}\\FS-GAN generates  MITM attacks for cross-\\domain interaction segment \cite{refII_B_3}.} \\

		\midrule
		\makecell[c]{ \textbf{Architecture}} & \makecell[l]{	\raisebox{-0.7ex}{\scalebox{2}{\textcolor{red}{\textbullet}}} \textbf{Poor Cross-Modal Detection:\quad \quad} \\ 
			Fixed architectures struggle to fuse\\ multimodal data.} 
		&
		\makecell[l]{\raisebox{-0.7ex}{\scalebox{2}{\textcolor{green}{\textbullet}}} \textbf{Superior Cross-modal Detection:} \\
			Modular design enables cross-Modal\\ correlation.\\
		} &\makecell[l]{
			\raisebox{-0.7ex}{\scalebox{2}{\textbullet}}{\textbf{ FmFormer for Cross-Modal Detection:}}\\
			FmFormer detects the multi-domain routing \\hijacking attacks in SAGINs \cite{refII_B_1_8}.
		} \\
		
		\midrule
		\makecell[c]{ \textbf{Data} \\ \textbf{Dependency} } & \makecell[l]{
			\raisebox{-0.7ex}{\scalebox{2}{\textcolor{red}{\textbullet}}} \textbf{Over-reliance on Databases and} \\  \quad \textbf{Possible Privacy Issues:}\\
			Rely on amounts of labelled data with \\ poor performance in few-shot scenarios.
		} &
		\makecell[l]{\raisebox{-0.7ex}{\scalebox{2}{\textcolor{green}{\textbullet}}} \textbf{Powerful Data Generation and }\\ \quad \textbf{Prospective Privacy Protection:}\\
			Synthetic data generation addresses \\labeling gaps and preserves privacy.
		} & \makecell[l]{
			\raisebox{-0.7ex}{\scalebox{2}{\textbullet}}{\textbf{ DCT-GAN for Few-Shot Preservation:}} \\
			DCT-GAN reduces detection latency and \\ balances privacy via simulation, to detect \\ stealthy attacks in data-scarce SAGINs \cite {refII_B_1_12}.
		}\\
		
		\midrule
		\makecell[c]{ \textbf{Dynamic}\\  \textbf{Adaptation}} & \makecell[l]{
			\raisebox{-0.7ex}{\scalebox{2}{\textcolor{red}{\textbullet}}} \textbf{Outdated Models Perform Poorly}\\ \quad \textbf{in New Threats Defense:}\\
			Static models fail to respond in real-\\time to dynamic threats.
		} &
		\makecell[l]{
			\raisebox{-0.7ex}{\scalebox{2}{\textcolor{green}{\textbullet}}} \textbf{Superior On-Demand Optimization}\\  \quad  \textbf{and Real-Time Anti-Jamming:}\\
			PE and ICL enable real-time strategy \\adjustments.
		} &\makecell[l]{
			\raisebox{-0.7ex}{\scalebox{2}{\textbullet}}{\textbf{ PE/ICL for Evolving Threats Defense:}} \\
			ICL-enhanced generative model for SAGINs'\\ zero-day defense\cite {refII_B_1_15} while PE-based \\anomaly detection through ChatGPT\cite {refII_B_1_16}.
		}\\
		
		\bottomrule[1pt]	
	\end{tabular}
	\label{table_GAI_AI}
\end{table*}

The integration of GAI with SAGINs represents a significant advancement in communication technology\cite{refIntro_1}.  The emergence of complex security threats, including cross-domain signal jamming and dynamic cyber attacks in SAGINs, highlights the potential application of GAI in threat modelling, real-time defence, and data privacy protection.  In this context, GAI-enabled SAGIN security enhancement will serve as a crucial factor in the evolution of AI-SAGIN.

\section{GAI for Authenticity and Confidentiality Security in SAGINs: Current Research}
\begin{figure*}[!t]
		\vspace*{-1em}
	\centering
	\includegraphics[width=5.5in]{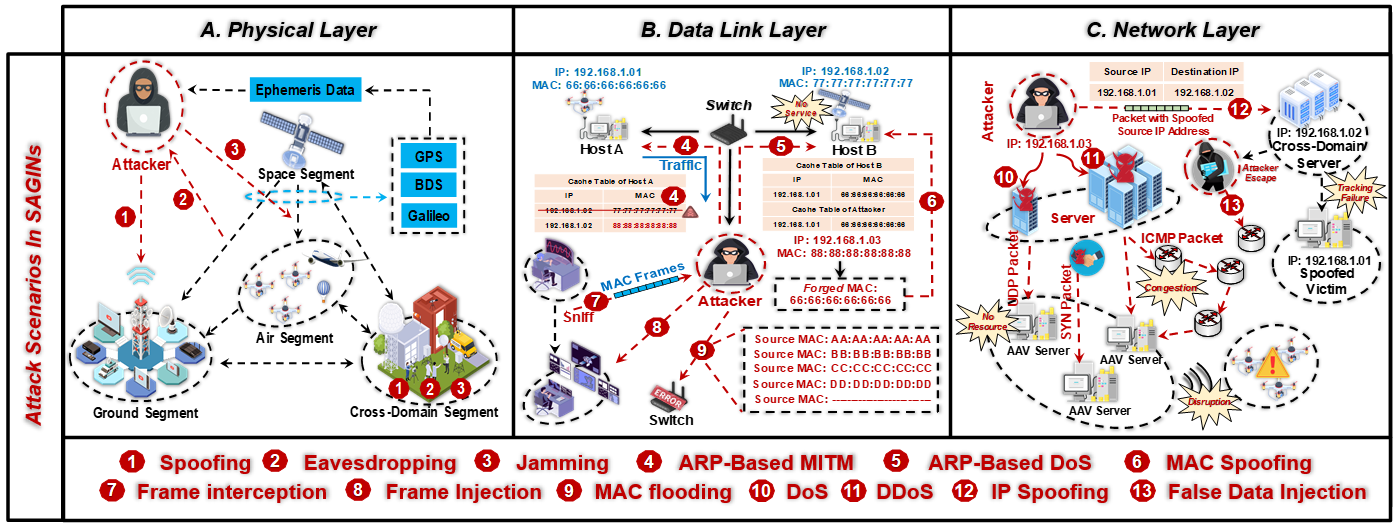}
	\caption{Overview of typical attack scenarios at different layers in SAGINs. The 13 typical attack methods summarized above significantly cause authentication failures, confidentiality breaches, integrity tampering, and availability disruptions in SAGIN communications.}
	\label{SectionIII_outline_p1}
	\vspace*{-1em}
\end{figure*}

Section III and IV review the role of GAI in mitigating SAGIN security threats through four security requirements as illustrated in Fig. \ref{fig_outline}. 
Meanwhile, this exploration incorporates the open systems interconnection (OSI) layered protocol layers (i.e., physical layer, data link layer, and network layer) under each requirement to bridge current research gaps. 
This is because different layers exhibit distinct vulnerabilities due to their underlying protocols as shown in Fig. \ref{SectionIII_outline_p1}.
In this section, we first summarize the authenticity failures and confidentiality breaches that may be encountered in SAGINs, and overview the corresponding GAI-based security solutions in detail.

\subsection{GAI for Authenticity Failures}
\subsubsection{Physical Layer}
Failures in authenticity at the physical layer in SAGINs primarily arise from signal spoofing attacks, wherein attackers transmit either fabricated signals or relevant real signals to target devices.
The openness of long-link signals, including satellite navigation signals like GPS \cite{refIII_A_1_1} and the Beidou navigation satellite (BDS) system \cite{refIII_A_1_2}, as well as air signals such as automatic dependent surveillance-broadcast (ADS-B) \cite{refIII_A_1_3}, makes them susceptible to spoofing.  Attackers utilise publicly accessible ephemeris data to spoof global navigation satellite system (GNSS) signals, leading to location inaccuracies and trajectory deviations in AAVs \cite{refII_A_22}.

GANs improve security by facilitating a competitive interaction between the generator and the discriminator.
This research aims to develop effective spearheads (spoofing jamming) and robust shields (spoofing defence) utilising GANs for anti-spoofing applications. 
In the air segment, the end-to-end FlightSense \cite{refIII_A_1_3} integrated GAN and CNN methodologies to identify ADS-B spoofing and perform aircraft identification using raw I/Q signals, resulting in a detection accuracy of 98.87\% and a classification accuracy of 99\% on synthetic datasets. 
In the space segment, as shown in Fig. \ref{SectionIII_outline_p2}, \cite{refIII_A_1_4} utilised GAN to attain high-precision signal detection in GNSS acquisition.
Although these methods demonstrate high precision, they generally concentrate on individual attack types and exhibit limited scalability. 
To address these limitations, \cite{refIII_A_1_5} developed a GAN-assisted contextual pattern-aware intrusion detection system that demonstrates high accuracy in identifying six types of attacks, including spoofing and replay, along with ultra-low latency in real-vehicle experiments.  This model demonstrates significant applicability potential for SAGINs.

Hybrid GAI approaches are emerging beyond the single GAN.  GenCoder \cite{refIII_A_1_6} integrated a five-layer deep neural network (DNN) with VAE to effectively produce adaptive training data for unidentified attacks.  Reference \cite{refIII_A_1_7} integrated VAE with Wasserstein GAN (WGAN) for the detection of spoofing signals, utilising the accurate reconstruction capabilities of VAE alongside the feature learning strengths of WGAN. 
Additionally, the capabilities of TBMs to capture spatio-temporal features and identify anomalous patterns through a multi-head attention mechanism are incorporated with GANs.  DroneDefGANt, as proposed by \cite{refIII_A_1_8}, represents a hybrid methodology that integrates GANs with transformer models to identify external threats, such as GPS spoofing, alongside internal failures, including actuator malfunctions.



Current research also explores GAI's integration with traditional authentication approaches such as RF fingerprinting (RFF), channel state information (CSI), and channel impulse response (CIR). 
For instance, \cite{refIII_A_1_99} introduced GANSAT by combining GAN and satellite constellation RFF for GPS spoof detection and location estimation. 
This approach overcame conventional methods' environment-sensitivity limitations, but required location-specific training data with unverified cross-domain generalization.
As a refinement, \cite{refIII_A_1_104} employed GAN-CNN approach by analyzing the RFF of LoRa system, performing 92.4\% authentication accuracy without channel compensation. 
In addition, \cite{refIII_A_1_9} exploited conditional VAE (CVAE) to adaptively compress and extract discriminative RFF features to improve accuracy. 
Based on the CSI, \cite{refIII_A_1_csi_supply} proposed a GDM-based secure sensing solution for ISAC system to extract the true CSI for sensing.
\cite{refIII_A_1_10} constructed a conditional GAN (CGAN) framework integrating long short-term memory (LSTM) and gated recurrent unit (GRU) networks to predict mobile channel responses.
It achieved high accuracy but degraded without CSI training data.
As a complement, \cite{refIII_A_1_11} presented HVAE, combining an AE for CIR characteristics extraction and a VAE for the enhancement of CIR feature representation and authentication results.

\subsubsection{Data Link Layer} 
The expiration of contextual authentication in dynamic SAGINs 
creates temporal gaps for attackers to forge temporary MAC identities. Typical threats include MAC address spoofing, address resolution protocol (ARP) spoofing, and frame injection attack, facilitating DoS attack, MITM attack, and session hijacking attack \cite{refII_A_31}.

MAC address serves as a unique physical credential for device access. Attackers can bypass the network authentication by cloning the legal MAC address \cite{refIII_A_2_1} \cite{refIII_A_2_2}. 
Although threshold-based sequence number analysis detects MAC spoofing, it yields false alarms due to frame loss.
To overcome this drawback, an artificial neural network (ANN)-based detection method was proposed in \cite{refIII_A_2_3} to analyze sequence number gaps and statistical distributions in MAC headers. 
The ANN framework could function as a pre-trained feature extraction layer for GAN discriminators or generators.
Additionally, randomized dynamic defense \cite{refIII_A_2_4} improve IoT security by adapting signal strength thresholds. Such methods are compatible with GAN integration for enhanced defense.

\begin{figure*}[!t]
		\vspace*{-1em}
	\centering
	\includegraphics[width=5in]{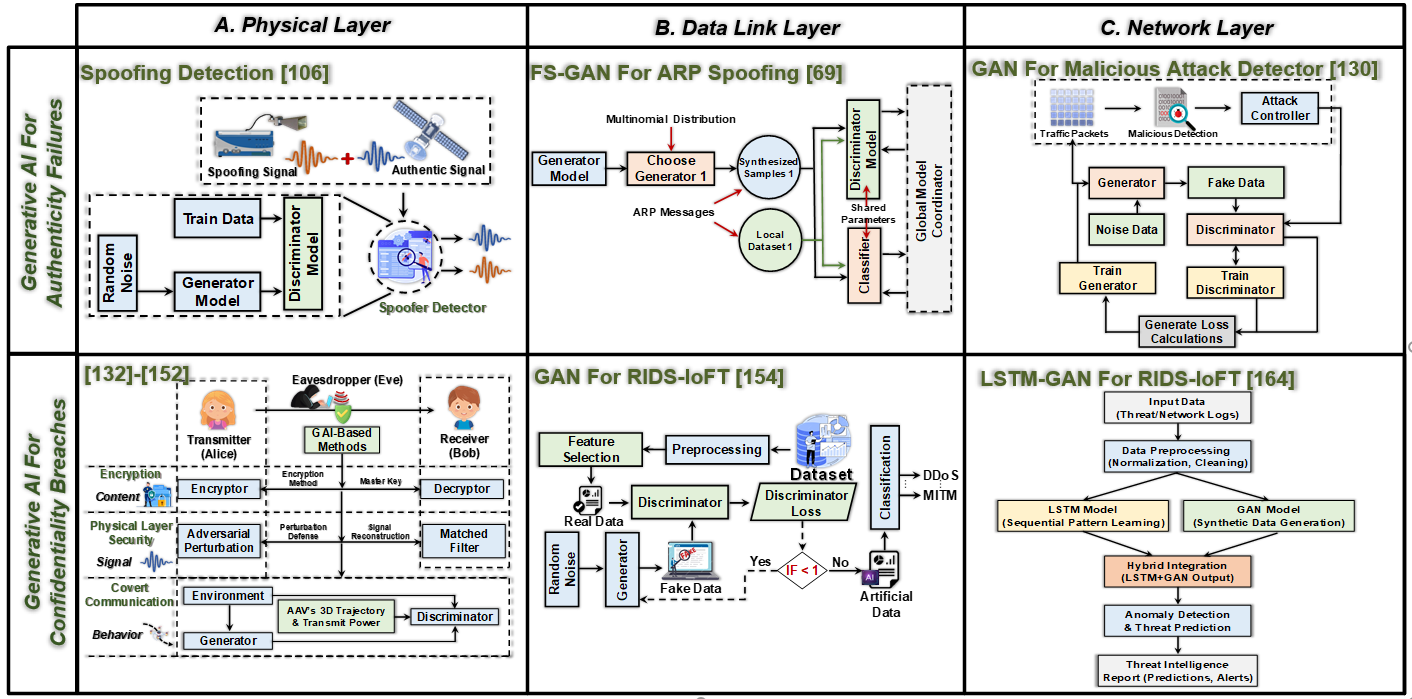}
	\caption{Typical GAI-based approaches for authenticity failures and confidentiality breaches in SAGINs. For authenticity failures, \cite{refIII_A_1_4} introduces a GAN-based GNSS spoofing framework; \cite{refII_B_3} illustrates a FS-GAN approach against ARP spoofing; \cite{refIII_Authenticity_network_9} presents a GAN-based malicious attack detector. For confidentiality breaches, \cite{refIII_Eva_2} - \cite{refIII_Eva_20} designs multiple physical layer anti-eavesdropping methods by combining GAI with encryption, physical layer security, and covert communication; \cite{refIII_Confidentiality_MAC_2} proposes a GAN-based IDS for IoFT; \cite{refIII_Confidentiality_Network_3} combines LSTM with GAN for IDS design.}
	\label{SectionIII_outline_p2}
	\vspace*{-1em}
\end{figure*}

ARP maps logical address (i.e., IP address) to corresponding physical address (i.e., MAC address) without verifying the authenticity of responses to ARP requests \cite{refIII_A_2_ARP}. This vulnerability  enables attackers to execute ARP spoofing (i.e., ARP cache poisoning). It is an active attack method involving the injection of forged IP-MAC mappings into devices via unsolicited ARP messages. Attackers manipulate ARP cache tables without awaiting target-initiated queries, facilitating subsequent MITM and denial-of-service (DoS) attacks. 
The FS-GAN in Fig. \ref{SectionIII_outline_p2} could
classify ARP-spoofed frames adopting distributed GANs for sample generation and FL for training  \cite{refII_B_3} .

In SAGINs, legitimate devices operate within open shared wireless media, allowing attackers to sniff MAC frames and forge spoofed frames for injection attacks such as false acknowledgment (ACK) and message frame injection \cite{refIII_A_2_6}. 
These frames disrupt user decisions and exhaust resources, causing communication blockages.
\cite{refIII_A_2_7} designed an airspace anomaly detection method combining flight plan data with GAN-LSTM model for ADS-B attack.
It could construct airspace image frames, utilizing GANs for temporal prediction and normalized cross-correlation for anomaly localization. Results demonstrated 92.3\% average detection accuracy with 11.9\% false positive rate and 6.2\% false negative rate, effectively identifying stealthy attacks like frame injection while maintaining compatibility with ATC operational constraints.

\subsubsection{Network Layer} 
Against threats like IP forgery and routing hijacks, GAI enhances end-to-end defense by synthesizing adversarial traffic for anomaly detection and generating dynamic authentication patterns to expose malicious actors.

In \cite{refIII_Authenticity_network_1}, the authors introduced a multi-architecture GAN framework aimed at mitigating data imbalance in network intrusion detection.  The application of synthetic data to the CIC-IDS2017 benchmark resulted in a 6.18\% enhancement in recall for Bot attack detection. 
However, real-time traffic in SAGINs is dynamic, and this method may have limitations.  
The Magteon-Turing L3TM framework in \cite{refIII_Authenticity_network_3} introduced a GAI-driven method to address identity spoofing and traffic camouflage.  The integration of Megatron-Turing NLG and Swarm OpenAI LLMs was achieved within a seven-layer architecture.  1) The data layer utilised GANs for the generation of synthetic hybrid traffic; 2) The traffic analysis layer implemented Langchain for semantic feature extraction, achieving 98.7\% accuracy in detecting SYN Flood and IP spoofing; 3) The behavioural validation layer employed Llama models for protocol logic verification, identifying anomalous session patterns with a false positive rate of 1.5\%. 
This study addresses the static rule limitations identified in \cite{refIII_Authenticity_network_1} by employing dynamic LLM ensembles, which facilitate adaptive responses with sub-second latency.

GAI is capable of predicting the time series of attacks in SAGINs via the generation of adversarial samples.   A hybrid deep learning framework integrating remaining useful life GAN (RUL-GAN) for time-series prediction of DoS attacks was proposed by \cite{refIII_Authenticity_network_4}. 
A domain-specific GAN mechanism was developed to synthesise distributed energy resource control messages (e.g., packet parameters, protocol features), revealing authenticity threats from adversarial message spoofing \cite{refIII_Authenticity_network_5}. 
The framework incorporated conditional value-at-risk for quantifying tail risk and employed an ensemble learning-based bagging method to identify synthetic identities, resulting in an attack vector generation accuracy of 95.7\%, a tail risk of 9.61\% (with 95\% stability), and an overall accuracy of 99\%. 

The GAN-DRL model presented in \cite{refIII_Authenticity_network_7} incorporated an intelligent routing model to tackle the issue of detecting minor-class attacks resulting from imbalanced data in healthcare-consumer IoT, while simultaneously optimising network-layer routing. 
The approach utilised GAN for generating synthetic data to identify rare attacks, such as smurf attacks, and employed DRL to dynamically adjust routing policies. This resulted in a 12\% improvement in throughput, a 20\% reduction in latency, and a 30\% increase in the probability of avoiding malicious attacks within an SDN architecture.
This approach demonstrates limited efficiency in generating high-dimensional traffic features.
To tackle this problem, \cite{refIII_Authenticity_network_8} proposed a lightweight GAN-DRL model compression technique aimed at enhancing the efficiency of high-dimensional traffic feature generation while mitigating the significant computational overhead in SDN-IoT environments.
In addition, 
\cite{refIII_Authenticity_network_9} 
proposed a distributed GAN training framework integrating a malicious attack detector (MAD) for multi-device collaborative novel attack detection as shown in Fig. \ref{SectionIII_outline_p2}.

\subsubsection{Summary and Lessons Learned}
As summarized in Table \ref{table_GAI_Authenticity}, 
GAI serves as a proactive tool for generating adversarial samples and a reactive shield for real-time anomaly detection, enabling adaptive defense in SAGINs.
GANs excel at detecting spoofing through signal discrepancy analysis.
Hybrid models integrating VAEs or transformers further enhance robustness against unknown attacks. 
Federated GANs and LLM-GAN hybrids demonstrate exceptional capabilities in mitigating MAC spoofing, IP forgery, and routing hijacks by leveraging synthetic traffic and semantic features to reduce false positives.
Despite these advancements, challenges remain:
\begin{itemize}
	\item[$\bullet$] \textit{Generalization-Specialization Trade-off:} Models like CVAE-ZSL achieve high zero-shot detection accuracy but degrade in cross-domain scenarios.
	\item[$\bullet$] \textit{Computational Overhead:} Computational overhead limits real-time high-dimensional traffic analysis. 
	\item[$\bullet$] \textit{Explainability Gaps:} Lack of explainable AI (XAI) integration undermines trust in hybrid model decisions.
	\item[$\bullet$] \textit{Adversarial Robustness:} GAI-enhanced attack tools exploit static defenses, requiring co-designed countermeasures.
\end{itemize}

\begin{table*}[htbp]
	\renewcommand{\arraystretch}{2}
	\centering
	\caption{Summary of GAI Solutions for Security Threats Affecting Authenticity in SAGINs, Where Multiple Attacks in Network Layer Refer to Traffic Injection, Spoofing, and Replay Attacks.}
	\newcolumntype{W}{!{\vrule width 0.5pt}}
	\setlength{\tabcolsep}{4pt}
	\renewcommand{\arraystretch}{1.2}
	\begin{tabular}{c WW c | c |c |c |p{7.6cm}}
		\rowcolor{gray!20}
		\Xhline{1pt}
		\multirow{1}{*}{\textbf{{Layer}}} &\multirow{1}{*}{\textbf{Ref.}}& \multirow{1}{*}{\textbf{Segment}} &\multirow{1}{*}{\textbf{Security Threat}} & \multirow{1}{*}{\textbf{GAI Approach}}& \makecell[c]{\textbf{Description}} \\
		
		\Xhline{0.7pt}
		\multirow{3}{*}[-53pt]{\shortstack{\textbf{Physical}\\ \textbf{Layer}}}&
		\makecell[c]{\cite{refIII_A_1_8}}&\makecell[c]{Cross-Domain}&\makecell[c]{GPS spoofing }  & \makecell[c]{TBM-GAN} & \makecell[c]{DroneDefGANt for external attacks and internal faults} 
		\\
		\cline{2-6}
		
		&\makecell[c]{\cite{refIII_A_1_10}} & \makecell[c]{Cross-Domain}& \makecell[c]{Authentication \\breach}& \makecell[c]{LSTM-CGAN} & \makecell[c]{ Physical-Layer authentication adopting CSI and CGAN-LSTM} 
		\\
		\cline{2-6}
		
		&\makecell[c]{\cite{refIII_A_1_4}}& \makecell[c]{Space}&\makecell[c]{GNSS spoofing}  & \makecell[c]{GAN} & \makecell[c]{High-precision
		detection against wrong timing and positioning} 
		\\
		\cline{2-6}
		
		&\makecell[c]{\cite{refIII_A_1_7}}& \makecell[c]{Space}&\makecell[c]{Unknown spoofing} & \makecell[c]{VAE-WGAN} & \makecell[c]{A GNSS spoof detection model comprising VAE-WGAN} 
		\\
		\cline{2-6}
		
		&\makecell[c]{ \cite{refIII_A_1_99}}	& \makecell[c]{Space}&\makecell[c]{GPS spoofing}   & \makecell[c]{GAN} & \makecell[c]{Combine GAN
		and RFF for detection and location estimation} 
		\\
		
		\cline{2-6}
		&\makecell[c]{\cite{refIII_A_1_3}}& \makecell[c]{Air}&\makecell[c]{ADS-B spoofing} & \makecell[c]{CNN-based GAN} & \makecell[c]{GAN-CNN
		for detecting ADS-B spoofing using raw I/Q signals} 
		\\
		\cline{2-6}
		
		&\makecell[c]{\cite{refIII_A_1_5}}& \makecell[c]{Ground}& \makecell[c]{Spoofing in IVN} & \makecell[c]{GAN} & \makecell[c]{A GAN-assisted contextual pattern-aware IDS} 
		\\
		\cline{2-6}
		
		&\makecell[c]{\cite{refIII_A_1_6}}& \makecell[c]{Ground}& \makecell[c]{Spoofing attack} & \makecell[c]{DNN-based VAE} & \makecell[c]{IDS to address the dynamic and evolving  cyberthreatsy in IoV} 
		\\
		\cline{2-6}
		
		&\makecell[c]{\cite{refIII_A_1_104}}
		& \makecell[c]{Ground}&\makecell[c]{Spoofing for IoT}  & \makecell[c]{CNN-based GAN} & \makecell[c]{Spoof detection and authentication using RFF-CWT-based GAN} 
		\\
		\cline{2-6}
		
		&\makecell[c]{\cite{refIII_A_1_9}}& 		 \makecell[c]{Ground}&\makecell[c]{Spoofing for AIS}  & \makecell[c]{Conditional VAE} & \makecell[c]{A RFFI-based CVAE detection framework} 
		\\
		\cline{2-6}
		
		&\makecell[c]{\cite{refIII_A_1_11}}& \makecell[c]{Ground}&\makecell[c]{Spoofing for IIoT}  & \makecell[c]{Hierarchical VAE} & \makecell[c]{CIR-based  authentication for static and mobile IIoT} 
		\\
		\cline{1-6}
		
		&\makecell[c]{\cite{refIII_A_2_3}}&\makecell[c]{Each Segment}& \makecell[c]{MAC spoofing} & \makecell[c]{ANN for GAN} & \makecell[c]{An ANN-based spoofing detection method} 
		\\
		\cline{2-6}
		
		\multirow{3}{*}[5pt]{\shortstack{\textbf{Data Link}\\ \textbf{Layer}}}&
		\makecell[c]{\cite{refII_B_3}}&\makecell[c]{Each Segment}& \makecell[c]{ARP spoofing} & \makecell[c]{GAN} & \makecell[c]{A federated self-supervised learning model for traffic analysis} 
		\\
		\cline{2-6}
		
		&\makecell[c]{\cite{refIII_A_2_7}}&\makecell[c]{Air}&\makecell[c]{ADS-B Spoofing} & \makecell[c]{LSTM-GAN} & \makecell[c]{An attack detection and marking model  based on LSTM-GAN} 
		\\
		\cline{2-6}
		
		&\makecell[c]{\cite{refIII_A_2_4}}&\makecell[c]{Ground}&\makecell[c]{MAC spoofing }  & \makecell[c]{Adversarial for GAN } & \makecell[c]{A randomized dynamic target defense framework for IoT system} 
		\\
		\cline{1-6}
		
		\multirow{3}{*}[-25pt]{\shortstack{\textbf{Network} \\ \textbf{Layer}}}&
		\makecell[c]{\cite{refIII_Authenticity_network_1}}&\makecell[c]{Each Segment}& \makecell[c]{Traffic injection} & \makecell[c]{LSTM-GAN} & \makecell[c]{A traffic augment solution for IDS system} 
		\\
		\cline{2-6}
		
		&
		\makecell[c]{\cite{refIII_Authenticity_network_3}} &\makecell[c]{Each Segment}&\makecell[c]{Traffic spoofing} & \makecell[c]{LLM-GAN} & \makecell[c]{A Magteon-Turing L3TM framework} 
		\\
		\cline{2-6}
		
		&\makecell[c]{\cite{refIII_Authenticity_network_4}}&\makecell[c]{Ground}&\makecell[c]{Multiple attacks}  & \makecell[c]{GAN} & \makecell[c]{Time-series prediction of cyber attacks in vehicle networks} 
		\\
		\cline{2-6}
		
		&
		\makecell[c]{\cite{refIII_Authenticity_network_5}}&\makecell[c]{Ground}& \makecell[c]{Replay attack} & \makecell[c]{GAN} & \makecell[c]{GAN as an attack generation for training defense systems} 
		\\
		\cline{2-6}
		
		&
		\makecell[c]{\cite{refIII_Authenticity_network_7}}
		&\makecell[c]{Ground}&\makecell[c]{Minor-Class attack} & \makecell[c]{DRL-based GAN} & \makecell[c]{An anomaly detection system for optimum routing in H-CIoT} 
		\\
		\cline{2-6}
		
		&
		\makecell[c]{\cite{refIII_Authenticity_network_8}}&\makecell[c]{Ground}&\makecell[c]{Traffic injection}  & \makecell[c]{GAN} & \makecell[c]{Lightweight
			GAN for traffic prediction and routing in SDN-IoT} 
		\\
		\cline{2-6}
		
		&
		\makecell[c]{\cite{refIII_Authenticity_network_9}} &\makecell[c]{Ground}&\makecell[c]{Multiple attacks} & \makecell[c]{GAN} & \makecell[c]{A GAN-based malicious attack detector for IoT systems} 
		\\

		\Xhline{1pt}	
	\end{tabular}
	\label{table_GAI_Authenticity}
\end{table*}


\vspace{-1em}
\subsection{GAI for Confidentiality Breaches}
\subsubsection{Physical Layer} 
Eavesdropping attacks, referred to as stealth attacks, are markedly exacerbated by the intrinsic openness of wireless channels and the architectural heterogeneity present in SAGINs. 
Attackers leverage vulnerabilities in satellite link broadcasts, AAV relays, and terrestrial network density to intercept data through signal capture, protocol parsing, or cross-layer traffic correlation.
Current countermeasures primarily depend on encryption methods for content protection, physical layer security for signal protection, and covert communication for behaviour concealment \cite{refIII_Eva_1}. These approaches face challenges related to dynamic adaptability, computational overhead, and cross-domain coordination.
Recently, GAI has improved these methods by incorporating dynamic learning and generative capabilities.

\textbf{\textit{GAI-assisted encryption:}} Conventional encryption methods depend on a pre-shared secret key between communicating entities, which are susceptible to elevated error rates and key exposure. 
A WGAN-GP adversarial autoencoder was proposed to dynamically extract cross-layer channel features, thereby enhancing key capacity and reducing errors \cite{refIII_Eva_2}. 

Additionally, the high spatio-temporal similarity of the CIR in high-density user scenarios enhances the likelihood of eavesdroppers successfully reconstructing keys.  In this context, \cite{refIII_Eva_3} proposed a GAN-assisted noise generation algorithm designed to inject adversarial noise into critical defence regions of dynamic LiFi networks. 
Moreover, frequent handovers in SAGIN elevate the risk of key leakage.  An intelligent soft handover method for UAV-enabled cellular networks is presented, utilising GAN in conjunction with blockchain and physically unclonable function for lightweight message encryption \cite{refIII_Eva_4}.

In addition to single-attacker scenarios, \cite{refIII_Eva_5} introduced a multiparty adversarial encryption model utilising GAN for secure multi-party communication.  The introduction of an enhanced adversarial training framework featuring four types of attackers (including ciphertext-only, key-leaked, and chosen-plaintext attackers) enabled the model to attain information-theoretic security via synchronised neural network parameters and secret keys, effectively resisting multimodal threats in complex SAGINs.

Furthermore, current methods focus on encryption algorithms designed for ``fixed eavesdroppers".
Eavesdroppers fail when they can flexibly adjust their methods based on the transmission policies of legitimate devices, a phenomenon we refer to as ``evolved eavesdroppers" possessing adaptive learning capabilities. 
According to \cite{refIII_Eva_6}, eavesdroppers adjust neural network parameters through a symmetric training process that emulates legitimate users, thereby increasing the risks to GAI-assisted SAGINs.
The authors proposed a secure autoprecoder (ASAP) framework based on adversarial learning for MIMO wiretap channels featuring ``evolved eavesdroppers." 
The two-stage adversarial training jointly optimises modulation and precoding, thereby enhancing the security-reliability trade-off.

Given the unique role of semantic communication in SAGINs \cite{refIII_Eva_7}, GAI-assisted encrypted semantic communication requires investigation.
A GAN-based encrypted semantic communication system was proposed by \cite{refIII_Eva_8}.  An effective defence against eavesdropping attacks was achieved in a generic semantic communication scenario with shared knowledge through the design of a symmetric encryption structure and an adversarial training algorithm. 
The system improves the balance between privacy-preserving capabilities and versatility in SAGINs.

\textit{\textbf{GAI-assisted physical layer security:}}  GAI enhances physical layer security by leveraging channel randomness and active defense strategies against sophisticated eavesdropping. 
For instance, an adversarial training framework was proposed to harden modulation classifiers against evasion attacks \cite{refIII_Eva_9}. By designing an iterative ``Least Likely"  white-box attack strategy, this active anti-eavesdropping method improved communication privacy by reducing the classification accuracy of eavesdroppers to random guessing levels (e.g., from 98\% to 4\%) in software-defined radio (SDR) tests. 

In addition, beamforming significantly contributes to physical layer security.
\cite{refIII_Eva_10} 
proposed an actor-critic GDM (AC-GDM) scheme optimizing precoding and IRS phase shifts, leveraging GDM's denoising capability in IRS-assisted IoT.
Another proactive defense method integrates perturbations into beamforming design \cite{refIII_bf-swipt}. 
As demonstrated in \cite{refIII_Eva_11}, an adversarial defense embedded waveform design (ADEWD) solution adopted GAN-generated amplitude-controlled perturbations superimposed on signals to disrupt eavesdropper identification while ensuring reliable communication.
In addition,\cite{refIII_Eva_12} extended this approach to cooperative beamforming in AAV swarm with a GDM-enabled twin delayed deep deterministic policy gradient (GDMTD3) method to tackle the non-convex, NP-hard dynamic problem and promote the secrecy rate. 

Beyond secrecy rate, secrecy energy efficiency (SEE) is critical for energy-constrained SAGINs \cite{refIII_Eva_2th}. 
\cite{refIII_Eva_13} designed a mixture of experts (MoE)-based GDM RL algorithm, improving SEE and resisting eavesdropping.
Additionally, the authors in \cite{refIII_Eva_14} further considered the energy constraints for HAP in SAGINs and proposed a GAN-empowered deep RL framework named Gen-DRL. 
Integrating bidirectional LSTM-enhanced GANs into the policy network, Gen-DRL dynamically predicted channel states and adapted to environment while capturing long-term temporal dependencies. It achieved 5.1\%-12.43\% higher SEE than benchmarks, exhibiting superior robustness against mobile user scenarios.

\textit{\textbf{GAI-assisted covert communication:}} Encryption and physical layer security cannot prevent adversaries from detecting communication behaviors, risking data reliability and exposing source locations. 
In SAGINs, covert communication outperforms by ensuring ``communication undetectability" (e.g., minimizing signal detection probability). \cite{refIII_Eva_15} developed a GAN framework optimizing transmit power against eavesdroppers with uncertain thresholds in uplink FL.
Extending this, the authors of \cite{refIII_Eva_16} proposed a model-driven GAN (MD-GAN) framework, featuring a GAN-based joint trajectory and power optimization algorithm when only partial channel information is available.
In addition, \cite{refIII_Eva_17} introduced an AE-based GAN for water-to-air communication scenario based on optical wireless communication (OWC), which optimized a signal generator and a decoder to produce covert signals statistically indistinguishable.

For AAV-assisted jamming, \cite{refIII_Eva_18} proposed a data-driven GAN (DD-GAN) framework to address the eavesdropping threats in downlink satellite-ground communication.
Targeting AAV-assisted jamming scenarios with partial environmental knowledge, the method integrated genetic algorithm-generated samples to co-optimize AAV jamming power and 3D trajectory through adversarial training.
For non-terrestrial networks (NTN) IoT, \cite{refIII_Eva_19} developed a tripartite collaborative-adversarial network (TCAN) based on GANs. This framework integrated a generator (emulating legitimate transmitters), a cooperative classifier (receiver), and an adversarial discriminator (attacker) to dynamically balance covert signal design and detection.
Beyond physical signals, text steganography is applied in covert communication. \cite{refIII_Eva_20} proposed StegAbb, a GPT-3-based linguistic steganographic method for SDRs. 
Leveraging GPT-3's language generation capabilities, StegAbb transformed cryptographic abbreviations into natural-looking cover texts, optimizing the security-textuality balance.

\subsubsection{Data Link Layer}
Eavesdropping at this layer takes advantage of unencrypted protocol frame structures, such as MAC addresses, and inherent protocol vulnerabilities, facilitating data interception, tampering, and impersonation. 
Intrusion detection systems (IDSs) function as essential safeguards for SAGINs, integrating with GAI to facilitate adaptive threat detection, thereby ensuring data confidentiality and integrity.

In \cite{refIII_Confidentiality_MAC_1}, the authors utilised WGAN to generate network traffic data, addressing data scarcity and class imbalance for multi-stage attack detection in distributed SAGINs.
To mitigate the training instability and adversarial vulnerability of distributed GANs, \cite{refIII_Confidentiality_MAC_2} proposed a GAN-based robust intrusion detection system for the security of the Internet of Flying Things (IoFT), addressing the issue of limited attack diversity in public datasets. 
To enhance few-shot detection, \cite{refIII_Confidentiality_MAC_3} introduced a semi-supervised GAN for anomaly detection. 
The method addressed redundant feature issues and data imbalance in high-dimensional traffic data targeting SAGINs by employing positive-sample-only training. 

Additionally, FL facilitates the improvement of model generalisation in generative artificial intelligence by enabling decentralised collaborative training while maintaining data privacy \cite{refIII_Confidentiality_MAC_4}.
A framework for federated learning (FedDWM) that incorporates conditional GAN (CGAN) was proposed in \cite{refIII_Confidentiality_MAC_5} to tackle data privacy and class imbalance issues in IDS for SAGINs.  
The proposed algorithm utilises CGAN for synthesising attack samples to achieve balanced training and incorporates dynamic weight-momentum aggregation, resulting in improved model convergence and enhanced detection robustness.  Evaluations of the CIC-IDS2017 and CSE-CIC-IDS2018 datasets revealed an accuracy of 95.74\% and an F1-score of 94.29\%.  GAI-FL introduces a new framework for secure distributed content generation in SAGINs.

In addressing the multi-class IDS, \cite{refIII_Confidentiality_MAC_8} introduced an autoencoder-based multi-task learning (MTL) framework aimed at IoT networks.  To address the limitations of single-task learning in the context of rare attacks and data imbalance, this method combined convolutional autoencoders for feature extraction with multi-task learning parameter-sharing mechanisms, further improved by stochastic weight averaging for model optimisation. 
The feature generation capability of the self-encoder serves as a foundation for botnet detection.  For instance, \cite{refIII_Confidentiality_MAC_9} combined traditional machine learning with GANs to detect dynamic botnet traffic, achieving an accuracy exceeding 98\% on the UNSW-NB15 dataset while minimising false positives.

To address the dynamic characteristics of SAGINs, \cite{refIII_Confidentiality_MAC_10} introduced a Wasserstein-distance-based composite GAN (WC-GAN) for dynamic access control in Internet of Vehicles (IoV) systems, effectively mitigating mode collapse and gradient vanishing issues present in conventional GANs.  The WC-GAN produced synthetic behavioural data to enhance limited training samples and integrated it with real datasets to train a neural network for real-time risk assessment.  The evaluations indicated that the hybrid-trained risk predictor achieved an accuracy of 87\% and a false-negative rate of 5.2\%.

\begin{table*}[htbp]
	\renewcommand{\arraystretch}{2}
	\centering
	\caption{Summary of GAI Solutions for Security Threats Affecting Confidentiality in SAGINs, Where Multiple Attacks in Data Link Layer Refer to Data tampering, Privacy Leakage, and Impersonation, and  Multiple Attacks in Network Layer Refer to Privacy Leakage and Traffic Eavesdropping.}
	\newcolumntype{W}{!{\vrule width 0.5pt}}
	\setlength{\tabcolsep}{4pt}
	\renewcommand{\arraystretch}{1.2}
	\begin{tabular}{c WW c | c | c |c |c |p{6.7cm}}
		\rowcolor{gray!20}
		\Xhline{1pt}
		\multirow{1}{*}{\textbf{{Layer}}}
		&\multirow{1}{*}{\textbf{{Type}}} &\multirow{1}{*}{\textbf{Ref.}}& \multirow{1}{*}{\textbf{Segment}} &\multirow{1}{*}{\textbf{Security Threats}} & \multirow{1}{*}{\textbf{GAI Approach}}& \makecell[c]{\textbf{Description}} \\
		
		\Xhline{0.7pt}
		&\multirow{3}{*}[-25pt]{\shortstack{\textbf{Encryption} }}&\makecell[c]{\cite{refIII_Eva_2}}&\makecell[c]{Each segment}&\makecell[c]{Eavesdropping} & \makecell[c]{WGAN-AE} & \makecell[c]{A physical
			layer key generation method} 
		\\
		\cline{3-7}
		
		& &
		\makecell[c]{\cite{refIII_Eva_5}}& \makecell[c]{Each segment}&\makecell[c]{Various attacks}  & \makecell[c]{GAN} & \makecell[c]{A multiparty
			adversarial encryption model} 
		\\
		\cline{3-7}
		
		& &
		\makecell[c]{\cite{refIII_Eva_6}}& \makecell[c]{Each segment}& \makecell[c]{Intelligent \\ Eavesdropping}  & \makecell[c]{AE for VAE} & \makecell[c]{A secure autoprecoder for MIMO wiretap channels} 
		\\
		\cline{3-7}
		& &
		\makecell[c]{\cite{refIII_Eva_8}}& \makecell[c]{Each segment}&\makecell[c]{privacy leakage} & \makecell[c]{GAN} & \makecell[c]{An encrypted semantic system for privacy preserving} 
		\\
		\cline{3-7}
		
		& &
		\makecell[c]{\cite{refIII_Eva_4}}& \makecell[c]{Air}&\makecell[c]{ Eavesdropping} & \makecell[c]{GAN}& \makecell[c]{An intelligent soft handover for UAV-enabled framework} 
		\\
		\cline{3-7}
		
		&	&
		\makecell[c]{\cite{refIII_Eva_3}}& \makecell[c]{Ground}&  \makecell[c]{Key leakage} & \makecell[c]{GAN} & \makecell[c]{Key defense in dynamic light-fidelity  networks} 
		\\
		\cline{2-7}

		\multirow{3}{*}[-18pt]{\shortstack{\textbf{Physical}\\ \textbf{Layer} }}& \multirow{3}{*}[-20pt]{\shortstack{\textbf{Physical Layer}\\ \textbf{Security}}}&
		\makecell[c]{\cite{refIII_Eva_9}}& \makecell[c]{Each segment}&\makecell[c]{Eavesdropping}  & \makecell[c]{GAN} & \makecell[c]{An enhanced and robust modulation classifier} 
		\\
		\cline{3-7}
		
		& &
		\makecell[c]{\cite{refIII_Eva_14}}& \makecell[c]{Satellite/Air}&\makecell[c]{Eavesdropping} & \makecell[c]{LSTM-GAN} & \makecell[c]{A GAI-based DRL framework for SEE} 
		\\
		\cline{3-7}
		
		& &
		\makecell[c]{\cite{refIII_Eva_12}}& \makecell[c]{Air}& \makecell[c]{Eavesdropping}& \makecell[c]{GDM} & \makecell[c]{Beamforming for multi-objective SEE optimization } 
		\\
		\cline{3-7}
		
		& &
		\makecell[c]{\cite{refIII_Eva_10}} & \makecell[c]{Ground}&  \makecell[c]{Eavesdropping} &\makecell[c]{GDM} & \makecell[c]{A actor-critic GDM-enhanced beamforming scheme} 
		\\
		\cline{3-7}
		
		& &
		\makecell[c]{\cite{refIII_Eva_11}}& \makecell[c]{Ground}& \makecell[c]{Eavesdropping}  & \makecell[c]{GAN } & \makecell[c]{An adversarial defense embedded waveform  design} 
		\\
		\cline{3-7}

		&&
		\makecell[c]{\cite{refIII_Eva_13}} & \makecell[c]{Ground}& \makecell[c]{Eavesdropping} & \makecell[c]{MoE-GDM} & \makecell[c]{A MoE-GDM-based  resource allocation strategy} 
		\\
		\cline{2-7}

		& \multirow{3}{*}[-20pt]{\shortstack{\textbf{Covert}\\ \textbf{Communication}}}&
		\makecell[c]{\cite{refIII_Eva_15}}& \makecell[c]{Ground}& \makecell[c]{Eavesdropping}& \makecell[c]{GAN} & \makecell[c]{An optimization framework to counter attackers} 
		\\
		\cline{3-7}
		
		& &
		\makecell[c]{\cite{refIII_Eva_16}}& \makecell[c]{Air}&\makecell[c]{Eavesdropping}   & \makecell[c]{GAN} & \makecell[c]{A joint trajectory and power optimization algorithm} 
		\\
		\cline{3-7}
		
		& &
		\makecell[c]{\cite{refIII_Eva_17}}& \makecell[c]{Water/Air}&\makecell[c]{Eavesdropping}& \makecell[c]{AE-GAN} & \makecell[c]{A covert signal generation scheme for OWC} 
		\\
		\cline{3-7}
		
		& &
		\makecell[c]{\cite{refIII_Eva_18}}& \makecell[c]{Satellite/Air}& \makecell[c]{Eavesdropping}  & \makecell[c]{DD-GAN} & \makecell[c]{Optimization for AAV's power and trajectory} 
		\\
		\cline{3-7}
		
		& &
		\makecell[c]{\cite{refIII_Eva_19}}& \makecell[c]{Air/Ground}&\makecell[c]{Eavesdropping}& \makecell[c]{GAN} & \makecell[c]{A covert signal design and detection framework} 
		\\
		\cline{3-7}
		
		& &
		\makecell[c]{\cite{refIII_Eva_20}}& \makecell[c]{Each segment}&\makecell[c]{Eavesdropping}   & \makecell[c]{LLM} & \makecell[c]{A GPT-3-based linguistic steganographic method} 
		\\
		\cline{1-7}
		
		& \multirow{3}{*}[-5pt]{\shortstack{\textbf{Intrusion}\\\textbf{Detection} }}&
		\makecell[c]{\cite{refIII_Confidentiality_MAC_3}}& \makecell[c]{Each segment}& \makecell[c]{Data tampering}   & \makecell[c]{GAN} & \makecell[c]{A hybrid high-dimensional anomaly detection model} 
		\\
		\cline{3-7}
		\multirow{3}{*}[-15pt]{\shortstack{\textbf{Data Link} \\ \textbf{Layer} }}& &
		\makecell[c]{\cite{refIII_Confidentiality_MAC_5}}& \makecell[c]{Space/Ground}&\makecell[c]{Privacy leakage} & \makecell[c]{CGAN} & \makecell[c]{CGAN-FL for data privacy and class imbalance} 
		\\
		\cline{3-7}
		
		&&
		\makecell[c]{\cite{refIII_Confidentiality_MAC_1}}&\makecell[c]{Ground}&\makecell[c]{Privacy leakage}  & \makecell[c]{WGAN} & \makecell[c]{A traffic data generation scheme} 
		\\
		\cline{3-7}
		
		& &
		\makecell[c]{\cite{refIII_Confidentiality_MAC_2}} & \makecell[c]{Ground}& \makecell[c]{Privacy leakage} & \makecell[c]{GAN} & \makecell[c]{A GAN-based robust IDS  for IoFT security} 
		\\
		\cline{2-7}
		
		&\multirow{3}{*}[-2pt]{\shortstack{\textbf{Multi-class}\\ \textbf{Intrusion}\\ \textbf{Detection} }}&
		\makecell[c]{\cite{refIII_Confidentiality_MAC_9}}& \makecell[c]{Each segment}&\makecell[c]{Impersonation}   & \makecell[c]{GAN} & \makecell[c]{A hybrid approach for
			dynamic botnet traffic detection} 
		\\
		\cline{3-7}
		
		& &
		\makecell[c]{\cite{refIII_Confidentiality_MAC_8}}& \makecell[c]{Ground}&\makecell[c]{Multiple attacks}  & \makecell[c]{VAE}& \makecell[c]{Autoencoder-based multi-task learning framework for IoT} 
		\\
		\cline{3-7}
		
		& &
		\makecell[c]{\cite{refIII_Confidentiality_MAC_10}}& \makecell[c]{Ground}&\makecell[c]{Multiple attacks}   & \makecell[c]{WC-GAN} & \makecell[c]{A wasserstein-distance-based composite GAN in IoV} 
		\\
		\cline{1-7}
		
		\multirow{3}{*}[-13pt]{\shortstack{\textbf{Network} \\\textbf{Layer}  }}& \multirow{3}{*}[-3pt]{\shortstack{\textbf{Traffic}\\\textbf{Encryption} }}&
		\makecell[c]{\cite{refIII_Confidentiality_Network_1}}&\makecell[c]{Each Segment}& \makecell[c]{Privacy leakage} &\makecell[c]{GAN}   & \makecell[c]{Traffic feature hiding model for encrypted traffic analysis} 
		\\
		\cline{3-7}
		
		& &
		\makecell[c]{\cite{refIII_Confidentiality_Network_6}}& \makecell[c]{Each segment}& \makecell[c]{Eavesdropping} & \makecell[c]{GAN} & \makecell[c]{A GAN-blockchain integrated secure routing framework} 
		\\
		\cline{3-7}
		
		& &
		\makecell[c]{\cite{refIII_Confidentiality_Network_2}}	& \makecell[c]{Ground}& \makecell[c]{Eavesdropping} & \makecell[c]{GAN} & \makecell[c]{A GAN-based chaotic
			logistic encryption method for IoV} 
		\\
		\cline{2-7}

		& \multirow{3}{*}[4pt]{\shortstack{ \textbf{Intrusion}\\ \textbf{Detection} }}&
		\makecell[c]{\cite{refIII_Confidentiality_Network_3}}& \makecell[c]{Each segment}&\makecell[c]{Multiple attacks} & \makecell[c]{ LSTM-GAN} & \makecell[c]{LSTM-GAN for dynamic defense and attack prediction} 
		\\
		\cline{3-7}
		
		& &
		\makecell[c]{\cite{refIII_Confidentiality_Network_8}}& \makecell[c]{Ground}& \makecell[c]{Multiple attacks} & \makecell[c]{ WC-GAN} & \makecell[c]{A proactive prediction solution for IoV} 
		\\
		
		\Xhline{1pt}	
	\end{tabular}
	\label{table_GAI_confidentiality}
\end{table*}

\subsubsection{Network Layer}
Breaches at this layer pose a risk of exposing sensitive data, such as user identities and interaction patterns, which directly contradicts SAIGN's objectives of preserving privacy. 
GDM can be utilized to synthesize data to replace the real data for privacy protection and VAE can be adopted to generate synthetic samples of rare class for training against malicious data injection \cite{refIII_Confidentiality_Network_supply}.

A GAN-based traffic feature hiding model (TFHM) was proposed in \cite{refIII_Confidentiality_Network_1}, utilising GRU-enhanced generators to maintain contextual dependencies in the context of encrypted traffic analysis.
Building on this, \cite{refIII_Confidentiality_Network_2} introduced a GAN-based chaotic logistic encryption technique for the protection of IoV trajectory data. 
This approach addresses the limitations of traditional chaotic encryption related to static key distribution and low plaintext-key correlation, while exhibiting resilience against noise, MITM, and differential attacks.
Furthermore, the integration of GAN with LSTM facilitates dynamic defence and attack prediction.  Fig. \ref{SectionIII_outline_p2} demonstrates that \cite{refIII_Confidentiality_Network_3} explored a hybrid LSTM-GAN model to mitigate the issues of high false positives and inadequate generalisation identified in \cite{refIII_Confidentiality_Network_2}. This approach combines LSTM for capturing sequential dependencies with GANs for generating adversarial scenarios.  Experimental results indicated a detection accuracy of 92.5\%, a 15\% decrease in false positives relative to standalone LSTM, and a 25\% enhancement in threat prediction.

Additionally, \cite{refIII_Confidentiality_Network_6} proposed a secure routing framework that integrates GAN and blockchain technologies for edge-assisted wireless sensor networks (WSNs). This framework facilitates the co-optimization of energy efficiency and confidentiality by employing dynamic bio-inspired clustering and reinforcement learning-based scheduling. 
Furthermore, \cite{refIII_Confidentiality_Network_8} enhanced security by transitioning from reactive defence to proactive prediction.  Utilising dynamic risk modelling and multi-policy collaboration with a Wasserstein distance-based combined GAN, communication efficiency was optimised while ensuring the confidentiality of IoV in SAGINs.

\subsubsection{Summary and Lessons Learned}

As summarized in Table \ref{table_GAI_confidentiality}, 
GAI enables dynamic encryption from channel characteristics, enhances physical layer security via noise-like signals, and optimizes covert communication parameters to evade detection. 
In addition, GAI synthesizes adversarial traffic for robust IDS, mitigates data scarcity in threat models, and obfuscates traffic patterns. 
The inherent adaptability of GAI models allows FL to learn and respond to the dynamic topologies, volatile channels, and diverse threat landscapes characteristic of SAGINs, offering a level of agility unattainable by static security solutions.
Despite these significant promise, challenges remain:
\begin{itemize}
\item[$\bullet$] \textit{Resource constraints:} 
High complexity and energy consumption hinder deployment on resource-limited edge/aerial/satellite nodes within SAGINs.
\item[$\bullet$] \textit{Fighting vulnerability:} GAI vulnerability to "evolved eavesdroppers" enables training mimicry and evasion attacks.
\item[$\bullet$] \textit{Cross-domain integration:} Heterogeneous segment integration (satellite/aerial/terrestrial) demands standardized frameworks and cross-domain coordination. 
\item[$\bullet$] \textit{High scalability:} Scalability to massive mobile networks requires real-time inference via algorithmic/system co-design.
\end{itemize}
\vspace{-1em}
\section{GAI for Integrity and Availability Security in SAGINs: Current Research}
This section summarizes integrity tampering and availability disruptions that may be encountered in SAGINs, and overviews the corresponding  GAI-based solutions in detail.

\vspace{-1.5em}
\subsection{GAI for Integrity Tampering}

\begin{figure*}[!t]
		\vspace*{-1em}
	\centering
	\includegraphics[width=5in]{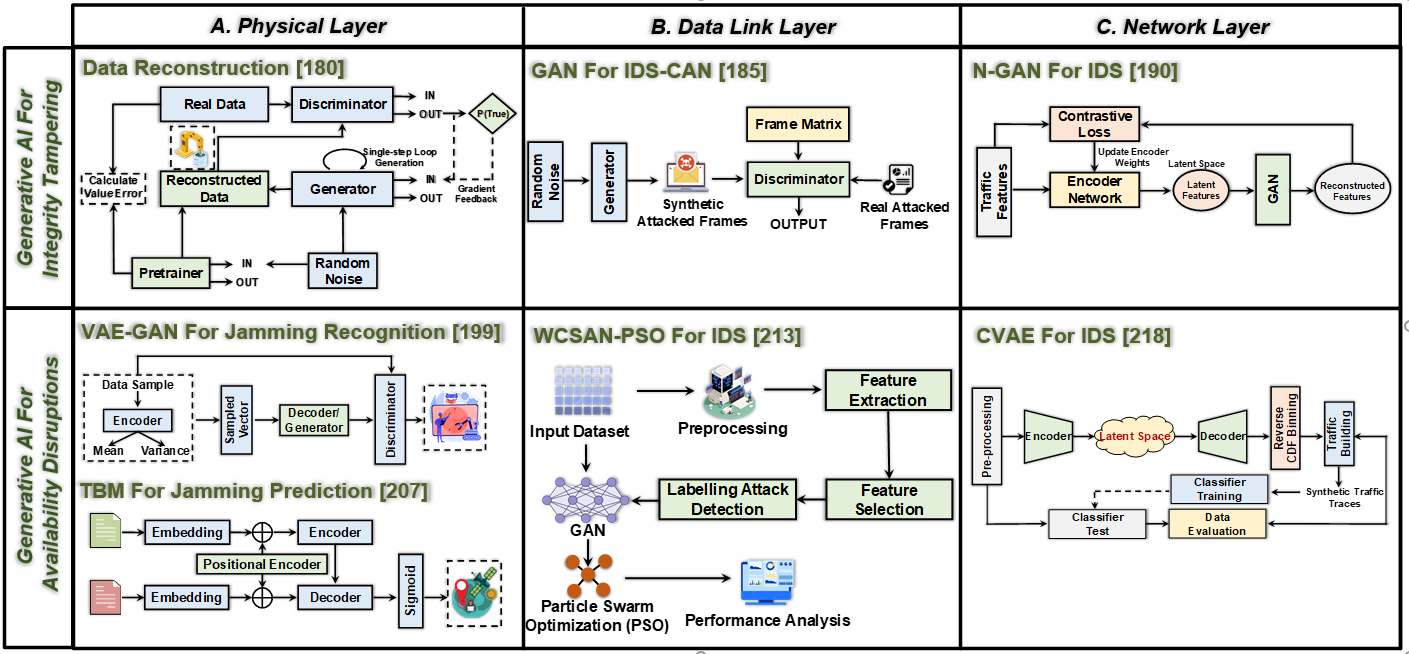}
	\caption{Typical GAI-based approaches for integrity tampering and availability disruptions in SAGINs. For integrity tampering, \cite{refIII_integrity_phy_15} demonstrates a data reconstruction method based on MTS-GAN; \cite{refIII_integrity_MAC_2} introduces a GAN-based IDS for CAN; \cite{refIII_integrity_Network_2} illustrates an encoder-based GAN for IDS. For availability disruptions, \cite{refIII_jamming_4} and \cite{refIII_jamming_12} present VAE-GAN-based and TBM-based approaches for jamming recognition and jamming prediction, respectively; \cite{refIII_Availability_MAC_3} integrates PSO with GAN for IDS; \cite{refIII_Availability_Network_5} introduces how to apply VAE in IDS.}
	\label{SectionIII_outline-p3}
		\vspace*{-1em}
\end{figure*}

\subsubsection{Physical Layer} Attacks such as jamming, spoofing, and tampering can all affect data integrity to some degree\cite{refIII_integrity_phy_1}, \cite{refII_A_Authentication}. Data anomaly detection and data reconstruction are the primary means of verifying and ensuring data integrity \cite{refI_B_7}.

\textit{\textbf{Anomaly detection:}} It identifies deviations from normal patterns in datasets.
GAI improves this capability by offering self-supervised anomalous data samples for training, facilitating high-dimensional anomaly detection and accurate localisation in complex and unfamiliar scenarios.  A discriminative autoencoding framework (Dis-AE) was introduced by \cite{refIII_integrity_phy_3}, which synergistically integrates GAN and autoencoders for semi-supervised anomaly detection. 
Dis-AE encounters implementation challenges in SAGINs owing to the complexity of high-speed spatio-temporal data. 
To tackle these challenges, \cite{refIII_integrity_phy_4} introduced an anomaly detection model that combines improved GAN (IGAN) with LSTM networks for ADS-B-based air transportation systems. 

In addition to high-precision detection, \cite{refIII_integrity_phy_5} facilitates accurate localisation of anomalous data.  The authors introduced a cross-correlation graph-based encoder-decoder GAN for the purposes of anomaly detection and localisation. 
Expanding upon \cite{refIII_integrity_phy_5}, \cite{refIII_integrity_phy_6} addressed the issue of pattern sensitivity in scenarios involving sample imbalance.  A hybrid model combining K-means, VAE, and Support Vector Data Description (SVDD) was developed to tackle anomaly detection in multidimensional spacecraft telemetry characterised by imbalanced and unlabelled samples.
In the context of contaminated datasets, frameworks based on normalising flows have been effective in reducing training biases \cite{refIII_integrity_phy_7}. Additionally, a convolutional variational autoencoder-GAN (CVAE-GAN) has facilitated zero-shot anomaly detection in AAV systems in the absence of contaminated data \cite{refIII_integrity_phy_8}. 

To enhance defence against multimodal threat coverage in AAV swarms, \cite{refIII_jamming_2} proposed a hybrid anomaly detection system that integrates GANs and FL.
The framework addresses in-flight anomalies, such as GPS jamming and spoofing, as well as network attacks including blackhole, grayhole, and flooding. It integrates an unsupervised stacked autoencoder with federated learning for flight anomaly detection and employs a supervised LightGBM classifier augmented by generative adversarial networks for data balancing.
To address data heterogeneity, \cite{refIII_integrity_phy_10} utilised deep federated learning (DFL) and VAEs as local models in satellite communication systems, focussing on mitigating privacy leakage and fairness issues inherent in centralised methods. 

Current methodologies primarily focus on recognising the information contained within the IP/TCP header, rather than directly analysing the payload.  The multimodal process of LLM interprets data content as a specific "language" of the device, facilitating deeper semantic feature extraction from network behavioural characteristics to content semantics.
A content-based IoT detection framework utilising ChatGPT 4.0 Turbo embeddings and LSTM temporal modelling attained an accuracy of 99.75\% through the extraction of semantic features from network behaviour \cite{refIII_integrity_phy_11}.

\textit{\textbf{Data reconstruction:}}  Data reconstruction in SAGINs faces challenges including heterogeneity and resource constraints.   GAI models exhibit the ability to reconstruct high-dimensional data, including channel states and images, from low-dimensional observations.
For instance, \cite{refIII_integrity_phy_12} presented an adversarial autoencoder (AAE) framework designed for simultaneous anomaly detection and signal reconstruction in sub-Nyquist sampled spectrum monitoring scenarios relevant to 6G and satellite systems. 
This architecture combines an autoencoder with adversarial discriminators to enable simultaneous signal recovery.
The AAE exhibits a delay in detection and has restricted capabilities for attack localisation. 
In response to these limitations, \cite{refIII_integrity_phy_13} introduced a CVAE-based deep learning approach for the detection of cyber-physical attacks in distribution systems, attaining a detection accuracy of 98\% on the BATADAL dataset. 

To improve high-precision data reconstruction, \cite{refIII_integrity_phy_14} established a framework based on a denoising diffusion probabilistic model (DDPM) for the compression of satellite monitoring data.   The proposed method mitigates the limitations of existing dictionary-dependent sparse representation techniques by utilising deep generative modelling to effectively capture the intricate distributions of vibration signals.
Similarly, as shown in Fig. \ref{SectionIII_outline-p3}, \cite{refIII_integrity_phy_15} proposed a high-precision reconstruction method utilising multi-component time series GAN (MTS-GAN) to address electromagnetic data loss caused by equipment failures in SAGINs.

Ensuring the integrity of image data is essential in SAGINs.
A multi-modality semantic-aware framework for vehicular networks was introduced by \cite{refII_B_GAN}, utilising GAI-based image generation and reconstruction.
A single-input multitask reconstruction framework utilising an efficient pyramidal GAN for remote sensing images was proposed by \cite{refIII_integrity_phy_16}.
This method is hindered by over-smoothed textures resulting from pixel-wise reconstruction loss.  A GAN-based super-resolution method (SRGAN) for AAV image enhancement was designed to address the critical challenge of low-resolution artefacts resulting from hardware constraints and motion blur \cite{refIII_integrity_phy_17}.  
SRGAN primarily emphasises pixel-wise feature mapping, thereby overlooking the spatial-spectral interdependencies present in multisensor data. 
In response to this challenge, \cite{refIII_integrity_phy_18} proposed TransGAN-CFR, a novel framework that combines transformer and GAN for cross-modal reconstruction.  The method utilised window-based multi-head self-attention and depthwise-convolution-enhanced feedforward networks to achieve accurate multispectral texture recovery.

\subsubsection{Data Link Layer}
Integrity at the data link layer requires that data frames remain unchanged and consistent throughout transmission, protecting against unauthorised alteration or corruption.  Threats predominantly leverage protocol frame structures, such as ethernet frame tampering, and vulnerabilities, including ARP/DHCP spoofing and VLAN hopping attacks, to forge, inject, or truncate frames.

The authors in \cite{refIII_integrity_MAC_1} introduced a GAN-enhanced hybrid machine learning framework aimed at the real-time detection of ARP spoofing attacks.  The approach combined dynamic ARP cache validation with CNN classifiers, utilising GAN-generated synthetic attack traffic to enhance model generalisation to previously unobserved attack patterns.  It relies on static feature engineering and lacks the capability to dynamically generate adversarial attack samples.

The enhanced GAN-based IDS for automotive CAN networks was proposed in \cite{refIII_integrity_MAC_2}, as illustrated in Fig. \ref{SectionIII_outline-p3}.  An enhanced discriminator was developed to detect data tampering through the verification of signal value ranges.  Experiments indicated enhanced efficacy in identifying DoS, injection, masquerade, and data tampering attacks, attaining 99.9\% precision and recall.

To address the dynamic and heterogeneous characteristics of SAGINs, \cite{refIII_integrity_MAC_3} proposed a GAN-enhanced anomaly detection framework that generates synthetic attack patterns, such as data tampering and frame spoofing.  The lightweight classifiers were developed to detect unauthorised alterations in network frames. 
This framework is limited as it does not incorporate proactive remediation mechanisms and functions exclusively at the threat identification stage.
In response, \cite{refIII_integrity_MAC_4} proposed a hybrid security evaluation framework for future IoV, which addresses integrity verification challenges in complex attack scenarios through the multidimensional integration of vehicle dynamics modelling. 
The authors in \cite{refIII_integrity_MAC_5} integrated a generative neural network for data imputation with blockchain technology to provide tamper-proof storage, facilitating the prediction of missing attributes and ensuring dataset completeness.
Evaluations demonstrated a 17\%-23\% improvement in imputation accuracy compared to traditional SVM method, while blockchain consensus mechanisms reduced tampering risks by 34\%.

\subsubsection{Network Layer}
The integrity of the network layer in SAGINs is crucial for ensuring reliable routing and packet delivery; however, it is susceptible to threats such as MITM attacks and malicious routing injection, particularly during dynamic topology changes.
GAI tackles these challenges by employing adversarial traffic generation for routing verification hardening, spatiotemporal anomaly reconstruction, and topology-aware integrity self-healing mechanisms, thereby establishing a proactive defence against multi-domain coordinated attacks.

In the context of anomaly detection, \cite{refIII_integrity_Network_1} proposed a GAN-based framework that utilised latent interactions to identify hardware trojans (HTs) affecting routing components and to classify attack types.  Evaluations indicated a 91
Additionally, \cite{refIII_integrity_Network_2} introduced N-GAN to address the high false alarm issue by integrating partial attack semantics via weak supervision, in contrast to the unsupervised GAN-based IDS presented in \cite{refIII_integrity_Network_1}. 
The study in \cite{refIII_integrity_Network_3} introduced a GAN-based method for generating synthetic attack traffic to mitigate issues of data scarcity and class imbalance in intrusion detection systems (IDS).  The authors utilised vanilla GAN, WGAN, and conditional tabular GAN (CTGAN) to generate high-fidelity Botnet attack samples from the CIC-IDS2017 dataset, resulting in an increase in the Botnet detection F1-score from 0.60 to 0.90, while ensuring robustness across various attack classes.
Furthermore, \cite{refIII_Authenticity_network_9} examined a GAN-based dynamic MAD, in which the generator produced varied attack patterns while the discriminator worked to improve detection robustness collaboratively.

Stealthy cyber attacks deliberately avoid detection by imitating legitimate network behaviours and conforming to protocol specifications while carrying out malicious payloads. 
In \cite{refIII_integrity_Network_4}, the authors developed a hybrid adversarial framework that integrates GAN and constrained optimisation to generate protocol-compliant attack samples in industrial control systems.  The approach focused on injection, function code, and reconnaissance attacks, implementing ICS-specific constraints such as immutable packet headers while modifying negotiable features. This resulted in over 80
Extending beyond single GANs, \cite{refIII_integrity_Network_5} designed a hybrid GAN-driven framework for secure cluster-based routing in ad-hoc networks, jointly addressing energy efficiency and malicious node resilience. 
Additionally, LLM can collaborate effectively with GANs.   A GAN-based intelligent fuzzer (DCGAN-MNFuzz) was introduced by \cite{refIII_integrity_Network_6}, which generated protocol-aware mutated payloads through adversarial training. This was combined with an LLM-powered dynamic risk assessment engine, validated in a smart airport IoT testbed.

In addition to IDS design, GAI enables protocol and packet encryption.  In this context, \cite{refIII_integrity_Network_7} proposed a GAN-based framework for generating balanced datasets to mitigate imbalanced attack detection in drone video analytics, specifically focussing on replay, packet injection, and physical capture threats.  The integration of MAVLink's continuous authentication methods, such as digital signatures, along with lightweight encryption, improved data integrity against network-layer tampering while preserving operational efficiency.

\subsubsection{Summary and Lessons Learned}

\begin{table*}[htbp]
	\renewcommand{\arraystretch}{2}
	\centering
	\caption{Summary of GAI Solutions for Security Threats Affecting Integrity in SAGINs. Multiple Attacks in Physical Layer Refer to jamming, spoofing, and
	tampering; Multiple Attacks in Data Link Layer Refer to ARP/DHCP Spoofing and VLAN Hopping Attack; Multiple Attacks in Network Layer Refer to MITM Attacks and Malicious Routing Injection.}
	\newcolumntype{W}{!{\vrule width 0.5pt}}
	\setlength{\tabcolsep}{4pt}
	\renewcommand{\arraystretch}{1.2}
	\begin{tabular}{c WW c | c | c |c |c |p{6cm}}
		\rowcolor{gray!20}
		\Xhline{1pt}
		\multirow{1}{*}{\textbf{{Layer}}}
		&\multirow{1}{*}{\textbf{{Type}}} &\multirow{1}{*}{\textbf{Ref.}}& \multirow{1}{*}{\textbf{Segment}} &\multirow{1}{*}{\textbf{Security Threats}} & \multirow{1}{*}{\textbf{GAI Approach}}& \makecell[c]{\textbf{Description}} \\
		\Xhline{0.7pt}
		\multirow{3}{*}[-65pt]{\shortstack{\textbf{Physical}\\ \textbf{Layer} }}&\multirow{3}{*}[-30pt]{\shortstack{\textbf{Anomaly }\\\textbf{Detection} }}&
		\makecell[c]{\cite{refIII_integrity_phy_3}}&\makecell[c]{Each segment}& \makecell[c]{Data tampering} & \makecell[c]{AE-GAN} & \makecell[c]{A semi-supervised anomaly detection model} 
		\\
		\cline{3-7}
				
		& &
		\makecell[c]{\cite{refIII_integrity_phy_5}}& \makecell[c]{Each segment}&\makecell[c]{Data tampering} & \makecell[c]{GAN} & \makecell[c]{A cross-correlation graph-based encoder-decoder} 
		\\
		\cline{3-7}	
%
		&&
		\makecell[c]{\cite{refIII_integrity_phy_6}} & \makecell[c]{Space}&\makecell[c]{Data tampering}& \makecell[c]{VAE}& \makecell[c]{A hybrid high-precision detection model} 
		\\
		\cline{3-7}
		
		& &
		\makecell[c]{\cite{refIII_integrity_phy_10}}& \makecell[c]{Space/Ground}&\makecell[c]{Data tampering}  & \makecell[c]{VAE} & \makecell[c]{A heterogeneous multidimensional detection model } 
		\\
		\cline{3-7}

		& &
		\makecell[c]{\cite{refIII_integrity_phy_4}}& \makecell[c]{Air}&\makecell[c]{Data tampering} & \makecell[c]{LSTM-GAN} & \makecell[c]{A detection model for ADSB-based system} 
		\\
		\cline{3-7}

		& &
		\makecell[c]{\cite{refIII_integrity_phy_8}}& \makecell[c]{Air}& \makecell[c]{Multiple attacks}& \makecell[c]{VAE-GAN} & \makecell[c]{A CVAE-GAN for zero-shot learning for AAV} 
		\\
		\cline{3-7}
		
		& &
		\makecell[c]{\cite{refIII_jamming_2}} & \makecell[c]{Air}& \makecell[c]{Multiple attacks } & \makecell[c]{GAN}  & \makecell[c]{Anomaly detection for  anomalies and attacks} 
		\\
		\cline{3-7}

		& &
		\makecell[c]{\cite{refIII_integrity_phy_11}}& \makecell[c]{Ground}& \makecell[c]{Multiple attacks} &\makecell[c]{LLM} & \makecell[c]{Anomaly detection  using
			ChatGPT 4.0 Turbo} 
		\\
		
		\cline{2-7}
		
		& &
		\makecell[c]{\cite{refIII_integrity_phy_15}}& \makecell[c]{Space/Air/Ground}&\makecell[c]{Multiple attacks}& \makecell[c]{GAN}  & \makecell[c]{A high-precision reconstruction method} 
		\\
		\cline{3-7}
		
		& &
		\makecell[c]{\cite{refIII_integrity_phy_16}}& \makecell[c]{Satellite/Ground}&\makecell[c]{Multiple attacks}  & \makecell[c]{GAN} & \makecell[c]{A multitask image reconstruction framework} 
		\\
		\cline{3-7}
		
		& \multirow{3}{*}[-4pt]{\shortstack{ \textbf{Data}\\ \textbf{Reconstruction} }}&
		\makecell[c]{\cite{refIII_integrity_phy_12}}& \makecell[c]{Space}& \makecell[c]{Multiple attacks} & \makecell[c]{VAE} & \makecell[c]{Joint anomaly detection and signal reconstruction} 
		\\
		\cline{3-7}
		
		& &
		\makecell[c]{\cite{refIII_integrity_phy_14}}& \makecell[c]{Space}&\makecell[c]{Multiple attacks}   & \makecell[c]{GDM} & \makecell[c]{High-precision  reconstruction for satellite data } 
		\\
		\cline{3-7}
		
		& &
		\makecell[c]{\cite{refIII_integrity_phy_18}}& \makecell[c]{Space}&\makecell[c]{Multiple attacks}   & \makecell[c]{TBM-GAN} & \makecell[c]{TransGAN-CFR for cross-modal  reconstruction} 
		\\
		\cline{3-7}
		& &
		\makecell[c]{\cite{refIII_integrity_phy_17}}& \makecell[c]{Air}&\makecell[c]{Multiple attacks}   & \makecell[c]{GAN} & \makecell[c]{Super-resolution for AAV image enhancement} 
		\\
		\cline{3-7}
		
		& &
		\makecell[c]{\cite{refIII_integrity_phy_13}}& \makecell[c]{Ground}& \makecell[c]{Multiple attacks}  & \makecell[c]{VAE} & \makecell[c]{A detection approach for cyber-physical attack} 
		\\
		\cline{3-7}
		
		& &
		\makecell[c]{\cite{refII_B_GAN}}& \makecell[c]{Ground}& \makecell[c]{Multiple attacks}  & \makecell[c]{GAI} & \makecell[c]{A multi-modality semantic-aware framework} 
		\\
		\cline{1-7}
		
		\multirow{3}{*}[-11pt]{\shortstack{\textbf{Data Link}\\ \textbf{Layer} }}&\multirow{3}{*}[5pt]{\shortstack{\textbf{Intrusion}\\\textbf{Detection} }} &
		\makecell[c]{\cite{refIII_integrity_MAC_1}}& \makecell[c]{Each segment}& \makecell[c]{ARP
			spoofing} & \makecell[c]{GAN} & \makecell[c]{A  hybrid
			ML framework for real-time detection} 
		\\
		\cline{3-7}
		
		& &
		\makecell[c]{\cite{refIII_integrity_MAC_2}}& \makecell[c]{Each segment}& \makecell[c]{Multiple attacks} & \makecell[c]{GAN} & \makecell[c]{A IDS for automotive CAN networks} 
		\\
		\cline{2-7}
		
		&\multirow{3}{*}[10pt]{\shortstack{\textbf{Prediction} }}  &
		\makecell[c]{\cite{refIII_integrity_MAC_5}}&\makecell[c]{Each segment}&\makecell[c]{Frame tampering}  & \makecell[c]{GAN} & \makecell[c]{Data imputation for secure blockchain} 
		\\
		\cline{2-7}
		
		& \multirow{3}{*}[5pt]{\shortstack{\textbf{Anomaly}\\\textbf{Detection} }}  &
		\makecell[c]{\cite{refIII_integrity_MAC_3}}& \makecell[c]{Ground}&\makecell[c]{Frame
			spoofing}  & \makecell[c]{GAN} & \makecell[c]{The lightweight frame classifier} 
		\\
		\cline{3-7}
		
		& &
		\makecell[c]{\cite{refIII_integrity_MAC_4}}& \makecell[c]{Ground}&\makecell[c]{Frame tampering}  & \makecell[c]{GAN} & \makecell[c]{A security evaluation framework for future IoVs} 
		\\
		\cline{1-7}

		\multirow{3}{*}[-35pt]{\shortstack{\textbf{Network}\\ \textbf{Layer} }}& \multirow{3}{*}[-35pt]{\shortstack{\textbf{Intrusion}\\\textbf{Detection} }}&
		\makecell[c]{\cite{refIII_integrity_Network_1}}&\makecell[c]{Each segment}& \makecell[c]{Packet dropping} &\makecell[c]{GAN}   & \makecell[c]{GAN-based
		 detection for hardware trojans} 
		\\
		\cline{3-7}
		
		& &
		\makecell[c]{\cite{refIII_integrity_Network_2}}& \makecell[c]{Each segment}&\makecell[c]{Packet dropping }  & \makecell[c]{GAN} & \makecell[c]{GAN for partial semantic attack} 
		\\
		\cline{3-7}
		
		& &
		\makecell[c]{\cite{refIII_integrity_Network_3}}& \makecell[c]{Each segment}&\makecell[c]{Multiple attacks}   & \makecell[c]{GAN} & \makecell[c]{A synthetic attack traffic generation approach} 
		\\
		\cline{3-7}

		&&
		\makecell[c]{\cite{refIII_integrity_Network_4}} & \makecell[c]{Each segment}&\makecell[c]{Stealthy attack }  &\makecell[c]{GAN}   & \makecell[c]{A  framework for protocol-compliant attack samples} 
		\\
		\cline{3-7}
		
		& &
		\makecell[c]{\cite{refIII_integrity_Network_7}}& \makecell[c]{Air}& \makecell[c]{Packet injection} & \makecell[c]{GAN} & \makecell[c]{GAN for authentication and lightweight encryption} 
		\\
		\cline{3-7}
		& &
		\makecell[c]{\cite{refIII_integrity_Network_6}}& \makecell[c]{Air/Ground}&\makecell[c]{Multiple attacks}  & \makecell[c]{LLM-GAN} & \makecell[c]{A GAN-based
			intelligent fuzzer} 
		\\
		\cline{3-7}
		
		& &
		\makecell[c]{ \cite{refIII_Authenticity_network_9}}& \makecell[c]{Ground}&\makecell[c]{Multiple attacks }  & \makecell[c]{GAN} & \makecell[c]{A GAN-driven dynamic malicious attack
			detector} 
		\\
		\cline{3-7}
		
		& &
		\makecell[c]{\cite{refIII_integrity_Network_5}}& \makecell[c]{Ground}& \makecell[c]{Stealthy attacks}  & \makecell[c]{GAN} & \makecell[c]{GAI for secure cluster-based routing} 
		\\
	
		\Xhline{1pt}	
	\end{tabular}
	\label{table_GAI_Integrity}
\end{table*}

As summarized in Table \ref{table_GAI_Integrity}, 
GAI enables proactive defense through two synergistic capabilities: high-dimensional anomaly detection and robust data reconstruction. 
Meanwhile, generative models transcend traditional threshold-based methods by enabling semantic-aware validation—interpreting payloads as domain-specific ``languages" and verifying contextual coherence beyond syntactic checks.
Despite these transformative advantages, critical challenges remain:
\begin{itemize}
	\item[$\bullet$] \textit{Resource-needs conflict:} Computational intensity of GAI models (e.g., transformer-GAN hybrids) conflicts with satellite/avionic latency/energy constraints.
	\item[$\bullet$] \textit{Over-reliance on dataset:} Training-data dependence creates vulnerability to adversarial poisoning that may induce model biases.
	\item[$\bullet$] \textit{Explainability deficit:} Explainability deficits in deep generative models hinder breach root-cause analysis.
	\item[$\bullet$]  \textit{Cross-Domain gaps:} Layer-specific defenses (e.g., physical-layer recovery, network-layer verification) lack holistic orchestration across SAGINs.
\end{itemize}


\subsection{GAI for Availability Disruptions}
\subsubsection{Physical Layer} 
Jamming attacks compromise communication availability by transmitting high-power signals that degrade legitimate signal-to-noise ratio (SNR). GAI-based research addresses jamming recognition and mitigation.

\textit{\textbf{Jamming recognition:}} 
Protection mechanisms in SAGINs initiate dynamic spectrum reconfiguration and topology adaptation for swift defence.
GANs utilise zero-sum game dynamics for the recognition of multi-attack signals, akin to spoofing detection \cite{refIII_A_1_4}.
Leveraging this advantage, \cite{refIII_jamming_1} introduced a robust IDS utilising GAN and adversarial sample regularisation, effectively addressing spoofing and jamming attacks encountered by AAVs that depend on GPS navigation. 
Nonetheless, a single GAN would face challenges in managing multi-dimensional data characterised by complex relationships, such as high-resolution images. 
To address this issue, \cite{refIII_jamming_102} developed a conditional tabular GAN (CTGAN) that synthesises data rows from discrete columns, effectively resolving unbalanced distributions.
Furthermore, TBM may be employed to improve the efficacy of GAN.  The DroneDefGANt, as discussed in \cite{refIII_A_1_8}, utilised the multi-head attention mechanism of transformers to address discriminator gradient challenges in GANs.  Furthermore, autoencoders enhance GANs in situations with limited labelled data, such as AAV swarm \cite{refIII_jamming_2}.

In contrast to GANs, VAEs generate interpretable features for jamming recognition by utilising latent space recognition.
In \cite{refIII_jamming_3}, the authors introduced a VAE-based unsupervised framework for detecting anomalous interference in MIMO-OFDM ISAC through reconstruction probabilities.
The advantages of VAEs are particularly evident in few-shot learning.  The study \cite{refIII_jamming_4} explored the latent space of small sample datasets through the use of VAE, subsequently sampling and decoding this latent space to facilitate dataset expansion prior to GAN discrimination.
Centralised VAEs encounter difficulties in distributed SAGINs because they are susceptible to jamming.
A federated augmented aggregate training algorithm was proposed by \cite{refIII_jamming_5}, which integrates spectral function feature extraction through CVAE to address the issue of unknown interference detection in distributed SAGINs. 
Additionally, in response to intelligent jammers that replicate legitimate signals, \cite{refIII_jamming_6} introduced a VAE that integrates a dynamic adaptive spectro-temporal resilient filter (DASTRF), further augmented by a vision transformer (ViT) and LSTM.  The proposed FL-ViT-LSTM-VAE enhances the detection of signal-replicating smart jammers through the optimisation of time-frequency distribution feature separation, thereby ensuring the safe operation of AAVs.

Jamming recognition models are primarily characterised by TBMs\cite{refIII_jamming_7} and GDMs\cite{refIII_jamming_8}.  For instance, \cite{refIII_jamming_7} introduced a distributed radar multi-interference identification method utilising a transformer network and adaptive beamforming to mitigate the decline in identification performance caused by the overlap of multiple interference sources in the time-frequency domain. 
Additionally, to address the issues of inefficient intrusion detection and privacy leakage in SAGIN with heterogeneous data, the authors in \cite{refIII_jamming_8} developed a STINIDF framework that collaboratively trains the conditional diffusion model (DP-CDM) via federated learning (FL) and produces global traffic data by integrating the differential privacy (DP) mechanism.

\textit{\textbf{Jamming mitigation:}} Jamming mitigation involves both active signal suppression and passive data reconstruction methods.  Active mitigation relies on the awareness of jammers and the surrounding environment. 
According to the available literature, \cite{refIII_jamming_9} introduced a collaborative reinforcement learning algorithm that employs a mixture Gaussian distribution model, integrating GAN localisation with time difference of arrival techniques. 
In situations characterised by unknown or incomplete information, \cite{refIII_jamming_10} developed an intelligent spectrum access algorithm that integrates GAN and DRL to mitigate interference, even in cases with up to 90\% missing data. 
Furthermore, \cite{refIII_jamming_11} introduced a vision transformer-based adaptive blind beamforming method (ViT-BF) aimed at jammer suppression.  A proactive jamming prediction method that integrates a pseudo-random algorithm with a transformer module to forecast jammer behaviour was proposed by \cite{refIII_jamming_12}.

Passive data reconstruction emphasises recovery on the receiver's side.  A conditional diffusion model (CDM)-based anti-jamming algorithm was implemented to ensure accurate reception of satellite navigation signals in complex SAGINs \cite{refIII_jamming_13}.
The algorithm incrementally introduced noise and learnt the noise distribution via the forward diffusion process, while systematically denoising the noise-embedded QPSK signal as a condition in the reverse diffusion process.
Furthermore, \cite{refIII_jamming_14} proposed GAN-inspired anti-jamming techniques for semantic communication, ensuring decoding consistency during attacks. 
Additionally, few-shot sample scenarios in IoT networks were examined in \cite{refIII_jamming_15}, highlighting that various jamming styles were rare and challenging to mitigate effectively.  The authors proposed a meta-learning and multi-task approach for source separation to address unknown interference.

\subsubsection{Data Link Layer}
Availability threats, such as MAC flooding and ARP-based DoS, encompass adversarial actions that interfere with protocol operations or deplete resources, consequently denying legitimate access to network services. 
GAI can improve the resilience of the data link layer protocols in SAGINs against jamming or resource-exhaustion attacks through the simulation of intricate adversarial scenarios and the development of adaptive defence strategies.

In \cite{refIII_Availability_MAC_1}, the authors introduced a GAN-enhanced IDS combined with a LSTM-based MAC protocol to mitigate availability threats in underwater wireless sensor networks (UWSNs).  The framework utilises GAN for real-time assessment of acoustic channel quality, including noise patterns and malicious signal interference, alongside LSTM-MAC for adaptive medium access control. This approach dynamically optimises contention access periods and implements reactive jamming to counter DoS attacks, resulting in a detection accuracy of 98.9\% with 5\% malicious nodes present.
Based on this, \cite{refIII_Availability_MAC_2} examined cross-layer attacks that encompass the data link and network layers, including flooding attacks that utilise IPv6 router advertisements, resulting in link layer congestion. 
The authors proposed a DL-based approach for detecting router advertisement flooding DDoS attacks in IPv6 networks.  Integrating feature ranking algorithms with a RNN addresses the vulnerability of the neighbour discovery protocol (NDP), specifically mitigating RA flooding attacks.

The aforementioned studies depend heavily on static feature libraries and are unable to manage adversarial perturbation traffic in real time.
To address this issue, \cite{refIII_Availability_MAC_3} proposed an adversarial attack detection framework (WCSAN-PSO) utilising an optimised weighted conditional stepwise adversarial network alongside particle swarm optimisation, as illustrated in Fig. \ref{SectionIII_outline-p3}.  The integration of GAN-generated adversarial samples with feature selection techniques enhanced the robustness of IDS.  The framework demonstrated an accuracy of 99.36\% for normal traffic and 98.55\% for malicious traffic when assessed using the CIC-IDS2017 dataset.  Utilising GAI to dynamically generate attack patterns and optimise model parameters enhanced the generalisation of IDS against evolving threats, consequently improving service availability.

\begin{table*}[htbp]
	\renewcommand{\arraystretch}{2}
	\centering
	\caption{Summary of GAI Solutions for Security Threats Affecting Availability in SAGINs, Where Multiple Attacks in Network Layer Refer to DDoS Attack and Routing Flooding.}
	\newcolumntype{W}{!{\vrule width 0.5pt}}
	\setlength{\tabcolsep}{4pt}
	\renewcommand{\arraystretch}{1.2}
	\begin{tabular}{c WW c| c | c |c |c |p{7cm}}
		\rowcolor{gray!20}
		\Xhline{1pt}
		\multirow{1}{*}{\textbf{{Layer}}}
		&\multirow{1}{*}{\textbf{{Type}}} &\multirow{1}{*}{\textbf{Ref.}} &\multirow{1}{*}{\textbf{Segment}} &\multirow{1}{*}{\textbf{Security Threats}} & \multirow{1}{*}{\textbf{GAI Approach}}& \makecell[c]{\textbf{Description}} \\
		
		\Xhline{0.7pt}
		\multirow{3}{*}[-75pt]{\shortstack{\textbf{Physical}\\ \textbf{Layer} }}&\multirow{3}{*}[-40pt]{\shortstack{\textbf{Jamming }\\\textbf{Recognition} }}&
		\makecell[c]{\cite{refIII_jamming_1}}&\makecell[c]{Space/Air}&\makecell[c]{GPS jamming}  & \makecell[c]{GAN} & \makecell[c]{A robust IDS based on GAN
			and sample regularization} 
		\\
		\cline{3-7}
		
		&&
		\makecell[l]{\cite{refIII_jamming_102}}& \makecell[c]{Air}& \makecell[c]{Jamming attack} & \makecell[c]{GAN} & \makecell[c]{Conditional tabular GAN to handle multi-dimensional data}
		\\
		
		\cline{3-7}
		
		&&
		\makecell[l]{\cite{refIII_A_1_8}}& \makecell[c]{Air}& \makecell[c]{Jamming attack} & \makecell[c]{TBM-GAN} & \makecell[c]{DroneDefGANt smooths
			gradient during backpropagation} 
		\\
		
		\cline{3-7}
		&&
		\makecell[l]{\cite{refIII_jamming_2}}& \makecell[c]{Air}& \makecell[c]{Jamming attack} & \makecell[c]{AE-GAN} & \makecell[c]{AE-GAN solves insufficient labelled data in AAV swarm}
		\\
		
		\cline{3-7}
		& &
		\makecell[l]{\cite{refIII_jamming_3}}& \makecell[c]{Ground}&\makecell[c]{Jamming attack} & \makecell[c]{VAE} & \makecell[c]{
			Unsupervised detection  for MIMO-OFDM ISAC
			system}
		\\
		
		\cline{3-7}
		&&
		\makecell[l]{\cite{refIII_jamming_4}}& \makecell[c]{Each segment}& \makecell[c]{Jamming attack} & \makecell[c]{VAE-GAN} & \makecell[c]{Jamming recognition based on AC-VAEGAN}
		\\
		
		\cline{3-7}
		&&
		\makecell[l]{\cite{refIII_jamming_5}}& \makecell[c]{Air/Ground}& \makecell[c]{Unknown jamming} & \makecell[c]{VAE } & \makecell[c]{A federated augmented
			aggregate training algorithm}
		\\
		
		\cline{3-7}
		&&
		\makecell[l]{\cite{refIII_jamming_6}}& \makecell[c]{Air/Ground}&  \makecell[c]{Intelligent jamming}& \makecell[c]{VAE } & \makecell[c]{A dynamic adaptive spectro-temporal resilient filter}
		\\
		
		\cline{3-7}
		&&
		\makecell[l]{\cite{refIII_jamming_7}}& \makecell[c]{Ground}& \makecell[c]{Jamming attack} & \makecell[c]{TBM} & \makecell[c]{A distributed radar multi-interference identification method}
		\\

		\cline{3-7}
		&&
		\makecell[l]{\cite{refIII_jamming_8}}& \makecell[c]{Space/Ground}& \makecell[c]{Privacy leakage} & \makecell[c]{GDM} & \makecell[c]{STINIDF through FL and generated traffic data}
		\\
		\cline{2-7}
		
		&\multirow{3}{*}[-24pt]{\shortstack{ \textbf{Jamming}\\ \textbf{Mitigation} }}&
		\makecell[l]{\cite{refIII_jamming_9}}& \makecell[c]{Air}& \makecell[c]{Jamming attack} & \makecell[c]{GAN} & \makecell[c]{A collaborative RL algorithm based on GAN}
		\\
		\cline{3-7}
		
		&&
		\makecell[l]{\cite{refIII_jamming_10}}& \makecell[c]{Ground}&\makecell[c]{Jamming attack}  & \makecell[c]{GAN} & \makecell[c]{An GAN-DRL-based spectrum access algorithm}
		\\
		\cline{3-7}
		
		&&
		\makecell[l]{\cite{refIII_jamming_11}}& \makecell[c]{Space}& \makecell[c]{Jamming attack} & \makecell[c]{TBM} & \makecell[c]{A vision transformer based blind beamforming method}
		\\
		\cline{3-7}
		
		& &
		\makecell[l]{\cite{refIII_jamming_12}}& \makecell[c]{Air}&\makecell[c]{Jamming attack} & \makecell[c]{ TBM} & \makecell[c]{Jamming prediction integrates pseudo-random with TBM}
		\\
		\cline{3-7}
		
		&&
		\makecell[l]{\cite{refIII_jamming_13}}& \makecell[c]{Space}& \makecell[c]{Signal jamming} & \makecell[c]{GDM } & \makecell[c]{A GDM-based anti-jamming algorithm}
		\\
		
		\cline{3-7}
		
		&&
		\makecell[l]{\cite{refIII_jamming_14}}& \makecell[c]{Each segment}& \makecell[c]{Jamming attack} & \makecell[c]{GAN} & \makecell[c]{A anti-jamming scheme for semantic communication}
		\\
		\cline{3-7}
		
		&&
		\makecell[l]{\cite{refIII_jamming_15}}& \makecell[c]{Ground}& \makecell[c]{Jamming attack} & \makecell[c]{ TBM} & \makecell[c]{A sample less source separation algorithm}
		\\
		\cline{1-7}
		
		\multirow{3}{*}[-1pt]{\shortstack{\textbf{Data Link}\\ \textbf{Layer} }}& &
		\makecell[c]{ \cite{refIII_Availability_MAC_1}}& \makecell[c]{Underwater}& \makecell[c]{MAC flooding} & \makecell[c]{LSTM-GAN} & \makecell[c]{A GAN-enhanced IDS integrated with MAC protocol} 
		\\
		\cline{3-7}
		
		& \multirow{3}{*}[10pt]{\shortstack{ \textbf{Intrusion}\\ \textbf{Detection} }}&
		\makecell[c]{ \cite{refIII_Availability_MAC_2}}& \makecell[c]{Cross-Domain}& \makecell[c]{Flooding DDoS} & \makecell[c]{GAN} & \makecell[c]{Detection for routing flooding in IPv6 networks} 
		\\
		\cline{3-7}
		
		& &
		\makecell[c]{ \cite{refIII_Availability_MAC_3}}& \makecell[c]{Each segemnt}& \makecell[c]{MAC flooding} & \makecell[c]{GAN} & \makecell[c]{An adversarial attack detection framework}
		\\
		\cline{1-7}
		\multirow{3}{*}[-30pt]{\shortstack{\textbf{Network}\\ \textbf{Layer} }}&\multirow{3}{*}[-30pt]{\shortstack{ \textbf{Intrusion}\\ \textbf{Detection} }} &
		\makecell[c]{ \cite{refIII_Availability_Network_1} }& \makecell[c]{Each segment }& \makecell[c]{DDoS attack } & \makecell[c]{VAE } & \makecell[c]{A VAE-powered two-phase DDoS mitigation framework } 
		\\
		\cline{3-7}
		
		& &
		\makecell[c]{\cite{refIII_Availability_Network_2}  }& \makecell[c]{Each segment  }& \makecell[c]{DDoS attack} & \makecell[c]{ VAE} & \makecell[c]{An unsupervised
			IDS using deep
			autoencoders}
		\\
		\cline{3-7}
		& &
		\makecell[c]{ \cite{refIII_Availability_Network_3} }& \makecell[c]{Each segment }& \makecell[c]{ DDoS attack} & \makecell[c]{VAE } & \makecell[c]{A FL-VAE
			scheme combining VAE with FL}
		\\
		\cline{3-7}
		& &
		\makecell[c]{\cite{refIII_Availability_Network_4}  }& \makecell[c]{ Each segment }& \makecell[c]{Multiple attacks } & \makecell[c]{VAE } & \makecell[c]{ VAE is combined with neural architectures}
		\\
		\cline{3-7}
		& &
		\makecell[c]{\cite{refIII_Availability_Network_5}  }& \makecell[c]{ Each segment}& \makecell[c]{Multiple attacks  } & \makecell[c]{VAE } & \makecell[c]{ A privacy-preserving traffic generation
			method}
		\\
		\cline{3-7}
		& &
		\makecell[c]{  \cite{refIII_Availability_Network_6}}& \makecell[c]{Each segment }& \makecell[c]{Multiple attacks } & \makecell[c]{ GAN} & \makecell[c]{A hybrid IDS integrating CGAN and gradient boosting }
		\\
		\cline{3-7}
		& &
		\makecell[c]{\cite{refIII_Availability_Network_7}  }& \makecell[c]{Each segment}& \makecell[c]{DDoS attack  } & \makecell[c]{LLM } & \makecell[c]{A GAI-driven framework using pre-trained transformers}
		\\
		\cline{3-7}
		& &
		\makecell[c]{ \cite{refIII_Availability_Network_8} }& \makecell[c]{Each segment }& \makecell[c]{DDoS attack} & \makecell[c]{LLM } & \makecell[c]{A GAI-powered real-time
			DDoS detection framework }
		\\
		
		\Xhline{1pt}	
	\end{tabular}
	\label{table_GAI_Availability}
\end{table*}

\subsubsection{Network Layer}
The availability of the network layer in SAGINs is compromised by dynamic topology-based DDoS attacks, resource depletion in high-latency links, and routing flooding attacks. 
GAI facilitates federated anomaly detection and supports privacy-preserving adversarial training.

VAE demonstrates effective performance in traffic anomaly detection within SAIGINs.  For instance, \cite{refIII_Availability_Network_1} presented a two-phase DDoS mitigation framework powered by a VAE, which incorporated cyclic queuing to optimise the balance between persistent resource consumption and detection efficiency for edge devices in dynamic SAGIN environments. 
Furthermore, \cite{refIII_Availability_Network_2} introduced an unsupervised intrusion detection system utilising deep autoencoders to capture spatiotemporal patterns of normal network flows within industrial control systems.  This method attained a detection accuracy of 98.8\% for DDoS attacks, accompanied by a false alarm rate of only 1.13\%. 
However, the method presented in \cite{refIII_Availability_Network_2} is limited by inadequate privacy protection when dealing with non-independent and identically distributed data (non-IID) and exhibits weak cross-device generalisation. 
In response to this issue, \cite{refIII_Availability_Network_3} examined a FL-VAE scheme that integrates variational autoencoders with momentum-accelerated FL, with a specific focus on DDoS mitigation in SAGINs.  The framework tackled feature heterogeneity among distributed nodes via dynamic client sampling and adaptive model retraining, resulting in a 99.97\% attack recognition precision on non-IID traffic and a 32\% reduction in cross-node data exchange.

VAE can be integrated with neural architectures to provide defence mechanisms against attacks in SAGINs. 
The study \cite{refIII_Availability_Network_4} utilised the anomaly detection capability of VAEs by employing probabilistic reconstruction of multi-dimensional traffic patterns.
Additionally, \cite{refIII_Availability_Network_5} introduced a privacy-preserving traffic generation method utilising CVAE, which tackles security availability issues in SAGINs arising from dynamic traffic patterns and device heterogeneity. 
The authors in \cite{refIII_Availability_Network_6} proposed a hybrid IDS that integrates conditional GAN (CGAN) and extreme gradient boosting (XGBoost) for small sample datasets.  This method decreased the physical layer deployment requirements in wireless sensor networks by utilising adversarial data generation.  The approach attained 99.9\% detection accuracy and a 1.83\% reduction in false alarms on the NSL-KDD and CICIDS2017 datasets by integrating GAI with lightweight classifiers, thereby improving network-layer anti-jamming capabilities, especially for resource-constrained nodes in SAGINs.

In the context of DDoS vulnerabilities, \cite{refIII_Availability_Network_7} presented PLLM-CS, a framework driven by GAI that utilises pre-trained transformers.  The model restructured network data into context-aware token sequences, effectively capturing spatiotemporal patterns in cyber threats via self-attention mechanisms, thereby overcoming the limitations of traditional IDSs in resource-constrained satellite environments.  
The centralised architecture presented in \cite{refIII_Availability_Network_7} is at odds with the distributed characteristics of SAGINs, facing challenges related to dynamic network traffic and the complexity of heterogeneous cross-domain protocols.
Therefore, \cite{refIII_Availability_Network_8} developed Llama2-Defender, leveraging LLMs for contextual reasoning across heterogeneous nodes to overcome generalization limitations in aerial-terrestrial networks.

\subsubsection{Summary and Lessons Learned}
As summarized in Table \ref{table_GAI_Availability}, 
GAI enables robust jamming recognition  by distinguishing stealthy adversarial patterns from legitimate transmissions. 
Meanwhile, diffusion models facilitate high-fidelity signal reconstruction from severely corrupted observations, while adversarial learning frameworks generate adaptive beamforming and spectrum access policies that dynamically circumvent interference. 
Crucially, GAI’s synthetic data generation capability overcomes the critical limitation of attack sample scarcity, enabling the training of detectors for novel or evolving threats without requiring extensive real-world datasets. 
Despite these advances, critical challenges remain:
\begin{itemize}
	\item[$\bullet$] \textit{Fidelity vs. deployability trade-offs:} While model distillation and FL offer partial mitigation, fundamental trade-offs persist between generative fidelity and edge deployability. 
	\item[$\bullet$] \textit{Intrinsic vulnerability:} Maliciously crafted inputs could poison generative training data or deceive detectors through perturbation attacks, potentially turning defensive systems into availability liabilities. 
	\item[$\bullet$] \textit{Cross-Layer coordination:} While GAI excels at layer-specific countermeasures (e.g., physical-layer beamforming and network-layer DDoS mitigation), holistic frameworks for orchestrating generative defenses across SAGIN’s protocol stack require development. 
\end{itemize}



\section{Tutorial 1: GDMTD3 for Multi-Objective Aerial Collaborative Secure Communication Optimization}

\subsection{Motivation}
The allocation of resources for secure communication in SAGINs presents a critical and complex challenge under various attacks.
This issue pertains to the dynamic allocation of constrained resources, such as transmit power, computing power, and routing, to optimise security metrics, including secrecy rate and SEE, while concurrently reducing operational costs, such as energy consumption and latency. 
The primary challenge is optimising problems characterised by high-dimensionality, non-convexity, and NP-hardness within the dynamic adversarial conditions and strict resource limitations of SAGINs.
For instance, \cite{refIII_Eva_12} presented a multi-objective problem aimed at optimising the excitation current weights and three-dimensional trajectories of UAVs to achieve a balance between secrecy rate and energy consumption in a dynamic high-dimensional space.
Traditional AI methods, such as DRL approaches, encounter significant challenges in accurately modelling the complex probability distributions associated with high-dimensional continuous action spaces. These methods often exhibit high policy variance, unstable convergence, and suboptimal Pareto frontiers in the context of secure resource allocation \cite{refII_B_buchong1} \cite{refIII_RSMA-PPO}.

GDMs offer a significant advantage compared to traditional DRL in addressing complex resource allocation issues. 
DRL utilises deterministic or simplified stochastic policy networks, whereas GDMs employ a progressive denoising mechanism to accurately capture and sample from complex, high-dimensional state-action distributions. 
This facilitates the production of varied and high-quality candidate solutions for DRL's policy networks, efficiently examining the complex trade-offs between opposing objectives (e.g., secrecy rate versus energy).
The integration of GDMs into the DRL framework significantly enhances the representational capacity of the policy, addressing the limitations of traditional DRL in navigating complex solution spaces and achieving robust convergence towards near-Pareto-optimal strategies under uncertainty.
The GDM-driven DRL framework creates a new and scalable model for secure and efficient SAGINs.

\subsection{System Description}
\begin{figure}[!t]
	\centering
	\includegraphics[width=2in]{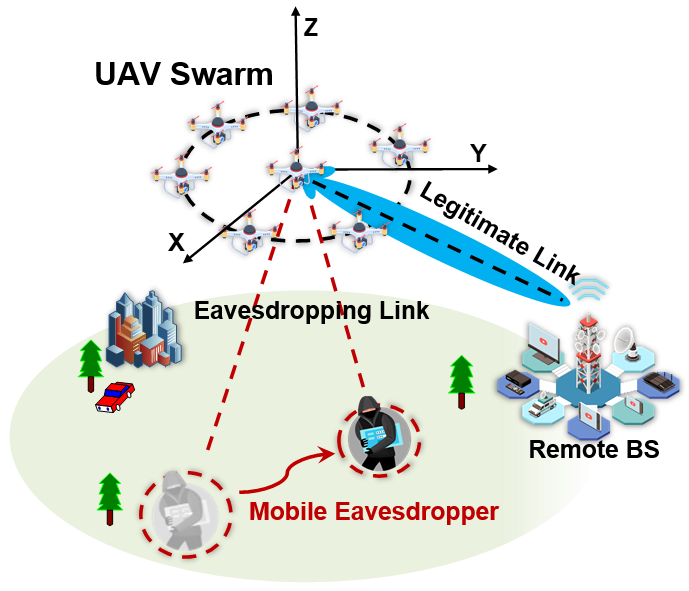}
	\caption{A UAV swarm-enabled surveillance network transmits sensitive data to a RBS, persistently threatened by a mobile eavesdropper attempting interception through time-varying wiretap channels \cite{refIII_Eva_12}.}
	\label{GDMTD3-MODEL}
	\vspace{-1em}
\end{figure}

Fig. \ref{GDMTD3-MODEL} illustrates that the UAV swarm-enabled secure surveillance network encounters ongoing threats from mobile eavesdroppers seeking to intercept communications via time-varying wiretap channels across discrete time slots.
Collaborative beamforming (CB) using UAV virtual antenna arrays (UVAAs) improves directional security but creates a significant energy-security trade-off. The continuous repositioning of UAVs to achieve optimal beam patterns significantly increases energy consumption in the pursuit of maximising secrecy rates.
Within this dynamic threat environment, a multi-objective optimisation problem for aerial secure communication and energy efficiency (ASCEE-MOP) is formulated to maximise the secrecy rate and minimise energy consumption as follows:
\begin{subequations}
	\label{e1}
	\begin{align}		
		\mathop {\max }\limits_{\textit{\textbf{I},\textbf{q}}} {\kern 1pt} {\kern 1pt} {\kern 1pt} {\kern 1pt} {\kern 1pt} {\kern 1pt} {\kern 1pt} {\kern 1pt} {\kern 1pt} {\kern 1pt} {\kern 1pt} {\kern 1pt} & \left( {\sum\limits_{n = 1}^N {{R_{SE}}[n]} , - \sum\limits_{n = 1}^N {E[n]} } \right)  \label {e1-a}\\
		\text {s.t.}{\kern 1pt} {\kern 1pt} {\kern 1pt} {\kern 1pt} {\kern 1pt} {\kern 1pt} {\kern 1pt} {\kern 1pt} {\kern 1pt} {\kern 1pt} {\kern 1pt} {\kern 1pt} &~ 0 \le I_k^U[n] \le 1,\forall k \in \{ 1,...,K\} , \label {e1-b}\\
		&{X_{\min }} \le x_k^U[n] \le {X_{\max }},{\kern 1pt} {\kern 1pt} \forall k \in \{ 1,...,K\} , \label {e1-c}\\
		&{Y_{\min }} \le y_k^U[n] \le {Y_{\max }},{\kern 1pt} {\kern 1pt} \forall k \in \{ 1,...,K\} , \label {e1-d}\\
		&{Z_{\min }} \le z_k^U[n] \le {Z_{\max }},{\kern 1pt} {\kern 1pt} \forall k \in \{ 1,...,K\} , \label {e1-e}\\
		&0 \le v_k^U[n] \le {V_{\max }},\forall k \in \{ 1,...,K\}, \label {e1-f}\\
		&||{q_{{k_1}}}[n],{q_{{k_2}}}[n]|| \ge D_{\min }^U,\forall {k_1},{k_2} \in \{ 1,...,K\}. \label {e1-g}
	\end{align}
\end{subequations}
where ${R_{SE}}[n]$ and $E[n]$ are the achievable secrecy rate and the flight energy consumption, respectively. $K$ and $N$ are the number of UAVs and time slots, respectively. \textit{\textbf{I}} = ${\{ I_k^U[n]\} _{k \in K,n \in N}}$ and \textit{\textbf{q}} = $ {\{ q_k^U[n]\} _{k \in K,n \in N}}$ are the excitation current weight matrix and the position matrix of UAVs at all time slots, respectively. 
$q_k^U[n] = (x_k^U[n], y_k^U[n], z_k^U[n])$ is the
the coordinate of UVAA center.
(\ref{e1-b}) expresses the range constraint of the excitation current weight. Moreover, Constraints (\ref{e1-c}), (\ref{e1-d}), and (\ref{e1-e}) restrict the flight area of the UAV. Constraint (\ref{e1-f}) is the speed constrain of the UAV, and Constraint (\ref{e1-g}) is imposed to guarantee the minimum distance between two UAVs.

This problem is non-convex, NP-hard, and dynamic due to eavesdropper mobility. 
Existing approaches for such optimization problems typically decompose them into separable convex subproblems solved iteratively. However, solution accuracy critically depends on the decomposition strategy. Moreover, dynamic factors, such as mobile eavesdroppers and time-varying channels, impose prohibitive computational overhead for real-time SAGIN operations.
DRL provides an efficient framework for adaptive sequential decision-making through policy gradient frameworks that encode environmental dynamics into a Markov decision process (MDP).
This paradigm efficiently navigates the solution space of ASCEE-MOP without explicit problem decomposition.
Furthermore, GDM can be adopted in policy network to generate high-quality solutions for excitation current weights and UAV localisation, mapping the state of the environment directly to the optimal action.
\vspace{-1em}
\subsection{GAI-Based Solutions} 
Generally, the GDM-driven DRL approach significantly improves the robustness and convergence of the policies by fusing the distributional modelling capability of diffusion models with the decision optimization framework of DRL.
It is divided into three main steps when solving complex single-objective and multi-objective optimization problems with high-dimensional state-action spaces.
Here, we take the GDMTD3 in \cite{refIII_Eva_12} as an example to illustrate.
\begin{figure}[!t]
	\centering
	\includegraphics[width=3in]{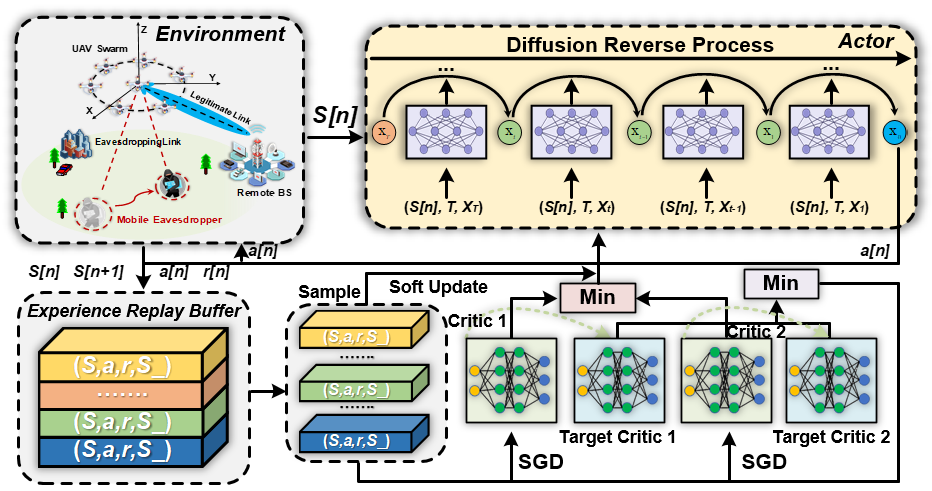}
	\caption{The proposed GDMTD3 framework, which integrates a GDM into actor network to capture complex state features and generate optimal actions from environmental states \cite{refIII_Eva_12}.}
	\label{GDMTD3}
		\vspace{-1em}
\end{figure}
\begin{itemize}
	\item[$\bullet$] \textit{Step 1:} Modelling the dynamic optimisation problems as MDP tuples is fundamental, where the MDP tuple is $ < {\cal S},{\cal A},{\cal P},{\cal R},\gamma  > $. 
	The environment state at slot $n$ is $\textit{\textbf{s}}[n]$. 
	The action set is $\textit{\textbf{a}}[n] = (\textit{\textbf{I}}[n],\textit{\textbf{q}}[n])$. 
	The reward function is defined as $r[n]$.
	${\cal P}$ is the transition probability of the state, and $\gamma$ is the discount factor. 
	This formulation transforms ASCEE-MOP into a tractable DRL problem where UAV swarms interacts with the environment to maximize cumulative discounted rewards.
	\item[$\bullet$] \textit{Step 2:} We integrate the GDM with the actor network of TD3, where the multi-layer perceptrons (MLPs)-based actor network struggles with the non-linear state space induced by mobile eavesdroppers and multi-objective trade-offs.
	GDMs address this limitation by modeling high-dimensional distributions for more balanced and optimized decisions in uncertain and dynamic environments. 
	As shown in Fig. \ref{GDMTD3}, GDMTD3 integrates GDM with the actor network for enhanced state-action distribution capture and a more diverse set of potential actions. The core is GDM-based action sampling.
	This process is similar to the Gen-DRL in \cite{refIII_Eva_14}, but this tutorial analyzes the more complex multi-objective optimisation rather than the single-objective optimisation in \cite{refIII_Eva_14}.
	Furthermore, while \cite{refIII_Eva_10}, \cite{refIII_Eva_13}, and \cite{refIII_Confidentiality_Network_6} consider resource optimisation in secure communication, their schemes address lower dimensions andthe algorithms may fail in dynamic and complex SAGINs. Using GDM-driven DRL to implement secure security resource allocation would support them being more applicable to real SAGINs to some extent.
	Specifically, in GDMTD3, based on the current state $\textit{\textbf{s}}[n]$, action \boldmath{$a[n]$}, and initialized Gaussian distribution ${x_T} \sim {\cal N}(0,I)$, a denoising distribution ${\varepsilon _{{\theta _d}}}({x_t},t,s[n])$ is deduced and then the mean of the reverse process $\kappa _{{\theta _d}}$ can be computed as follows:
	\begin{align}		
		{\kappa _{{\theta _d}}}({x_t},t,s[n]) = \frac{1}{{\sqrt {{\alpha _t}} }}\left({x_t} - \frac{{{\beta _t} \cdot {\varepsilon _{{\theta _d}}}}}{{\sqrt {1 - {{\overline \alpha  }_t}} }}\right),
	\end{align}
	where ${{\beta _t}}$ is the variance function of variance preserving stochastic differential equations and ${\alpha _t} = 1 - {\beta _t}$. ${\overline \alpha  _t} = \prod\nolimits_{k = 1}^t {{\alpha _k}} $. Then, a reparameterization trick that facilitates differentiable sampling is employed to compute the distribution $x_{t-1}$ as follows:
	\begin{align}		
	{x_{t - 1}} = {\kappa _{{\theta _d}}}({x_t},t,s) + {({\widetilde \beta _t}/2)^2} \odot \varepsilon ,
	\end{align}
	where ${\widetilde \beta _t} $ is a predetermined variance factor and $\odot$ is the operator of Hadamard product.
	Finally, we can obtain the generative distribution $ 	{p_{{\theta _d}}}({x_0})$ as follows:
	\begin{align}		
	{p_{{\theta _d}}}({x_0}) = p({x_T})\prod\limits_{t = 1}^T {{p_{{\theta _d}}}({x_{t - 1}}|{x_t})},
	\label{e4}
	\end{align}
	where $p({x_T})$ represents a standard normal distribution. Once the generative distribution ${p_{{\theta _d}}}({x_0})$ is successfully trained, we can sample the action $x_0$ from (\ref{e4}).
	
	\item[$\bullet$] \textit{Step 3:} Now, we can conduct the training and execution process. The RBS governs the training process via the actor-critic framework of GDMTD3. During this step, environment interactions of the UAV swarm are continuously recorded and cached in a replay buffer. Upon completion of the training cycle, the actor network is deployed across the UAV swarm, enabling autonomous real-time adaptation to dynamic threats for sustained secure communication during operational execution.
\end{itemize}

\subsection{Numerical Results} 
As shown in Fig. 6 of \cite{refIII_Eva_12}, the proposed GDMTD3  achieves a 20\% higher average reward per episode compared to benchmarks (TD3, PPO, DDPG, SAC, and transformer-based TD3 methods). This performance gain stems from its generative diffusion mechanism, which enhances exploration-exploitation trade-offs in high-dimensional state-action spaces, thereby maximizing cumulative rewards.

\begin{figure}[!t]
	\centering
	\includegraphics[width=3in]{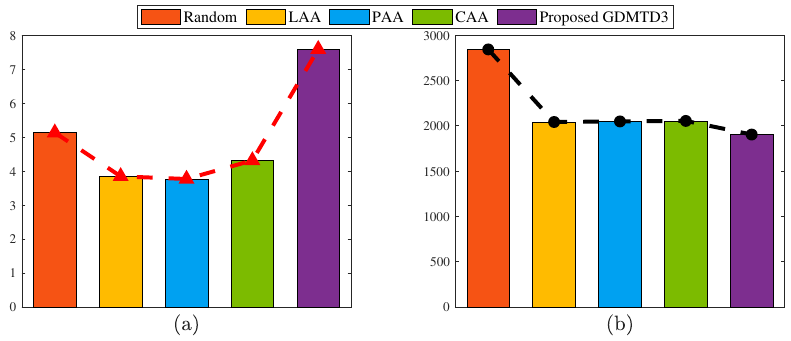}
	\caption{Comparison results of the proposed GDM-enabled DRL approach and other four deployment policies. (a) Average secrecy rate per step [bps/Hz]. (b) Average flight energy consumption per step [J] \cite{refIII_Eva_12}.}
	\label{GDMTD3_OBJECT}
	\vspace{-1em}
\end{figure}

Fig. \ref{GDMTD3_OBJECT} demonstrates the superior performance of GDMTD3 in terms of average secrecy rate and average energy consumption compared to baselines. 
This gain stems from dynamic joint optimisation of UAV excitation currents and spatial configurations, effectively countering eavesdropping threats.  It exhibits 18\% lower average energy consumption than conventional strategies, which is attributed to the joint optimisation of beamforming parameters and trajectory planning. 
Collectively, these results confirm that the GDM-DRL framework uniquely balances two competing objectives: maximizing secrecy rates while minimizing energy consumption.
We can further conclude that the proposed GDM-DRL approach still performs well when the security resource optimisation is more complex, e.g., cross-layer or cross-domain resource optimisation, as considered in \cite{refII_A_20}, \cite{refIII_Eva_2}, and \cite{refIII_Availability_MAC_2}. Traditional DRL, on the other hand, fails further.

\begin{figure}[!t]
	\centering
	\includegraphics[width=2in]{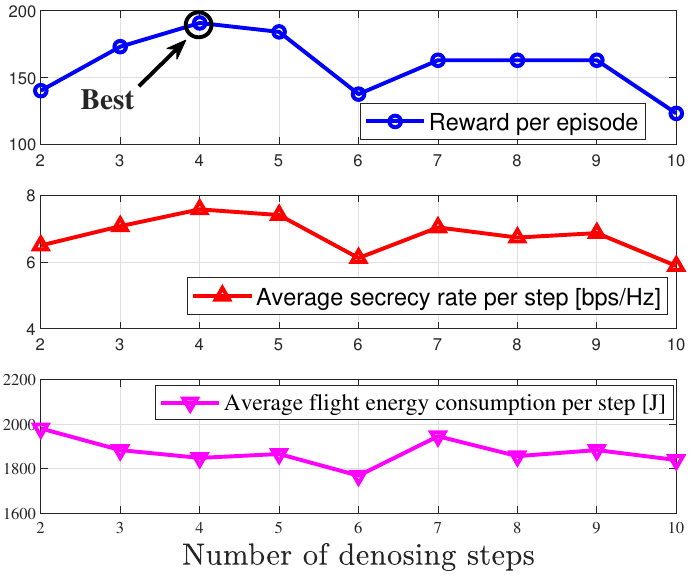}
	\caption{Comparison of curves of GDMTD3 with different denoising steps \cite{refIII_Eva_12}. When the number of denosing steps is 4, the reward reaches its maximum value and the average security rate is the highest and the energy consumption is the lowest. As the number of denosing steps is raised or lowered, the performance starts to decrease.}
	\label{GDMTD3_DENOISE}
		\vspace{-1em}
\end{figure}

In addition, the denoising step count critically modulates noise suppression efficacy and overfitting vulnerability in GDMTD3. As validated in Fig. \ref{GDMTD3_DENOISE}, performance peaks at 4 steps for ASCEE-MOP. Beyond this threshold, marginal utility attenuation occurs due to noise-pattern overfitting, inducing oscillatory artifacts in action generation. This establishes 4-step diffusion as the optimal robustness-efficiency trade-off.

\vspace{-1em}
\subsection{Lessons Learned}
GDM-driven DRL robustly models high-dimensional state-action spaces to resist adversarial perturbations and data noise. Their integration replaces conventional policy networks with iterative diffusion processes, generating high-quality action sequences via multi-step refinement and overcoming traditional DRL's local convergence in complex spaces.
With this mechanism, the challenge of \cite{refIII_Eva_10}, \cite{refIII_Eva_13}, and \cite{refIII_Confidentiality_Network_6} not being able to cope with high-dimensional dynamic security optimization problems is addressed. It also shows vast potential in higher dimensional cross-layer and cross-domain optimization problems in \cite{refII_A_20}, \cite{refIII_Eva_2}, and \cite{refIII_Availability_MAC_2}.

\section{Tutorial 2: GDM for Privacy-Preserving Mobile Crowdsensing}
\subsection{Motivation}
The collection and sharing of extensive sensing data in SAGINs, including satellite telemetry, AAV trajectories, and user behaviours, poses significant risks to privacy. 
Attackers can acquire raw data through eavesdropping on wireless links and compromising edge servers or IoT devices, resulting in the exposure of user identity, location, and behavioural information.
Traditional privacy-preserving methods exhibit significant limitations: static encryption techniques struggle to accommodate the dynamic topology of SAGIN, resulting in increased latency \cite{refIII_Eva_1}; DP necessitates a trade-off between data utility and anonymity, thereby reducing the effectiveness of data analytics \cite{refIII_jamming_8}; and FL utilising discriminative AI safeguards local data but suffers from slow convergence and accuracy degradation in heterogeneous device environments \cite{refIII_Confidentiality_MAC_4}. 
None of these approaches can effectively protect privacy while simultaneously addressing the constraints of dynamism, utility preservation, and real-time requirements.

GAI, particularly GDM, overcomes the aforementioned limitations via paradigm reconstruction.  Sensitive privacy attributes are removed from the original data using a forward diffusion process, followed by the generation of synthetic data that retains equivalent statistical features while lacking privacy information, based on backward denoising \cite{refIII_Confidentiality_Network_supply}.
In contrast to DRL, GDM mitigates issues related to high variance and sub-optimal policies in complex privacy-utility trade-offs in reinforcement learning \cite{refII_B_buchong1}. 
In comparison to GAN and VAE, GDM offers a more stable training process and superior generation quality, effectively addressing the pattern collapse associated with GAN and the ambiguous reconstruction issues of VAE \cite{refII_B_5} \cite{refII_B_GDM}.

The typical application involves secure and privacy-preserving mobile crowdsensing (SPPMCS) as discussed in \cite{refIII_Confidentiality_Network_supply}.
This system employs GDM to substitute raw sensing data, such as vehicle images, with synthetic samples that do not compromise privacy. 
This approach effectively eliminates potential privacy leakage pathways and quantitatively enhances the equilibrium between privacy and task accuracy via the privacy-preserving utility index (PPUI). Consequently, it equips the SAGIN with a dynamic privacy mitigation mechanism unattainable by conventional AI methods.
\vspace{-1em}
\subsection{System Description}
\begin{figure}[!t]
	\centering
	\includegraphics[width=2in]{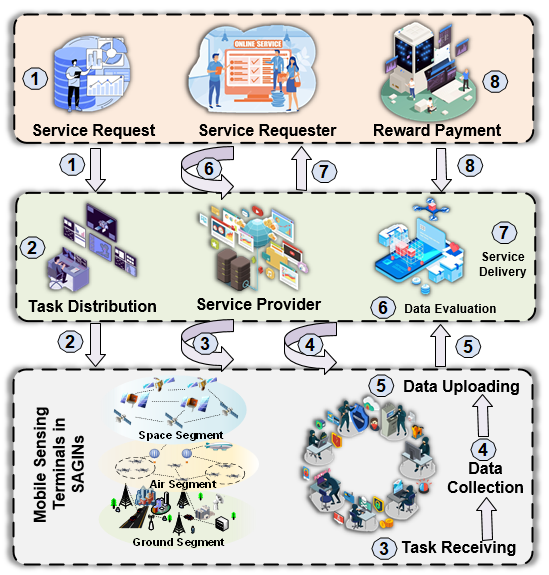}
	\caption{The privacy preservation framework, where GAI-enabled SPPMCS can alleviate the privacy preservation concerns on sensing data, and MST’s identification and location information \cite{refIII_Confidentiality_Network_supply}.}
	\label{SPPMCS-MODEL}
		\vspace{-1em}
\end{figure}
As  illustrated in Fig. \ref{SPPMCS-MODEL}, SPPMCS comprises three core entities: a service requester (SR) for task generation and specification, a service provider (SP), and distributed mobile sensing terminals (MSTs) performing SP-coordinated data acquisition. 
Specifically, this system operates in  three phases: \textit{i) Task Generation and Allocation:} The SR generates sensing tasks (specifying data type, volume, and quality) and distributes them to SPs. SPs then authenticate MSTs using broadcast encryption and assign decryption keys. However, dynamic task adjustments (e.g., for real-time data needs) incur significant key-management overhead and delays under traditional encryption, degrading system efficiency.
\textit{ii) Data Collection and Submission:}
MSTs submit collected data to SPs. While conventional discriminative AI techniques (e.g., federated learning) preserve local data privacy, they suffer from slow convergence and reduced accuracy due to edge device heterogeneity (varying computational and communication capabilities).
\textit{iii) Result Evaluation and Reward Payment:}
SPs evaluate data quality and reward MSTs. Some RL and blockchain-based approaches, face high computational costs and slow convergence, hindering real-time verification of reward transactions.
\vspace{-1em}
\subsection{GAI-Based Solutions}
\begin{figure}[!t]
	\centering
	\includegraphics[width=2.8in]{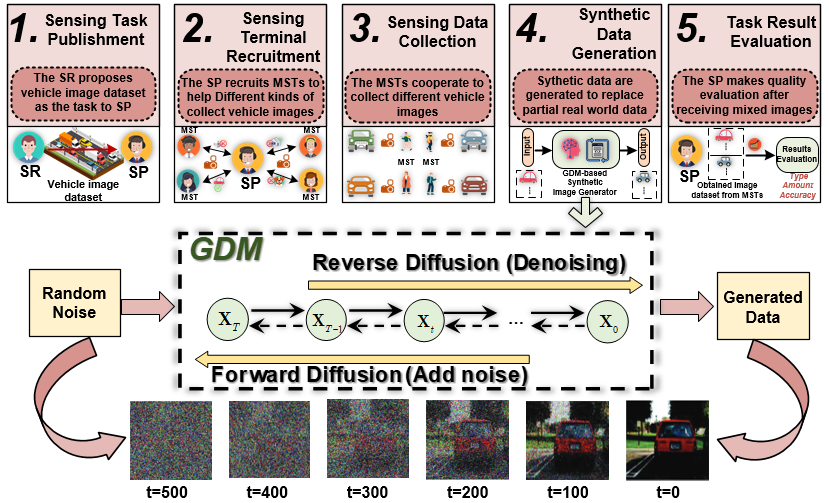}
	\caption{A GDM-enabled SPPMCS framework for sensing data content protection, where GDM is adopted to synthesize vehicle images. The GDM process extracts features from original images by adding noise during the forward diffusion stage. A subsequent denoising process in the reverse diffusion stage then generates the target synthetic image. By replacing real-world images with these privacy-free synthetic counterparts, the framework effectively alleviates sensing data privacy concerns \cite{refIII_Confidentiality_Network_supply}.}
	\label{SPPMCS-GDM}
	\vspace{-1em}
\end{figure}

GDM-driven privacy preservation mechanisms contribute to robust security defenses against threats like malicious data injection, unauthorized access, and spectrum manipulation, while simultaneously enhancing protection for both data content and terminal identification/location in SAGINs. 
Here, we introduce the GDM-driven SPPMCS scheme to compensate for the lack of dynamic adaptability of static encryption \cite{refIII_Eva_8}, poor accuracy of differential privacy data \cite{refII_B_1_16}, and slow convergence of FL-based traditional AI \cite {refIII_Confidentiality_MAC_4} \cite{refIII_Confidentiality_MAC_5}.
Main steps in Fig. \ref{SPPMCS-GDM} are as follows:
\begin{itemize}
	\item[$\bullet$] \textit{Step 1:} In sensing task publishment, SRs outsource application-specific data requirements to SPs, leveraging SP resources to overcome collection and analysis limitations.
	
	\item[$\bullet$] \textit{Step 2:} In sensing terminal recruitment, the SP recruits qualified MSTs based on SR specifications utilizing a utility-driven reward model where compensation scales with data quality and objective fulfillment.
	
	\item[$\bullet$] \textit{Step 3:} In sensing data collection,
	MSTs collect data per SR specifications. While high-capacity MSTs fulfill multi-type tasks independently, others form coalitions to achieve full coverage. This cooperation enhances task completeness but necessitates proportional reward redistribution.
	
	\item[$\bullet$] \textit{Step 4:} We conduct GDM to finish synthetic data generation. 
	Specifically, the GDM process involves two stages: (1) a forward diffusion phase that incrementally adds noise to original training images to learn latent features; (2) a reverse diffusion (denoising) phase that reconstructs realistic synthetic images from noise. 
	By replacing a portion of real-world data with these synthetic samples during data submission, MSTs mitigate privacy exposure without degrading downstream task performance (e.g., vehicle detection). 
	GDM eliminates the lack of dynamic adaptability inherent in encryption in \cite{refIII_Eva_8}. Synthetic data requires no cryptographic keys or complex key-distribution protocols, enabling seamless adaptation to new data types/privacy requirements.
	This GDM mechanism can also be adopted to enhance DP with poor data accuracy in \cite{refII_B_1_16}, by generating structurally consistent, high-fidelity synthetic data.
	In addition, GDM avoids the slow convergence of FL-based privacy in \cite{refIII_Confidentiality_MAC_5}. It operates locally or at the SP without iterative parameter exchanges, reducing latency. Synthetic data submission incurs minimal overhead compared to FL’s multi-round coordination.

	\item[$\bullet$] \textit{Step 5:} Hybrid data submissions from MSTs undergo SP-led quality attestation—measuring critical dimensions, which directly determines reward distribution via mechanism-defined payment rules anchored in quality-effectiveness proof.
\end{itemize}

\vspace{-1em}
\subsection{Numerical Results} 
Since GDM can generate new synthetic data to replace the original data to be analyzed and processed, it can lower data attacks and privacy leakage risks for the original data. Simulations are based on the YOLOv3 model.
The GDM parameters are configured with 500 iterations, a mini-batch size of 32, a learning rate of 0.005, and a squared gradient decay factor of 0.9999. 

\begin{figure}[!t]
	\centering
	\includegraphics[width=2in]{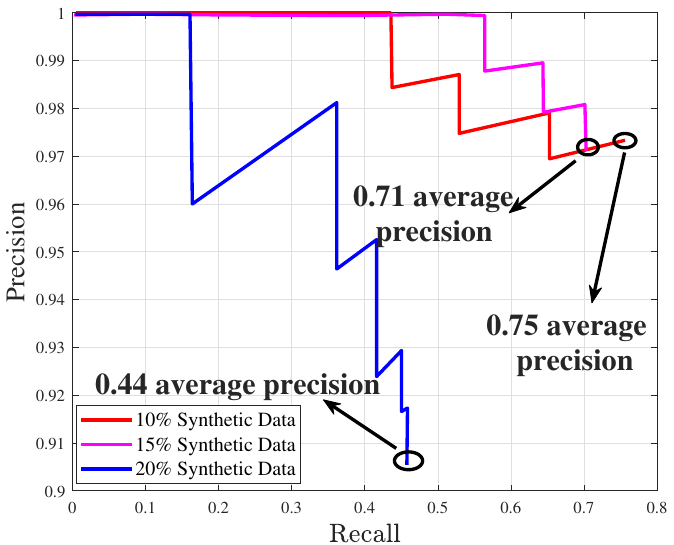}
	\caption{Precision recall performance for YOLOv3 detection model with different percentage of synthetic data for training dataset \cite{refIII_Confidentiality_Network_supply}.}
	\label{GDMTD3_p2}
\end{figure}

\begin{figure}[!t]
	\centering
	\includegraphics[width=2in]{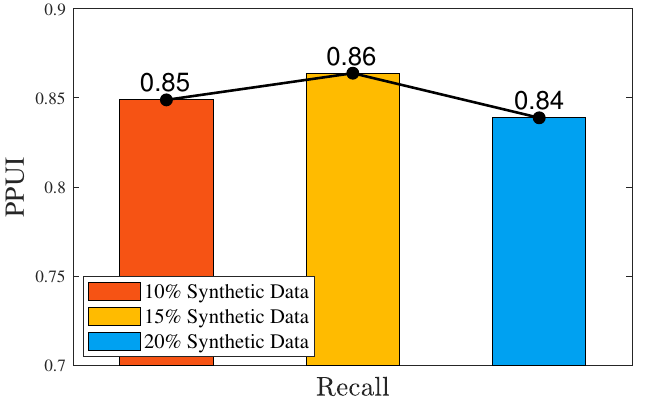}
	\caption{Privacy-preserving utility index performance with different percentage of synthetic data for training dataset \cite{refIII_Confidentiality_Network_supply}.}
	\label{GDMTD3_p3}
\end{figure}

Fig. \ref{GDMTD3_p2} illustrates the performance of YOLOv3 when trained with varying proportions of GDM-generated synthetic data. Increased integration of synthetic data correlates with a reduction in YOLOv3 detection accuracy. While synthetic data effectively mitigates privacy leakage from raw data, its inclusion inevitably compromises downstream task performance. 
To address this issue, Fig. \ref{GDMTD3_p3} introduces the privacy protection utility index (PPUI), quantifying the balance between privacy preservation and data utility. PPUI is calculated as a weighted sum of the synthetic data proportion in the training dataset and the average accuracy of the downstream task model, both normalized to the range [0, 1]. 
The index demonstrates an inverse relationship with synthetic data proportion: higher proportions enhance privacy protection but degrade model accuracy, thereby reducing PPUI. The strategy maximizing PPUI achieves an optimal privacy-accuracy equilibrium.

\subsection{Lessons Learned}
GDM-driven privacy-preserving mechanisms that generate high-fidelity synthetic data for high-altitude platform remote sensing data, UAV swarm cooperative path data, and SAGIN resource scheduling data. 
It generates statistical attributes that are consistent with, but free of, privacy properties. 
It can effectively deal with the privacy preservation challenges encountered in encryption \cite{refIII_Eva_8}, differential privacy \cite{refII_B_1_16}, and FL-based traditional AI \cite {refIII_Confidentiality_MAC_4} \cite{refIII_Confidentiality_MAC_5}.

\section{Tutorial 3: GAI for Multi-Modality Semantic-Aware Communications in SAGINs}
\subsection{Motivation}
Real-time secure transmission and defense requirements face great challenges in dynamic and vulnerable SAGIN environments. For example, high-resolution image transmission delays cannot meet millisecond security decision-making needs \cite{refII_B_GAN}. 
The static models employed in traditional AI require periodic retraining to adapt to dynamic threats in SAGINs, resulting in delayed responses to emerging risks \cite{refII_B_1_13} \cite{refII_B_1_14}. In addition, 
data generation and reconstruction based on a single textual prompt can deviate significantly from real-world scenarios (e.g., CNN-based image recognition)  \cite{refII_B_1_7}.

The GAI approaches enhanced by multi-modal data generation and semantic compression can effectively break through the above limitations to achieve real-time adaptive security communication and defense \cite{refIII_integrity_phy_11} \cite{refIII_jamming_14}. 
Semantic compression-based VAEs constribute to  probabilistic compression of SAGIN data, saving SAGIN resource. 
In cross-modal alignment, the text-image semantic space can be unified, eliminating semantic ambiguities caused by dynamic environment changes. 
GDM balances generation quality and latency by adjusting the number of denoising steps, which significantly improves environment adaptability.
A representative GAI-based multi-modality semantic-aware framework is proposed in \cite{refII_B_GAN} to promote the service quality of GAI for SAGINs. In this case, 
text and image data are exploited to create multimodal content to provide more reliable and secure communications for vehicle networks, mainly including GAI-assisted skeleton-semantic co-generation and DRL-assisted co-optimization.

\subsection{System Description}
We take the vehicle-to-vehicle (V2V) networks as an example to demonstrate the advantages of the GAI-based multi-modality semantic-aware framework.
The Fig. 1 in \cite{refII_B_GAN} illustrates
the GAI-enabled V2V networks, where multi-camera systems are employed to capture real-time road imagery for safety applications, such as collision alerts during accidents.
Other major applications include navigation and route optimization, insurance and risk assessment, traffic simulation and prediction, and driving data generation.
However, conventional generative models relying solely on textual descriptors exhibit high uncertainty, often generating inaccurate visual reconstructions (e.g., misrepresenting accident scenes or contextual details like license plates). 
Therefore, while the application of GAI in V2V shows great potential, it also faces critical challenges, including real-time data processing and decision making and adapting to dynamic and unpredictable environments.
These unreliability challenges necessitate a robust generative architecture capable of preserving semantic fidelity to ensure safety-critical decision-making in V2V networks.

\subsection{GAI-Based Solutions}
\begin{figure}[!t]
	\centering
	\includegraphics[width=2.8in]{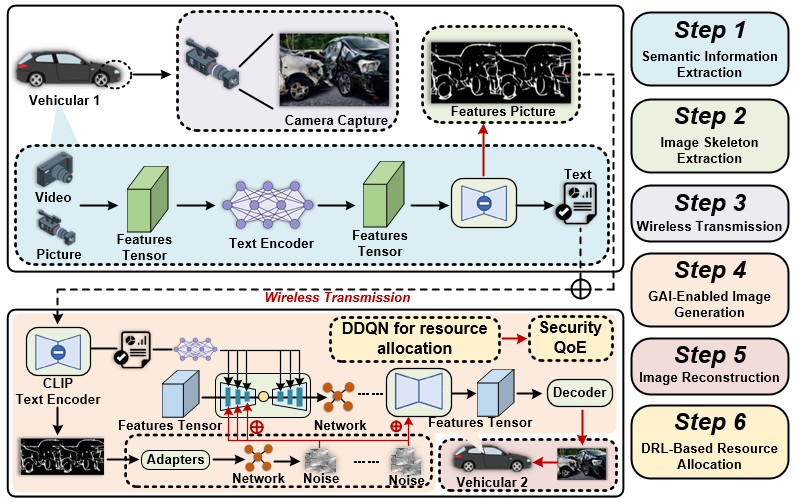}
	\caption{Proposed GAI-enabled multi-modality semantic-aware communication framework for V2V networks\cite{refII_B_GAN}.}
	\label{TUTORIAL3}
	\vspace{-1.5em}
\end{figure}

We conduct a multimodal semantic-aware GAI framework for V2V networks, leveraging complementary data modalities (text and images) to enhance situational awareness. 
Multimodal fusion mitigates ambiguities inherent in unimodal data, yielding superior environmental perception. 
As depicted in Fig. \ref{TUTORIAL3}, semantic skeletons derived from textual and visual inputs enable generative models to produce highly accurate reconstructions of road scenes.
In addition, By fusing textual semantics and structural skeletons, this framework establishes tamper-evident environmental fingerprints, ensuring trustworthy perception against malicious attacks in  \cite{refIII_Availability_MAC_1} and \cite{refIII_Availability_MAC_3}.

\begin{itemize}
	\item[$\bullet$] \textit{Step 1: Semantic information extraction:} Textual semantics are extracted from road images via a secure encoder, detecting objects (vehicles, pedestrians) and attributes (position, trajectory). 
	
	\item[$\bullet$] \textit{Step 2: Image skeleton Extraction:} A streamlined structural skeleton is derived by identifying salient edges and contours within the road scene. This compact representation preserves critical topological features, enabling efficient downstream processing (e.g., object recognition, lane detection) while discarding superfluous details.
	
	\item[$\bullet$] \textit{Step 3: Wireless transmission:} Semantic text and structural skeletons are fused into a lightweight data package for V2V broadcast. This approach reduces bandwidth consumption by $>$ 90\% compared to raw image transmission, ensuring reliable delivery of essential environmental data under constrained network resources.
	
	\item[$\bullet$] \textit{Step 4: GAI-enabled image generation:} A generative model synthesizes road scenes by integrating the structural skeleton (spatial foundation) with semantic descriptors (contextual attributes). This dual-input approach ensures high-fidelity reconstructions across diverse scenarios (e.g., weather variants, traffic densities). Model efficiency is tunable via diffusion-step optimization. In addition, this enhanced GAI can be utilized to generate adversarial scenarios to proactively identify hazards \cite{refIII_Availability_MAC_3} and contribute the traffic real-time generation in \cite{refIII_Authenticity_network_1}. Cross-modal validation detects synthesized deepfakes targeting SAGIN perception.
	
	\item[$\bullet$] \textit{Step 5: Image reconstruction:} 
	Onboard intelligence compares GAI-reconstructed scenes with real-time feeds. Potential hazards (e.g., obstacles, accidents) are identified through real-time scene comparison, triggering multimodal (auditory/visual) alerts to prompt driver intervention. The proposed framework can alleviate the problems of insufficient accuracy and real-time performance in \cite{refIII_integrity_phy_12}, \cite{refIII_integrity_phy_14}, and \cite{refIII_integrity_phy_17}.
	
	\item[$\bullet$] \textit{Step 6: DRL-based resource allocation:} For the proposed GAI framework, we propose a double deep Q-network (DDQN)-based approach to optimize the security quality of experience (QoE) in V2V networks within the constraints of the transmission power budget and the probability of successful transmission for each vehicle. More specific process can be found in \cite{refII_B_GAN}.

\end{itemize}
\vspace{-1em}
\subsection{Numerical Results}
\begin{figure}[!t]
	\centering
	\includegraphics[width=2in]{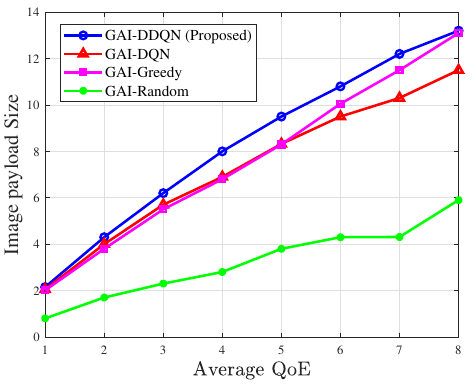}
	\caption{Average QoE of different methods versus image payload size \cite{refII_B_GAN}.}
	\label{TUTORIAL3_plot}
	\vspace{-1em}
\end{figure}

We evaluate the performance of the proposed GAI-DDQN algorithm for multi-modality semantic-aware framework. In Fig. 4 (a) of \cite{refII_B_GAN}, the proposed DDQN strategy shows superior convergence dynamics versus benchmarks (e.g., DQN-based, greedy-based, and random-based GAI approaches). 
In addition, as shown in Fig. \ref{TUTORIAL3_plot},
we analyze the variation of QoE with the size of image payload.
The QoE metric, evaluated per timestep in unconstrained environments, quantifies generative image fidelity critical for security-sensitive perception. 
The DDQN-based approach consistently outperforms benchmarks, demonstrating its efficacy in adversarial SAGIN conditions. Crucially, increased image payload enhances system QoE by conveying richer environmental signatures, enabling more robust anomaly detection against spoofing or data tampering attacks in \cite{refIII_integrity_Network_1}, \cite{refIII_Availability_Network_1}, and \cite{refIII_Availability_Network_4}. This payload-QoE synergy directly strengthens spatial-temporal integrity verification across ground-air-space links.

\vspace{-1em}
\subsection{Lessons Learned}
The transmission delays impeding real-time security decisions in SAGINs, and traditional AI fails to address due to their latency and single-modality inaccuracies.
The proposed GAI-based multi-modality semantic-aware framework overcomes these limitations through cross-modal fusion and semantic compression, enabling tamper-evident environmental fingerprints and adaptive diffusion-step optimization for real-time threat response. It can effectively enhance cross-domain spoofing detection in \cite{refIII_A_1_10}, data tampering resilience in \cite{refIII_integrity_Network_1}, and integrity verification of SAIGN links in \cite{refIII_integrity_MAC_4}. 
Ultimately, the co-design of generative semantics and DRL resource allocation establishes a new paradigm for security-assured dynamic perception in safety-critical SAGIN operations.
\vspace{-1em}
\section{Open Issues and Future Research Directions}
This section explores the open issues and potential research directions for the application of GAI in SAGIN security.
\vspace{-1em}
\subsection{Resource Constraints and Lightweight Deployment Issues}
The high computational overhead inherent in current GAI models is in fundamental conflict with the limited resources (i.e., computing power and energy) of the SAGIN nodes. Existing compression techniques (e.g., knowledge distillation) often struggle to balance real-time security reasoning and threat detection accuracy with dynamic topologies. 
Advancing theoretical efficiency boundaries for edge-deployed models remains imperative.
\subsubsection{Neural Architecture Search (NAS)-Driven GAI Model Compression}
NAS leverages ML methods to automate the neural network architecture design in a data-driven manner \cite{direction_1}, eliminating extensive manual effort required to explore vast network structure spaces for optimal task-specific efficiency. 
Within SAGINs, NAS can automatically generate lightweight GAI models with high compression ratio and low latency (e.g., adversarial sample generator and anomalous traffic synthesiser) to run real-time threat detection directly on satellite/IoT devices, reducing cloud dependency.
Novel NAS frameworks should automate compression of GAI models for SAGIN-edge devices, optimizing layer pruning and quantization while preserving anomaly detection accuracy.
\subsubsection{MoE-Driven Lightweight GAI Model}
The MoE-driven GAI addresses SAGIN deployment challenges through modular task-specialization \cite{refII_B_15}. 
Its dynamic routing mechanism activates only relevant lightweight expert modules (e.g., compact models for generating defensive data), significantly reducing computation load on satellites. 
The modular design facilitates cross-platform adaptation and independent updates, enabling collaborative threat analysis without raw data transfer. 
This architecture delivers complex security capabilities to resource-constrained nodes through elastic deployment.
\subsubsection{Edge-Cloud Collaborative Generative Frameworks} These frameworks mitigate resource constraints through layered task offloading and semantic communication \cite{refI_B_9}. 
Edge devices execute lightweight submodules (e.g., compressed diffusion models) for localized tasks, while dynamically offloading compute-intensive steps (e.g., multi-step denoising) to the cloud. 
Transmitting latent features instead of raw data significantly reduces bandwidth consumption.
Federated knowledge distillation facilitates continuous compression of cloud-trained large models into micro-experts deployed at the edge, supporting offline generation of security decoys. 
\vspace{-1em}
\subsection{Adversarial Robustness and Trustworthy Mechanisms Issues}
Current GAI-driven SAGIN security defenses are vulnerable to dynamically evolving adversarial attacks.
Static defence models cannot counter adaptive attackers such as ``evolutionary eavesdroppers", which
continuously analyze defenses and adjust attack strategies in real-time, utilizing the static nature of defenses and the flaws of black-box decision-making.

\subsubsection{Explainable Generative Frameworks}
Explainable generative frameworks are expected to address the opaque decision-making of black-box models in SAGIN security \cite{refII_B_1_1}.
These frameworks employ multi-modal explanation techniques to trace adversarial samples through cross-domain feature attribution.
In addition, domain knowledge constraints, such as communication protocols and physical-layer features, are embedded into generative models to restrict output spaces and suppress adversarial perturbations.
Robustness verification tools mathematically certify explanation integrity against attacks, ensuring auditability for critical missions.
Integrated into detection pipelines, they provide interpretable alerts through XAI interfaces.
\subsubsection{Digital Twin-GAI Fusion for Proactive Defense}
Digital twin (DT)-GAI synergy establishes a dynamic cyber-physical testing environment for SAGIN resilience \cite{direction_2}.
Using GANs, the DT synthesizes high-fidelity attack scenarios by leveraging historical attack data and threat intelligence.
The DT to iteratively optimizes defense policies through millions of simulated attack cycles, employing techniques like OpenAI's ``red teaming" to stress-test robustness. 
Crucially, concept drift adaptation mechanisms inspired by ``digital cousins" that generalize physical environments—enable extrapolation to unseen attack variants, enhancing cross-domain generalization. 
\subsubsection{Continual Learning-Driven GAI for Sustainable Adaptive Defense}
This framework mitigates catastrophic forgetting in SAGINs by synergizing adversarial robustness with trustworthy mechanisms through continual learning \cite{direction_3}. 
Federated continual learning enables edge devices to locally train adversarial detectors on emerging cross-domain threats. 
Stable knowledge retention leverages Hessian-based regularization to freeze critical weights from prior tasks, mathematically certifying that new adaptations do not degrade defenses against known threats, thus ensuring robustness continuity. 
This end-to-end approach bridges adaptive defense and trustworthy operation for SAGINs' dynamic threat landscape.
\vspace{-1em}
\subsection{Cross-Domain Coordination and Governance Compliance Issues}
In SAGINs, GAI-enabled cross-domain defense faces ``governance-efficiency paradox": multi-source heterogeneous data integration challenges and governance-compliance barriers. 
Cross-domain attacks (e.g., ``LEO relay with UAV poisoning") require fused information, but inter-domain data sovereignty conflicts and regulatory fragmentation create ``governance silos"
among satellite/aerial/ground nodes, hindering security policy synchronization.

\subsubsection{Semantics-Driven Policy Generation}
Multilevel generative models establish an end-to-end semantics-to-policy mapping framework to resolve defense policy fragmentation challenges in cross-domain SAGIN  \cite{refII_B_8}. 
LLMs such as GPT-4, parse threat descriptions and extract semantic feature vectors, and GANs map semantic vectors to specific defense rules. 
In addition, a predefined security policy ontology library can also be constructed to map heterogeneous protocol semantics—including satellite authentication rules, airborne encryption policies, and ground access control policies—into a unified framework.
Through FL, GAI aggregates localized policy features from each domain and synthesizes cross-domain coordinated policies based on semantic similarity. 
This approach unifies heterogeneous protocol semantics, enabling panoramic attack chain cognition.

\subsubsection{Neuro-Symbolic Quantum-Driven GAI Architectures}
This direction fuses the dynamic learning capability of neural networks with the interpretable rule engine of symbolic systems , combining quantum computing to accelerate the cross-domain verification process and crack the governance compliance and real-time contradiction \cite{direction_4}.
It can be achieved through three key capabilities: 
(1) A neuro-symbolic layer encoding governance rules (e.g., spectrum allocation) for interpretable compliance while learning multi-domain threats;
(2) A GAI module simulates cross-domain conflicts (such as SAGIN resource contention) to draft real-time protocols and optimize scheduling; 
(3) A quantum engine accelerates multi-domain policy verification, slashing audit time, and concurrently secures communications via quantum encryption.
This integration enables dynamic cross-domain policy alignment and real-time  collaborative governance.

\section{Conclusion}
This review presents an overview of GAI-enabled secure communications for SAGINs, emphasizing the enhanced effectiveness of GAI compared to traditional AI in protecting SAGINs. 
This analysis thoroughly examines the architecture of SAGINs and the specific security challenges faced. 
The efficacy of different GAI models in addressing security issues has been examined. 
This survey provides a comprehensive analysis of GAI-based methods addressing authenticity failures, confidentiality breaches, integrity tampering, and availability disruptions in SAGIN communications.
The three tutorials conducted provide a detailed exploration, highlighting the superior efficiency of GAI in addressing security threats in SAGINs relative to traditional AI.
We have identified open issues and associated research directions, offering insights into the future of GAI-enabled SAGIN security.


\newpage

%
%
%
%
%

\vfill

\end{document}